\documentclass[12pt,notitlepage,a4paper]{article}

\pdfoutput=1

\usepackage{a4wide}
\usepackage{epsfig}
\usepackage{color,graphicx}
\usepackage{float}
\usepackage{cite}
\usepackage{amssymb,amsmath}
\usepackage{breqn}
\usepackage{jheppub}
\usepackage{ytableau}
\ytableausetup{boxsize=0.7em, centertableaux}

\newcommand{\be}{\begin{equation}}
\newcommand{\ee}{\end{equation}}
\newcommand{\bea}{\begin{eqnarray}}
\newcommand{\eea}{\end{eqnarray}}
\newcommand{\SO}{\text{SO}}
\newcommand{\SU}{\text{SU}}
\newcommand{\U}{\text{U}}
\newcommand{\USp}{\text{USp}}

\usepackage{tikz}
\usetikzlibrary{decorations.markings, arrows.meta}
\tikzset{
    mid arrow/.style={
        postaction={decorate},
        decoration={
            markings,
            mark=at position 0.5 with {\arrow{Stealth}}
        }
    }
}

\begin{document}

\begin{center}  

\vskip 2cm 

\begin{center}
{\Large\bfseries\boldmath Dualities, symmetries and the elliptic genus\\ of 2d $(0,2)$ SCFTs\par}
\end{center}
\vskip 1cm

\renewcommand{\thefootnote}{\fnsymbol{footnote}}

   \centerline{
	  {\large \bf Salomon Malka${}^{a}$} \footnote{salomonmalkah@gmail.com}
     {\bf and}
    {\large \bf Gabi Zafrir${}^{b,c}$}\footnote{gabi.zafrir@oranim.ac.il}}

\vspace{1cm}
\centerline{{\it ${}^a$ Department of Physics, Technion, Israel Institute of Technology}} 
\centerline{{\it Haifa, 32000, Israel}}
\centerline{{\it ${}^b$ Department of Physics, University of Haifa at Oranim, Kiryat Tivon 36006, Israel}}
\centerline{{\it ${}^c$ Haifa Research Center for Theoretical Physics and Astrophysics, University of Haifa,}}
\centerline{{\it Haifa 3498838, Israel}}
\vspace{1cm}

\end{center}

\vskip 0.3 cm

\setcounter{footnote}{0}
\renewcommand{\thefootnote}{\arabic{footnote}}   
   
\begin{abstract}

While naively simpler than its higher dimensional cousins, 2d physics still has its own share of puzzles. An example of this is instances of dualities of $(0,2)$ theories, passing all consistency checks save for a mismatch in global symmetry. This is common when said duality can be motivated by dimensional reduction of a duality of 4d $\mathcal{N}=1$ theories where said symmetry emerges in 2d, often from an anomalous 4d symmetry. It is commonly believed that these symmetries are broken in the 2d theory, though often the mechanism for such breaking is not understood. We examine various appearances of this phenomenon and show that in these cases the elliptic genus shows signs of the presence of marginal operators that can break these symmetries. Interestingly, these operators do not appear perturbatively but emerge from the calculation, raising the question of what operators the $(0,2)$ elliptic genus actually counts. This provides a potential mechanism to overcome the discrepancy in symmetry, though we point out some potential subtleties.      

\end{abstract}
 
 \newpage
 
\tableofcontents

\section{Introduction}
\label{sec:intro}

An important ingredient of any quantum field theory is its spectrum of local operators, which provides the basic observables in the theory. In more formal approaches, like axiomatic quantum field theory, the spectrum of local operators and the algebra they form is used to describe the theory. In conformal field theory, the spectrum of operators, their scaling dimensions and OPE coefficients provide the basic data that define the CFT. As such determining the operator spectrum of a given theory already provides us with interesting data about it.

Understanding the local operator spectrum of a theory can be a non-trivial task, even if the theory is given in terms of a Lagrangian, for a variety of reasons. First, certain properties of operators, like scaling dimension, can change due to quantum corrections, and can be difficult to determine if the theory reaches strong coupling. Second, many QFT systems allow for disorder operators, meaning that the full operator spectrum of the theory can be larger than what is naively expected from the Lagrangian.

The situation is easier in supersymmetric theories. In this case one can concentrate on the special case of BPS multiplets. These are short representations of the supersymmetry algebra and in general the shortening condition provides restrictions on their charges. Still, counting the full spectrum of BPS operators can be quite involved, as short representations can merge to form longer ones, a process that can occur as the couplings or energy scale is varied. In superconformal theories, these problems can be alleviated by considering the spectrum of BPS operators up to possible merging to form long representations. This is usually addressed by the superconformal index \cite{Kinney:2005ej,Bhattacharya:2008zy}, which is a counting of all the BPS operators in the theory, performed such that combinations of short representations that can form a long representation sum up to zero. As such, the superconformal index holds ample information regarding the spectrum of BPS operators and their representation under the global symmetries of the theory. Furthermore, as the BPS spectrum can only change through the merger of short representations to long ones, the index is invariant under continuous transformations, and notably under RG flow. This makes it readily computable if the theory has a weakly coupled description, and the result then holds even at strong coupling. This also makes the index quite useful for understanding the strong coupling behavior of the theory.

In general the index can be computed by counting the BPS operators of the theory. Alternatively, the index can be represented as the partition function of the theory on $S^{d-1} \times S^1$, which is then amenable to computation through localization. In most cases, both approaches can be done and give compatible results, assuming one can identify all BPS local operators. The latter often involves non-perturbative contributions like monopole operators in 3d and instanton operators in 5d. There is however an exception to this, the case of two dimensions and the elliptic genus. The computation of the elliptic genus of a 2d $(0,2)$ gauge theory has been worked out using localization in \cite{Benini:2013nda,Benini:2013xpa}, but the resulting expression differs subtly from the naive counting of perturbative operators. However, it is not exactly clear what the perturbative counting is missing. As we review below, the difference is not in the presence of additional sectors, but rather in a change of contour. This raises the question of what are the additional operators that the elliptic genus counts but are not apparent perturbatively? How can these be understood directly from the gauge theory?

As highlighted above, the operator spectrum is one of the fundamental properties of a QFT system. The puzzle introduced in the preceding paragraph suggests that we do not properly understand the BPS spectrum of 2d $(0,2)$ Lagrangian theories. While this is observed in supersymmetric theories, it is possible that these additional operators also appear in non-supersymmetric theories, in which case this puzzle is relevant to more general 2d CFTs. It is then natural to seek to remedy this by better understanding the operator spectrum of 2d $(0,2)$ gauge theories.  

One can further wonder whether this discrepancy has interesting physical manifestations. A natural place to look for such is in dualities as these often map perturbative operators to non-perturbative ones. Interestingly, there are interesting puzzles that have been observed in dualities of $(0,2)$ gauge theories that descend from the $S^2$ reduction of a dual pair of 4d $\mathcal{N}=1$ gauge theories (see for instance \cite{Gadde:2015wta,Amariti:2024usp,Amariti:2025jvi}). Specifically, given a pair of dual theories in 4d, one can reduce both on $S^2$, with a topological twist to preserve $(0,2)$ SUSY in 2d, to get a dual pair of 2d gauge theories. The technical details were worked out in \cite{Gadde:2015wta} using the reduction of the 4d $S^2\times T^2$ partition function on the $S^2$. The duality of the original theories then guarantees that their 't Hooft anomalies and $S^2\times T^2$ partition functions agree. The 't Hooft anomalies of the 2d theory should be determined by integrating the anomaly polynomial of the 4d theory on the $S^2$ \cite{Benini:2009mz,Alday:2009qq,Hosseini:2020vgl}, and the $S^2\times T^2$ partition function should reduce to the elliptic genus of the 2d theory. This then suggests that the 't Hooft anomalies and elliptic genus of the proposed 2d dual pair agree, as expected if they are dual.

Nevertheless, there is a subtlety here. Sometimes the 2d theory manifests additional symmetries not observed in 4d. This happens for instance when a symmetry is broken by an ABJ anomaly in 4d and then emerges in 2d where the anomaly does not exist. Alternatively, certain fields in 4d may reduce to multiple fields in 2d, in which case there is a symmetry rotating the multiple copies that did not exist in 4d. The 't Hooft anomalies of these symmetries and the charges of operators under them are not constrained by the properties of the 4d theory\footnote{The $so(3)$ isometry of the sphere sometimes manifests in the 2d theory as a subgroup of the symmetry rotating multiple 2d copies of the same 4d field. Certain properties of this symmetry, like 't Hooft anomalies can be inferred from the properties of the 4d theory so it can be partly regarded as coming from the dimensional reduction.}, and often the anomalies involving said symmetries and elliptic genera refined by them do not match between the expected 2d dual theories. This then seems to put the duality proposal in peril. Similar behavior has also been observed when one reduces dual pairs of 4d $\mathcal{N}=1$ gauge theories on a circle to 3d, but there it can be circumvented by the introduction of monopole operators into the superpotential that break the additional symmetries \cite{Aharony:2013dha}. A natural expectation then is that something similar also occurs in the 2d case, but often an operator with the necessary charges is not apparent in the spectrum of the 2d theory. This results in a puzzle engulfing these proposed 2d dualities\footnote{There is a parallel story in the circle reduction of 3d supersymmetric theories to 2d where a similar apparent mismatch in symmetry emerges in 2d \cite{Aharony:2017adm}.}.

The above discussion now suggests that the two puzzles may be related. Specifically, that there are indeed additional 2d local operators which are not apparent from the perturbative spectrum and which break the additional 2d symmetries obstructing the duality. In this way the duality should really be understood as between the 2d theories deformed by these additional superpotentials, mimicking the 3d discussion in \cite{Aharony:2013dha}. Here we explore this issue by studying examples where this case arises and comparing their elliptic genus computed through the localization prescription against the naive perturbative counting. We show that in many cases there is indeed an additional operator which has the right charges to break the additional 2d symmetries, while preserving the symmetries originating from 4d. This provides evidence that the 2d duality indeed holds and should be thought of as between the theories deformed by the superpotential.

An important question is how to identify whether a given operator preserves the 4d symmetries, notably $(0,2)$ supersymmetry and the R-symmetry originating from 4d. Here we use the fact that such an operator should appear as a marginal operator when the 4d R-symmetry is the superconformal one. Generally the index is a counting of the BPS operators in the theory, and as marginal operators are also BPS, they contribute to the index. This allows their identification from certain contributions to the index. In theories with 4 supercharges, there is an interesting relation between marginal operators and conserved currents, where a marginal operator can become marginally irrelevant by absorbing a conserved current multiplet. This is manifested in the index as both operators contribute with the same fugacities under the superconformal symmetries but with opposite signs. In 3d and 4d, one can show under certain reasonable assumptions\footnote{Notably no higher spin currents.} that these two multiplets are the only ones contributing with this combination of fugacities of the superconformal symmetries, which then allows for their identification from the index \cite{Beem:2012yn,Razamat:2016gzx}. We show that similar behavior holds for the elliptic genus of $(0,2)$ theories, where marginal operators and conserved currents contribute similar terms but with opposite signs. For $(0,2)$ theories though, the identification is not unique and there can be other operators with the similar contributions that are not marginal operators or conserved currents. For $(2,2)$ theories the situation seems more similar to the 3d and 4d case though we do not prove it. 

The above then is used to identify operators that allow for the breaking of the accidental 2d symmetries while preserving the 4d one. Specifically, we look for an operator in the elliptic genus whose contribution is similar to a marginal operator and which carries charges only under accidental 2d flavor symmetries\footnote{We note that the contribution needs to be similar to that of a marginal operator when the elliptic genus is evaluated with the 4d R-symmetry. However, this R-symmetry is usually computationally inconvenient so we exploit the fact that the operator should be uncharged under the 4d flavor symmetry and mix them with the 4d R-symmetry to get a more convenient one. It is important though not to mix the 2d accidental symmetries in the R-symmetry used to compute the elliptic genus compared to the 4d one.}. This operator then has the right charges to break the accidental symmetries while preserving the ones originating from the 4d theory. We do note several things though. First, while this is consistent with the operator being a marginal one it does not prove that it is, as there might be other operators with similar contributions, and even if it is marginal it might be marginally irrelevant. Finally, the above only argues that the operator might be marginal if the R-symmetry used in the elliptic genus is the superconformal one. However, there can be mixing with the accidental 2d symmetries which then changes the scaling dimension of the operator at the IR fixed point. In fact, we actually expect this to happen such that these operators are actually relevant. In that case the deformation leads to a new fixed point, where the accidental symmetries are broken, and it is this fixed point that is identified under the dualities. One can use c-extremization \cite{Benini:2012cz,Benini:2013cda} to determine the R-symmetry at the fixed point, under certain assumptions, and try to check whether this is indeed the case. Unfortunately, c-extremization is subtle in theories with non-compact moduli which all our examples possess. Taking this into account, we indeed find that in all examples the accidental 2d symmetries mix with the R-symmetry but in such a way that these operators become irrelevant. If that is the case then deforming by them would just lead us back to the same fixed point where the accidental symmetries reemerge. As such, we cannot claim to fully resolve the puzzle engulfing these 2d dualities. We find the presence of these operators to strongly suggest that the duality does work and that perhaps the c-extremization results are too naive, given the subtlety of the non-compact moduli. We hope to return to this issue in future work.   

Overall, the results observed here appear to support the prevailing conjecture that the $(0,2)$ theories participating in the dualities should be modified by the addition of a superpotential of non-perturbative origin (modulo the subtlety surrounding the c-extremization procedure). Besides the interest in the fate of the dualities and the implication to the dimensional reduction of dualities, this is also of interest for the dynamics of the 2d theories themselves. The results suggest the presence of non-perturbative operators whose 2d interpretation is not understood. It would be of considerable interest to understand the origin of these operators in 2d, and the results observed here regarding their charges in certain models might be of use in this endeavor. Finally, we note that even though the discussion here was concerned with $(0,2)$ supersymmetric theories, this issue of understanding the operator spectrum of 2d theories might also be relevant for less supersymmetric 2d theories.  


The structure of the paper is as follows: in section \ref{sec:EGa2dI} we give a review of the elliptic genus and its physical interpretation as a counting of operators. In \ref{sec:examples}, we compute the elliptic genus on a wealth of examples where the aforementioned puzzle is expected. By comparing the elliptic genus with the result expected perturbatively, we isolate the non-perturbative contributions. We find that these non-perturbative contributions provide an avenue for explaining the apparent symmetry mismatch across the different dualities. Finally, several technical details regarding the computations are discussed in the appendices.  

\section{Elliptic genus and the 2d superconformal index}
\label{sec:EGa2dI}

The elliptic genus refers to the partition function of the 2d theory on $S^1 \times S^1$. Here we are mainly interested in supersymmetric theories, where this object also plays the role of the superconformal index. As such, it is a counting of the BPS operators in the theory, up to recombination, and so holds significant information regarding the spectrum of short representations of the theory. Additionally, it is invariant under continuous transformations, notably RG, and so is quite useful in understanding strong coupling behavior.

When thought of as the superconformal index (or as the Witten index), the partition function is evaluated with periodic boundary conditions for the fermions on the temporal circle. Unlike in higher dimensions, in 2d we also have the spatial circle and need to choose boundary conditions for it as well. One option is to give periodic boundary conditions for the fermions on the spatial circle, which is referred to as the Ramond sector, while another is to give anti-periodic boundary conditions, which is referred to as the NS sector. The resulting expressions are related to one another so it is not particularly important which one is chosen. Here we find it more convenient to work in the NS sector. 

For a $(0,2)$ superconformal 2d field theory the elliptic genus in the NS sector can be defined as \cite{Gadde:2013dda}:

\be \label{EGNSdef}
\mathcal{I}_{\text{NS}} = \text{Tr}_{\text{NS}} (-1)^F q^{H_L}\bar{q}^{H_R - \frac{1}{2}J_R}\prod_i a^{f_i}_i ,
\ee
where the trace is taken over the Hilbert space of the theory on a circle with anti-periodic boundary conditions for the fermions. Here $H_L$ and $H_R$ are the left and right moving dimensions, $J_R$ is the right moving R-charge and $f_i$ refer to charges under flavor symmetries. The $(-1)^F$ ensures that the fermions have periodic boundary conditions on the temporal circle.

The index is independent of $\bar{q}$ and so acquires contributions only from BPS states obeying $2H_R=J_R$. The fugacity $q$ keeps track of the left moving dimension while $a_i$ keep track of the flavor charges. We shall often expand the index in a power series in $q$, where the terms stand for various BPS operators in the theory.   

We note that a similar definition holds also for a $(2,2)$ SCFT, as the latter is also a $(0,2)$ SCFT, but now one can refine by an additional fugacity associated with the left-moving R-symmetry. This is usually done by inserting the term $y^{J_L}$ in the trace in \eqref{EGNSdef}. 

The elliptic genus can be computed through localization, see \cite{Gadde:2013dda,Benini:2013nda,Benini:2013xpa}. It is instructive to compare the computation to the one for the 4d superconformal index of an $\mathcal{N}=1$ gauge theory. In addition to the localization computation, the latter can be evaluated by an explicit counting of the operators (see \cite{Rastelli:2016tbz} for a review on the 4d superconformal index). Specifically, the main constituents of the gauge theory are chiral and vector multiplets. For each one we can list the components that contribute to the index and enumerate their contribution. This builds the so-called one particle index. More general operators can now be built by taking all possible products of the basic operators, while taking due note of their symmetry properties. This is generally achieved by taking the plethystic exponent of the one particle index. Finally one must project to gauge invariants, which is generically achieved by integrating over the gauge fugacities against the Haar measure. The resulting expression is known to agree with the exact evaluation done using localization. We can interpret this as the statement that in 4d "what you see is what you get": the 4d BPS (local) operators are all made of basic fields.

It is interesting to contrast this with the 3d case. Specifically, consider a 3d $\mathcal{N}=2$ gauge theory. As in 4d, we can attempt to count the spectrum of local operators by evaluating the contribution of the chiral and vector multiplets, and combine them in the one particle index. Following this, we take the plethystic exponential of the one particle index and project to gauge invariants. This results in the contribution of all BPS operators that can be built from the perturbative matter. However, this does not give the complete superconformal index of the theory \cite{Kim:2009wb,Imamura:2011su,Kapustin:2011jm}. In 3d gauge theories there are non-perturbative disorder operators one can build, called monopole operators, that are defined by inserting a magnetic flux around the location of the operator in some $\U(1)$ gauge sector. These operators can be BPS and contribute to the superconformal index. The correct expression for the index then involves a summation over all allowed monopole flux. The sector with zero monopole flux is equivalent to the spectrum of the perturbative local operators described previously, but this is only a part of the full answer which also contains the contribution of non-perturbative operators: here monopole operators\footnote{There is a similar story in 5d. The index of a 5d $\mathcal{N}=1$ gauge theory is also given by a sum, now over the allowed instanton number of the gauge group. The zero sector of the sum again agrees with the perturbative contribution, but needs to be supplemented by the non-perturbative contribution \cite{Kim:2012gu}. These are described by disorder operators, called instanton operators, that similarly to monopole operators are defined by the insertion of instanton flux surrounding their location \cite{Lambert:2014jna}.}.  

Next we turn to our main case of interest: 2d. We can treat the 2d case in a similar fashion as in 4d, by counting the spectrum of local operators \cite{Gadde:2013dda,Dimofte:2017tpi}.  Specifically, a $(0,2)$ (or $(2,2)$) gauge theory is made of chiral, Fermi and vector multiplets (chiral, twisted chiral and vectors in the $(2,2)$ case). For each one we can write its contribution to the elliptic genus, building the one particle index. The plethystic exponential of that will generate the multi-particle index, and finally integrating over the Haar measure should project to gauge invariants resulting in the elliptic genus.       

To illustrate this consider a $(0,2)$ theory. The basic multiplets are the chiral and Fermi multiplets. A free chiral contains a complex scalar $\phi$ of dimension $(0,0)$ and R-charge $0$ and a right moving fermion $\psi_+$ of dimension $(0,\frac{1}{2})$ and R-charge $-1$. One can see that $\phi$, $\overline{\phi}$ and $\overline{\psi}_+$ obey the shortening condition $2H_R=J_R$. Additionally, we can act on the operator with the derivatives $\partial_+$ and $\partial_-$. These would raise the corresponding dimension by $1$ but leave the R-charge unchanged. As such, we see that acting with $\partial_-$ preserves the shortening condition, while acting with $\partial_+$ does not. Finally, we recall that the free fields obey the Klein-Gordon and Dirac equation, $\partial_+ \partial_- \phi = 0$ and $\partial_- \psi_+ = 0$, respectively. The Dirac equation in particular then implies that only the zero modes of $\overline{\psi}_+$ contribute to the elliptic genus. In other words only the state $\overline{\psi}_+$ will contribute and there are no contributions from the states $\partial_-^n \overline{\psi}_+$ as they are vanishing. Finally, we can also take arbitrary products of the above operators. This amounts to taking at the end the Plethystic exponent of the one-particle index. For convenience, we summarize in the following table, the contribution for every state of the free chiral multiplet to the one-particle index:

\begin{center}
\begin{tabular}{c|c c c c c} 

 & $\phi$ & $\overline{\phi}$ & $\psi_{+}$ & $\overline{\psi}_{+}$ & $\partial_-$\tabularnewline
\hline 

$H_{L}$ & $0$ & $0$ & $0$ & $0$ & $+1$\tabularnewline

$H_{R}$ & $0$ & $0$ & $\frac{1}{2}$ & $\frac{1}{2}$ & $0$\tabularnewline

$J_{R}$ & $0$ & $0$ & $-1$ & $+1$ & $0$\tabularnewline

$f$ & $+1$ & $-1$ & $+1$ & $-1$ & $0$\tabularnewline

$H_R - \frac{1}{2} J_R$ & $0$ & $0$ & $1$ & $0$ & $0$\tabularnewline
\hline 
Contribution & $\sum_ {n\geq 0}q^{n}a$ & $\sum_ {n\geq 0}q^{n}a^{-1}$ & none & $-a^{-1}$ & $-$\tabularnewline

\end{tabular}
\end{center}

Summing it all up, and using \eqref{EGNSdef}, we find the contribution of a free chiral to be:

\be
\mathcal{I}^{(0,2)}_{\text{chiral}} = \text{PE}\left[\frac{a+\frac{q}{a}}{1-q} \right] = \frac{1}{\theta(a;q)} .
\ee

Here $\text{PE}$ stands for the plethystic exponent and we use the fugacity $a$ for the flavor $\U(1)_a$ symmetry acting on the chiral multiplet. The function $\theta(a;q)$ is defined as: $\theta(a;q) = (a;q)_\infty(\frac{q}{a};q)_\infty$ where $(a;q)_\infty$ is the $\infty$-Pochhammer symbol, $(a;q)_\infty = \prod^{\infty}_{i=0} (1-a q^i)$. 

Similarly, a free Fermi multiplet contains a left moving fermion of dimension $(\frac{1}{2},0)$ and R-charge $0$. More specifically, a free Fermi multiplet has the superspace expansion (see for instance \cite{Tachikawa:2018sae}):

\be
\Psi = \lambda_- + \overline{\theta}_+ E(\Psi) ,
\ee
where $E(\Psi)$ is some holomorphic function of the chiral fields of the theory. A non-trivial function $E(\Psi)$ then adds interactions that are usually referred to as E-terms. Here for the most part we shall assume that $E(\Psi)=0$ so that the Fermi multiplet contains only the left moving fermion $\lambda_-$ \footnote{Even if $E(\Psi)$ is non-zero, it can still be ignored as it should be equal to some function of the chiral multiplets which is already counted from their contribution.}.

As before, it is straightforward to see that the operators contributing to the elliptic genus are $\lambda_-$, $\overline{\lambda}_-$, their $\partial_-$ derivatives and all possible products of the above. The states and their contributions to the index are:

\begin{center}
    \begin{tabular}{c|c c c }

 & $\lambda_{-}$ & $\overline{\lambda}_{-}$ & $\partial_-$\tabularnewline

\hline 
$H_{L}$ & $\frac{1}{2}$ & $\frac{1}{2}$ & $+1$\tabularnewline

$H_{R}$ & $0$ & $0$ & $0$\tabularnewline

$J_{R}$ & $0$ & $0$ & $0$\tabularnewline

$F$ & $+1$ & $-1$ & $0$\tabularnewline

$H_R - \frac{1}{2}J_R$ & $0$ & $0$ & $0$\tabularnewline
\hline 
Contribution & $\sum_ {n\geq 0}-q^{n+\frac{1}{2}}a$ & $\sum_ {n\geq 0}-q^{n+\frac{1}{2}}a^{-1}$ & $-$\tabularnewline

\end{tabular}
\end{center}

From \eqref{EGNSdef}, we see that the free Fermi multiplet contributes to the elliptic genus as: 

\be \label{EGfreeFermi}
\mathcal{I}^{(0,2)}_{\text{Fermi}} = \text{PE} \left[-\frac{(a+\frac{1}{a})q^{\frac{1}{2}}}{1-q}\right] = \theta(a q^{\frac{1}{2}};q) ,
\ee
where we again use $a$ as the fugacity for a $\U(1)_a$ flavor symmetry, now acting on the Fermi multiplet.

Here we have used the free R-charge appropriate in the UV, but often the fields would acquire an R-charge in the IR from mixing of the R-symmetry with $\U(1)_a$. This can be taken into account in the elliptic genus by shifting $a \rightarrow a q^{\frac{r}{2}}$ for $r$ the IR R-charge of the field. This leads to the commonly cited expressions for the contribution of the two multiplets to the elliptic genus in the NS sector:

\be
\mathcal{I}^{(0,2)}_{\text{chiral}} = \frac{1}{\theta(a q^{\frac{r}{2}};q)} \; , \; \mathcal{I}^{(0,2)}_{\text{Fermi}} = \theta(a q^{\frac{r+1}{2}};q) . 
\ee


Finally, we need the contribution of the vector multiplet. This is a little trickier as the vector is not a free representation of a conformal theory. However, the gauge field can be conveniently represented by a field strength multiplet which in 2d is a Fermi multiplet of R-charge $1$. It has the following superspace expansion:

\be
W = \lambda_- + \overline{\theta}_+ (D + i F_{01}) ,
\ee
with $D$ a real scalar and $F_{01}$ the associated component of the field strength. Note that here we have an effective E-term that is independent of the other fields and so should also be counted. To find the contribution of the vector multiplet, we first regard $W$ as a free Fermi multiplet of R-charge $0$, where we can use \eqref{EGfreeFermi}, but we need to add the contribution of the E-term, $D + i F_{01}$. Note, that for a free Fermi multiplet the E-term has dimension $(\frac{1}{2},\frac{1}{2})$ and R-charge $-1$ and so its complex conjugate contributes to the elliptic genus. However, due to the equation of motion of $F$, only the zero mode of $F_{01}$ should contribute. We can then take the R-charge into account by setting $a \rightarrow q^{\frac{1}{2}}$. The states and their contributions are thus:

\begin{center}
\begin{tabular}{c|c c c c}
 
 & $\lambda_{-}$ & $\overline{\lambda}_{-}$ & $F_{01}$ & $\partial_-$\tabularnewline

\hline 
$H_{L}$ & $\frac{1}{2}$ & $\frac{1}{2}$ & $\frac{1}{2}$ & $+1$\tabularnewline

$H_R$ & $0$ & $0$ & $\frac{1}{2}$ & $0$\tabularnewline

$J_R$ & $0$ & $0$ & $1$ & $0$\tabularnewline

$R$ & $1$ & $-1$ & $0$ & $0$\tabularnewline

$H_R - \frac{1}{2}J_R$ & $0$ & $0$ & $0$ & $0$\tabularnewline
\hline 
Contribution & $\sum_ {n\geq 0}-q^{n+\frac{1}{2}}q^{\frac{1}{2}}$ & $\sum_ {n\geq 0}-q^{n+\frac{1}{2}}q^{-\frac{1}{2}}$ & $+q^{\frac{1}{2}}q^{-\frac{1}{2}}$ & $-$\tabularnewline

\end{tabular}
\end{center}

This then leads to the following expression for the contribution of the vector multiplet to the elliptic genus:

\be
\mathcal{I}^{(0,2)}_{\text{vector}} = \text{PE}\left[-\frac{(a+\frac{1}{a})q^{\frac{1}{2}}}{1-q} + \frac{q^{\frac{1}{2}}}{a} \right] \rightarrow \text{PE}\left[-\frac{2q}{1-q}\right] = (q;q)^2_\infty . 
\ee

In a non-abelian gauge theory the vector is in the adjoint of the gauge group and the expression becomes:

\be
\text{PE}\left[-\frac{2q}{1-q} \chi({\bf A}_{\mathfrak{g}})\right] = (q;q)^{2 r_{\mathfrak{g}}}_\infty \prod_{\alpha} (z^{\alpha}_i q;q)^{2}_\infty = (q;q)^{2 r_{\mathfrak{g}}}_\infty \prod_{\alpha} (z^{\alpha}_i q;q)_\infty (z^{-\alpha}_i q;q)_\infty, 
\ee
where we use $\chi[{\bf A}_{\mathfrak{g}}]=r_{\mathfrak{g}} + \sum_{\alpha} z^{\alpha}_i$ for the character of the adjoint representation of the gauge algebra ${\mathfrak{g}}$, with $r_{\mathfrak{g}}$ the rank of ${\mathfrak{g}}$ and $\alpha$ goes over the root lattice of ${\mathfrak{g}}$. Here we use $z_i$ as a collective notation for fugacities of the maximal torus of the gauge symmetry. In the last transition we used the fact that the adjoint character is real to rewrite the product.

Finally, we need to project to gauge invariants. This can be done by integrating the maximal torus over the Haar measure. This takes the form:

\be
 \frac{1}{|W|} \int\left[\prod_i\frac{d z_i}{2\pi z_i}\right] \prod_{\alpha} (1-z^{\alpha}_i)^2 ,
\ee
where $|W|$ stands for the order of the Weyl group. Since the vector is always accompanied by integrating over the Haar measure, the two are often combined and rewritten as:

\be
\mathcal{I}^{(0,2)}_{\text{vector+Haar}} = (q;q)^{2 r_g}_\infty \prod_{\alpha} (z^{\alpha}_i q;q)_\infty (z^{-\alpha}_i q;q)_\infty (1-z^{\alpha}_i)^2 = (q;q)^{2 r_g}_\infty  \prod_{\alpha} \theta( z^{\alpha}_i;q)
\ee

Finally, as the plethystic exponent is a linear function, given a $(0,2)$ gauge theory we just need to combine the contributions from the chiral, Fermi and vector multiplets, and integrate over the Haar measure. This should give the contribution to the elliptic genus from gauge invariant operators constructed using the basic fields in the theory.  

Before comparing with the localization result, we want to say a few words regarding the $(2,2)$ case. The elliptic genus of a 2d $(2,2)$ theory can be evaluated in the same manner by thinking of it as a $(0,2)$ one. However, a $(2,2)$ theory also has the left-moving R-symmetry, which is a global symmetry from the $(0,2)$ viewpoint. As such we can add an additional fugacity for it, and here we adopt the convention of \cite{Gadde:2015wta}, and use the fugacity $y$ for it. The contribution of the $(2,2)$ matter fields can now be worked out by combining the appropriate contributions of the $(0,2)$ multiplets. For completeness, we shall briefly summarize them here.

The basic matter fields of a $(2,2)$ gauge theory are chiral and vector multiplets. A chiral multiplet has equal left and right moving R-charges, say $(r,r)$, and decomposes into a $(0,2)$ chiral of the same R-charges, and a Fermi multiplet of R-charge $(r-1,r)$. As such, it contributes to the elliptic genus as:

\be
\mathcal{I}^{(2,2)}_{\text{chiral}} = \frac{\theta(a q^{\frac{r+1}{2}}y^{r-1};q)}{\theta(a q^{\frac{r}{2}}y^r;q)} . 
\ee

Similarly, the $(2,2)$ vector can be described by a twisted chiral multiplet of R-charges $(-1,1)$ and decomposes to a $(0,2)$ chiral of the same R-charges, and the $(0,2)$ vector. As such, it contributes to the elliptic genus as:

\be
\mathcal{I}^{(2,2)}_{\text{vector}} = (q;q)^{2 r_g}_\infty \prod_{\alpha} \frac{(z^{\alpha}_i q;q)_\infty (z^{-\alpha}_i q;q)_\infty}{\theta(z^{\alpha}_i q^{\frac{1}{2}}y^{-1};q)} . 
\ee

As before, one often combines the vector contribution with the Haar measure to the single expression:

\be
\mathcal{I}^{(2,2)}_{\text{vector+Haar}} = (q;q)^{2 r_g}_\infty \prod_{\alpha} \frac{\theta(z^{\alpha}_i;q)}{\theta(z^{\alpha}_i q^{\frac{1}{2}}y^{-1};q)} . 
\ee

Twisted chirals and vectors are described by similar expressions, but with $y \rightarrow \frac{1}{y}$. The contribution to the elliptic genus of the basic matter fields can now be evaluated by multiplying the total matter contribution and integrating over the maximal torus to project to gauge invariants. 

The computation outlined above should result in the list of BPS (local) operators one can build from the basic building blocks of the theory. However, unlike in the 4d case, in 2d this result differs from the exact result through localization performed in \cite{Benini:2013nda,Benini:2013xpa}. Furthermore, the results deviate in a peculiar way. Unlike in 3d there are no additional sectors, and the elliptic genus is still given by integrating over the contribution of the individual multiplets over the gauge fugacities exactly as outlined above. However, the contour of integration is different. When projecting to gauge invariants, one usually takes a contour around the unit circle for each gauge fugacity. Yet, the localization result uses a different contour that is deformed so to pick only certain poles determined through the Jeffrey–Kirwan (JK) residue prescription. 

It is easiest to describe the JK residue prescription when the gauge group has rank 1 and there is a single integral. In that case one can separate the charged matter into those carrying positive charges and those carrying negative charges. The JK prescription involves taking only residues associated with one set, say the positively charged chirals. The result should be independent of which set of chirals is chosen. The contour is slightly trickier for multiple integrals, that is when the gauge group has rank greater than one. In that case one introduces an auxiliary charge vector and takes only combinations of poles for which said vector lies in the positive cone. The resulting elliptic genus should be independent of the chosen charge vector.      

In general the evaluation through the JK prescription gives a slightly different result than the one from the projection to gauge invariants. This implies that the BPS spectrum of 2d operators is not made solely from products of the basic fields. Rather there are additional operators that enter through this difference in contour. 

Another notable implication of this difference is that the computation of the 2d elliptic genus generally requires a Lagrangian description. In contrast, in 4d, given the index of an SCFT one can immediately compute the index of more general theories containing said SCFT, notably through gauging certain continuous symmetries. However, this is not possible in 2d as the JK prescription, which is required to understand the contour one should use, requires knowledge of the charges of the basic fields. As such, giving the elliptic genus of a 2d SCFT is not sufficient to compute the elliptic genus of more general theories made by gauging continuous symmetries of said SCFT. Rather one needs to supplement additional information so that the contour can be determined.    

\subsection{Contribution of marginal operators and conserved currents to the elliptic genus}

For a 2d $(0,2)$ SCFT the elliptic genus plays the role of the superconformal index and as such should hold information about the BPS operators in the theory. This includes various operators of physical interest like conserved currents and marginal operators. An interesting question then is how can these be identified in the index. For example, in 4d and 3d conserved currents and marginal operators have distinct contributions to the superconformal index \cite{Beem:2012yn,Razamat:2016gzx}, which in many cases allow for their extraction from the index. Specifically, in both cases the operators appear as an entry of the form (marginal $-$ currents) at a certain order in the fugacities associated with the superconformal group ($p q$ for 4d and $x^2$ for 3d where we follow the conventions of \cite{Rastelli:2016tbz,Kapustin:2011jm}). That is, marginal operators and currents give similar contribution to the index but differ by sign with marginal operators contributing positively and conserved currents negatively.

There is an interesting reason for this behavior. In 4d $\mathcal{N}=1$ SCFTs it is known that a marginal operator can only become a long multiplet by absorbing a conserved current and in the process the current becomes non-conserved. This is manifested in the index by the two multiplets having similar contributions but with opposite signs. This behavior also leads to the statement of \cite{Green:2010da}, that a marginal operator can fail to be exactly marginal only if it breaks a global symmetry. Because of this the structure of the conformal manifold is generally given by a Kahler quotient of the marginal operators by the global symmetry. The same structure is also expected to arise in theories with similar susy structure, notably 3d $\mathcal{N}=2$ SCFTs leading to the above mentioned observation in the index. Interestingly, this statement is expected to hold also for 2d $(2,2)$ SCFTs \cite{Green:2010da}. This leads us to expect a similar structure in the elliptic genus of $(2,2)$ SCFTs. Next we shall investigate how marginal operators and conserved currents appear in the elliptic genus first for the $(0,2)$ case and then for the $(2,2)$ case.  

Consider a $(0,2)$ gauge theory. We can introduce interactions using a superpotential\footnote{It is also possible to introduce interactions by modifying the shortening condition of a Fermi multiplet, the so-called E-terms. We shall not consider the effect of these here.} which generically takes the form:

\be
W = \Psi J (\Phi) ,
\ee
with $\Psi$ a Fermi multiplet and $J$ a function of the chiral fields in the theory, here collectively denoted as $\Phi$. As usual, for an SCFT the relevancy of an operator under insertion to the superpotential is determined by its R-charge under the superconformal R-symmetry. Also note that the lowest component of the operator will be a fermion. As such, for an operator to be marginal it needs to be a Fermi multiplet of R-charge $1$. The lowest component of such a Fermi multiplet of R-charge $R$ should contribute to the elliptic genus as $-q^{\frac{R+1}{2}}$. Therefore, marginal operators in 2d $(0,2)$ SCFTs are expected to contribute to the elliptic genus as $-q$.

Next we move to conserved currents. These should be charged in the adjoint of the flavor symmetry they generate but should otherwise be uncharged under other symmetries, notably the R-symmetry. Their scaling dimension is $D-1$ which for $D=2$ reduces to $(1,0)+(0,1)$. Consider the $(1,0)$ component. It carries $0$ R-charge and left-moving scaling dimension, and so contributes to the elliptic genus. Since it's a boson with left-moving scaling dimension $1$, it should contribute $+q$ to the elliptic genus (multiplied by the adjoint character of the generated flavor symmetry). As such, we see that in 2d $(0,2)$ SCFTs marginal operators and conserved currents contribute at order $q$ with the former contributing negatively and the latter contributing positively. Note that unlike in 3d and 4d their contribution does not appear to be unique. For instance, the lowest component of a chiral field with R-charge $R$ contributes to the elliptic genus as $q^{\frac{R}{2}}$, and thus, a chiral field of R-charge $2$ gives the same contribution as a conserved current. Indeed we shall see in the examples below that the positive terms of order $q$ usually do not form the adjoint of some group suggesting that not all of them are conserved currents. Nevertheless, we do find the presence of terms in the adjoint of the expected global symmetry.    

Finally, we can consider the case of $(2,2)$ SCFTs. Naturally, the above is also valid for this case as these are also $(0,2)$ SCFTs. However, in this case marginal operators and conserved currents can also be seen at lower orders in an expansion by $q$ of the elliptic genus (this is similar to how conserved currents and marginal operators of 4d $\mathcal{N}=2$ SCFTs can be inferred from the moment map and dimension 2 Coulomb branch operators respectively). In $(2,2)$ marginal operators appear as chiral operators inserted in the superpotential. To be marginal they need to carry R-charge $(1,1)$ under the left and right moving R-symmetry. As such they should contribute to the elliptic genus as: $+y q^{\frac{1}{2}}$ \cite{Rastelli:2025nyn}. Alternatively, we can also turn on twisted marginal operators by inserting a twisted chiral with R-charges $(-1,1)$ into the superpotential. These contribute to the elliptic genus as: $+y^{-1} q^{\frac{1}{2}}$. These contributions are in addition to those at order $q$. Indeed a $(2,2)$ chiral field of R-charge $(R,R)$ decomposes to a $(0,2)$ chiral with the same charges and to a $(0,2)$ Fermi multiplet of R-charge $R$ and left-moving R-charge $R-1$. In the case of $R=1$ the Fermi gives the $(0,2)$ marginal operator which contributes as $-q$ while the chiral gives an elliptic genus contribution of $+y q^{\frac{1}{2}}$.  

A similar story applies for conserved currents. These are expected to be at the top of a short multiplet so their annihilation by the supercharges leads to the current conservation. As mentioned the current with dimension $(1,0)$ contributes to the elliptic genus. In a $(2,2)$ theory, we expect it to be a descendant under the left moving supercharge of a fermion with scaling dimension $(\frac{1}{2},0)$ and R-charges $(1,0)$ or $(-1,0)$ depending on whether it's $Q$ or $\bar{Q}$. This would then contribute to the index as $-y q^{\frac{1}{2}}$ or $-y^{-1} q^{\frac{1}{2}}$. This can also be seen explicitly in a $(2,2)$ gauge theory. Consider a chiral field $\Phi$. It contains several components that contribute to the index, notably the scalar $\phi$ and left-moving fermion $\psi_+$ \cite{Gadde:2013dda}. Note that the combination $\phi \bar{\psi}_+$ can always be made gauge invariant, is in the adjoint of the flavor symmetry and contribute as $-y q^{\frac{1}{2}}$. Given a chiral field we have a flavor symmetry $\Phi \rightarrow e^{i\theta} \Phi$. The associated left-moving part of the Noether current is schematically: $J_L \sim \phi \partial \bar{\phi} + \bar{\psi}_+ \psi_+$. First, we note that this contributes to the elliptic genus as $+q$ as expected. Second, since we have the schematic susy variations: $\delta \phi \sim \psi_+$, $\delta \psi_+ \sim \partial \phi$, we should have that $\delta (\phi \bar{\psi}_+) \sim J_L$ so that $J_L$ is a descendant of $\phi \bar{\psi}_+$ as expected. If we were to start with a twisted chiral instead we would just get a similar result but now contributing as $-y^{-1} q^{\frac{1}{2}}$.  

Overall, we see that in $(2,2)$ SCFTs marginal operators and conserved currents also contribute at order $q^{\frac{1}{2}}$ in addition to their contribution at order $q$. An interesting question then is whether one can uniquely identify the presence of marginal operators and conserved currents for the elliptic genus of a $(2,2)$ theory. Indeed, we have mentioned that this holds (under certain assumptions) for 4d $\mathcal{N}=1$ and 3d $\mathcal{N}=2$ as in those cases marginal operators can only fail to be marginal by absorbing a conserved current. A similar statement is also expected to hold in 2d $(2,2)$ SCFTs prompting such an expectation. For instance, we noted that a $(0,2)$ chiral of R-charge $2$ contributes similarly to a conserved current, but a $(2,2)$ chiral of R-charge $2$ contributes as $y^2 q$, which can be distinguished by its $y$ charge (as well as by the presence of the lower order contribution $-y q^{\frac{1}{2}}$). We leave a more in-depth study of this issue to future work.

\section{Evaluation in specific examples}
\label{sec:examples}  

Here we shall use the above observations to study proposed 2d dualities which are motivated by 4d dualities, but appear obstructed by the emergence of new symmetries in 2d. The idea is to evaluate the elliptic genus, refined by all compatible fugacities including those for the emergent symmetries, in two ways. One is in the standard way using the JK residue prescription, which should give the correct elliptic genus. The second way is to expand the one loop determinant in $q$ and take the residue at 0 for the gauge fugacities. This procedure essentially implements the projection to gauge invariants. This gives the naive expectation for the BPS operator spectrum of the theory based on the perturbative matter. Comparing the two of them allows us to infer the presence of "non-perturbative" operators, that contribute to the correct elliptic genus but do not appear in the naive perturbative counting.

Below we will show that in all cases we consider, said comparison reveals the presence of "non-perturbative" operators contributing as $- a^c q$, with $a$ the fugacity for the emergent symmetry, and $c$ is some number depending on the model. This suggests that if this operator is inserted into the superpotential it will preserve the R-symmetry and all flavor symmetries save for the emergent symmetry, which is broken to a discrete group depending on $c$. We show that this breaking is sufficient for the duality to work, that is for anomalies and the elliptic genus to match. 

Interestingly, in all the examples considered below the deviation between the elliptic genus and the perturbative result show a certain regular structure. Specifically, when expanded in a power series in $q$ the difference between the JK result and the perturbative one ($I_{JK}-I_{pert.}$) starts with $-q^{\frac{1}{2}\epsilon_0}\prod a^{-b_i}_i$, where $a_i$ are fugacities for $U(1)$ flavor symmetries. The terms $\epsilon_0$ and $b_i$ are quite similar to the R-symmetry and flavor charges, respectively, acquired by a 3d monopole operator with charges $(1,-1,0,0,...,0)$ due to fermionic zero modes (see \cite{Kapustin:2011jm} for instance for the contribution of monopole operators to the 3d index). For example, for the case of an $\SU(2)$ gauge group with fundamental matter, these read: $\epsilon_0=\sum_{\Phi} (1-R_{\Phi}) -\sum_{\Psi}R_{\Psi}-2$, $b_i = \sum q_i$, where the first sum in $\epsilon_0$ goes over the fundamental chiral fields whose R-charge is denoted by $R_{\Phi}$ and the second sum goes over the fundamental Fermi fields whose R-charge is denoted by $R_{\Psi}$. Similarly, the sum in $b_i$ goes over the fundamental chiral and Fermi fields with $q_i$ denoting their charge under the flavor symmetry associated with the fugacity $a_i$ (in all cases each $\SU(2)$ fundamental is counted once). Furthermore, in the examples in section \ref{section:isometry}, where we also observed higher order terms in the expansion of $I_{JK}-I_{pert.}$, these are remarkably similar to $-q^{\frac{1}{2}\epsilon_0}\prod a^{-b_i}_i \bar{I}_{pert.}$, with the bar conjugating flavor fugacity but not $q$. This relation almost holds, up to occasional differences in the power of $q$ of certain terms. This is again somewhat similar to 3d monopole operators, where the monopole ground state can be dressed by the perturbative operators though there are differences in the power of the fugacity coupling to the dimension of the operator. One intriguing difference though is the presence of the complex conjugation. All the above may suggest that the additional 2d operators observed in the elliptic genus have a similar origin to monopole operators in 3d.   

Before going into the specific examples, we want to comment on the limitations of what is shown here:

\begin{itemize}   
 \item We only show that the operator is consistent with being a marginal operator, that is it contributes to the elliptic genus as one.
\item Even if the operator is marginal, it is not clear whether it is not marginally irrelevant. The main issue is what happens to the theory deformed by said operator. Specifically, say we do not turn on this operator. The theory should flow to some fixed point. Next we want to deform this fixed point by this operator. There are three possible behaviors depending on the dimension of this operator in the new fixed point. Specifically, if the operator is relevant it will initiate an RG flow to a new fixed point with the emergent symmetry broken to a discrete one. If it is exactly marginal, it will break the emergent symmetry to a discrete one without initiating an RG flow. However, if it is irrelevant or marginally irrelevant then the deformation will flow back to zero and we return to the same fixed point. The first and second cases will resolve the puzzle of the mismatching symmetry: the correct 2d duality is between the theories deformed by the additional superpotential. However, the third case will not, as both sides will just flow back to the case without the superpotential.
\item One can try to address the previous point as follows. We can consider the RG flow without the additional superpotential. The superconformal R-symmetry can be determined by c-extremization \cite{Benini:2012cz,Benini:2013cda} where we allow mixing with the accidental symmetries. Assuming the additional operator is a superpotential term, one can determine its dimension from its R-charge under the superconformal R-symmetry, which can be determined from knowledge of the mixing and its charges under the flavor symmetry. This information can then be used to say whether said operator is relevant, irrelevant or marginal with respect to the fixed point. Performing the analysis in all the examples studied below, it appears that the operator is irrelevant with respect to the fixed point the theory flows to when the accidental symmetries are unbroken. As such, turning it on would just lead us back to the original fixed point and not to a new fixed point without the symmetry, and the puzzle engulfing the duality remains unresolved. However, there are certain subtleties involving the 2d c-extremization procedure when non-compact moduli are present, which is basically the case in all the examples considered here. It is thus possible that the c-extremization result is too naive and fails due to the presence of unbounded moduli. As such, we believe it is still too early to determine whether the operator is irrelevant or not, and leave further study of this issue for future work.

\end{itemize}   

\subsection{2d reduction of Seiberg duality of $\SU$ groups}

Here we consider the reduction of 4d Seiberg duality of $\SU$ groups to get a duality between 2d $(0,2)$ gauge theories \cite{Gadde:2015wta}. The idea is to consider an $\mathcal{N}=1$ $\SU(N_c)$ gauge theory with $N_f$ fundamental and anti-fundamental chiral fields. We can reduce it to 2d on $S^2$, where to preserve susy we need to introduce a topological twist. In general the resulting 2d theory is a direct sum of theories, due to the need to sum over magnetic fluxes on the $S^2$. However, if one makes a choice of R-charge under which the chirals carry R-charge of only $0$, $1$ or $2$, then the sum truncates and only a single 2d theory, associated with the zero flux sector remains (see \cite{Gadde:2015wta} for further details). The chiral fields of R-charge $0$ and $2$ reduce to 2d chiral and Fermi multiplets respectively, while those of R-charge $1$ do not lead to a 2d multiplet. The end result is a 2d $(0,2)$ gauge theory.

This procedure can be applied to the two sides of the duality and leads to a similar duality in 2d. The construction then guarantees that 't Hooft anomalies and the elliptic genus would match between the two sides. However, there is a subtlety here. The 4d theory has an anomalous symmetry, broken by the $\SU(N_c)$ ABJ type anomaly. However, in 2d this symmetry is not anomalous and so reemerges. If at least one 4d chiral has R-charge $1$ under the twisted R-symmetry, then this emergent 2d symmetry can be mapped to a non-anomalous 4d symmetry that also acts on this chiral. This follows as said chiral does not survive in 2d so we can just choose its charge such that the 4d mixed gauge-flavor anomaly vanishes. However, if there are no 4d chirals of R-charge $1$ then this 2d symmetry has no 4d uplift. In this case the matching of 't Hooft anomalies associated with this symmetry and the elliptic genus refined under it is no longer guaranteed from the 4d construction, and indeed often fails as we show below. The duality then cannot hold unless there is a way to break the emergent 2d symmetry. 

This subsection is devoted to exploring this issue for certain low values of $N_c$ and $N_f$.        

\subsubsection{$N_c=2$, $N_f=3$}

\begin{figure}
    \centering
    \begin{tikzpicture}[
        flavor/.style={draw, rectangle, minimum size=10mm, inner sep=2pt, font=\Large},
        gauge/.style={draw, circle, minimum size=10mm, inner sep=2pt, font=\Large},
        blackdot/.style={fill=black, circle, minimum size=2mm, inner sep=0pt},
        >={Stealth[scale=1.2]}
    ]

        \node[font=\large] at (-0.2, 2.7) {$\mathcal{T}_A$:};
        
        \node[flavor] (3fA) at (0, 0) {$3_f$};
        \node[blackdot] (dot1A) at (3, 0) {};
        \node[blackdot] (dot2A) at (1.5, 2.2) {};
        \node[flavor] (2hA) at (5.5, 0) {$2_h$};

        \draw[->, thick] (3fA) -- node[below, yshift=-2pt] {$\Phi$} (dot1A);
        \draw[->, thick] (dot1A) -- node[right, xshift=2pt] {$P$} (dot2A);
        \draw[thick, dashed] (dot2A) -- node[left, xshift=-2pt] {$\Gamma$} (3fA);
        \draw[thick, dashed] (dot1A) -- node[below, yshift=-2pt] {$\Psi$} (2hA);

        \node[font=\large] at (6.5, 1.1) {$\cong$};

        \node[font=\large] at (7.8, 2.7) {$\mathcal{T}_B$:};
        
        \node[flavor] (3fB) at (8, 0) {$3_f$};
        \node[gauge] (2B) at (11, 0) {$2$};
        \node[flavor] (2hB) at (9.5, 2.5) {$2_h$};
        \node[blackdot] (dotB) at (13.5, 0) {};

        \draw[->, thick] (3fB) -- node[below, yshift=-2pt] {$\hat{\Phi}$} (2B);
        \draw[->, thick] (2B) -- node[right, xshift=2pt] {$\hat{P}$} (2hB);
        \draw[thick, dashed] (2hB) -- node[left, xshift=-2pt] {$\hat{\Gamma}$} (3fB);
        \draw[thick, dashed] (2B) -- node[below, yshift=-2pt] {$\hat{\Psi}$} (dotB);

    \end{tikzpicture}
    \caption{Quiver diagram of a gauge - WZ duality. We adopt the common notation of solid lines for chiral multiplets and dashed lines for Fermi multiplets.}
    \label{fig:GaugeWZDuality_quiver}
\end{figure}
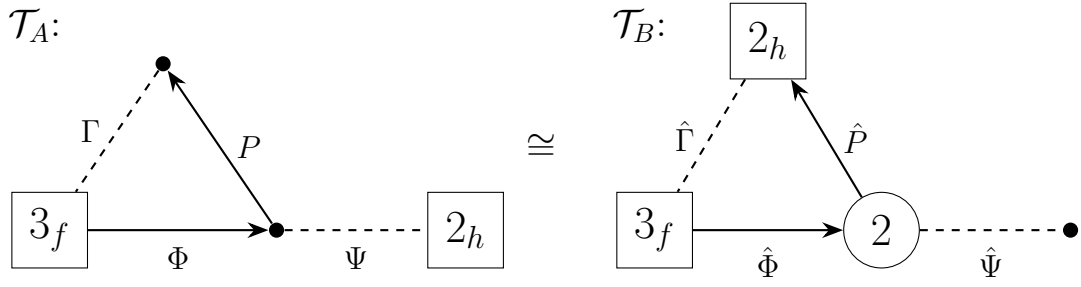

We begin with the case of $\SU(2)$ with three flavors. This 4d $\mathcal{N}=1$ gauge theory is dual to a collection of $15$ free chiral fields. As such, the 4d reduction leads to a 2d duality between an $\SU(2)$ gauge theory and a WZW model. Quiver diagrams of the two dual theories are given in \ref{fig:GaugeWZDuality_quiver} and a numeration of the charges of the operators under the global symmetries is given in table \ref{table:GaugeWZDuality_table}. Here, the non-abelian flavor symmetries, $\U(1)_b$ and $\U(1)_c$ come from 4d while $\U(1)_d$ emerges in 2d from the 4d axial symmetry broken by the anomaly. The gauge theory has a superpotential of the form: $W = \hat{\Gamma}\hat{P}\hat{\Phi}$ while the WZW model has the superpotential: $W=\Gamma P \Phi$.

\begin{table}

\begin{centering}
\begin{tabular}{c|c c c c c c c}

 & $\SU(2)_{h}$ & $\SU(3)_{f}$ & $\U(1)_{b}$ & $\U(1)_{c}$ & $\U(1)_{d}$ & $\U(1)_{R}^{\text{4d}}$ & $\U(1)_{R}^{\text{mix}}$\\
\hline 

$\Phi$ & $1$ & $\Box$ & $1$ & $0$ & $1$ & $0$ & $\frac{1}{6}$\\

$P$ & $1$ & $1$ & $-1$ & $2$ & $1$ & $0$ & $\frac{1}{2}$\\

$\Psi$ & $\Box$ & $1$ & $-1$ & $-1$ & $1$ & $1$ & $\frac{1}{2}$\\

$\Gamma$ & $1$ & $\overline{\Box}$ & $0$ & $-2$ & $-2$ & $1$ & $\frac{1}{3}$\\
 
\end{tabular}
\par \vspace{1.5ex}
$\cong$ \ %
\begin{tabular}{c|c|c c c c c c c}

 & $\SU(2)_{\text{gauge}}$ & $\SU(2)_{h}$ & $\SU(3)_{f}$ & $\U(1)_{b}$ & $\U(1)_{c}$ & $\U(1)_{d}$ & $\U(1)_{R}^{4d}$ & $\U(1)_{R}^{\text{mix}}$\\
\hline 
$\hat\Phi$ & $\Box$ & $1$ & $\overline{\Box}$ & $\frac{1}{2}$ & $0$ & $\frac{1}{2}$ & $0$ & $\frac{1}{12}$\\

$\hat P$ & $\Box$ & $\Box$ & $1$ & $-\frac{1}{2}$ & $1$ & $\frac{1}{2}$ & $0$ & $\frac{1}{4}$\\

$\hat\Psi$ & $\Box$ & $1$ & $1$ & $-\frac{1}{2}$ & $-2$ & $\frac{1}{2}$ & $1$ & $\frac{1}{4}$\\

$\hat\Gamma$ & $1$ & $\Box$ & $\Box$ & $0$ & $-1$ & $-1$ & $1$ & $\frac{2}{3}$\\

\end{tabular} 
\par\end{centering}
\caption{Table of gauge and symmetry group representations of the gauge - WZ duality. $\U(1)_R^{4d}$ is the R-symmetry inherited from the 4d theories. $\U(1)_R^{\text{mix}}=\U(1)_R^{4d} + \frac{1}{6} \U(1)_b + \frac{1}{3} \U(1)_c$ is an R-symmetry taken so that the R-charges of the fields are strictly positive. Note that $\U(1)_R^{\text{mix}}$ does not involve mixing with $\U(1)_d$, compared to $\U(1)_R^{4d}$, so the duality should hold for it as well.}
\label{table:GaugeWZDuality_table}
\end{table}

One can see that all anomalies that do not involve $\U(1)_d$ match across the duality. However, there is a mismatch in the $\U(1)_d$ anomaly:

\bea
&& Tr(\U(1)^2_d)_{WZW} = -10 \neq Tr(\U(1)^2_d)_{gauge} = -4 \; , \\ \nonumber 
&& Tr(\U(1)_d \U(1)_b)_{WZW} = 4 \neq Tr(\U(1)_d \U(1)_b)_{gauge} = 1 \; , \\ \nonumber
&& Tr(\U(1)_d \U(1)_c)_{WZW} = -8 \neq Tr(\U(1)_d \U(1)_c)_{gauge} = -2 \; .
\eea

Note though that in all cases the mismatch is a multiple of 3. This is not surprising as the 4d axial symmetry is broken to $\mathbb{Z}_6$ by the ABJ anomaly, which can be reduced to $\mathbb{Z}_3$ when quotienting by the center of the gauge symmetry, so the anomalies should match if $\U(1)_d$ is broken to $\mathbb{Z}_3$. This suggests that the dual theories should be modified by the addition of a superpotential breaking $\U(1)_d$ to $\mathbb{Z}_3$. However, there does not appear to be any perturbative superpotential that can implement this breaking leading to a puzzle engulfing this duality. To better understand this issue we next consider the elliptic genus.

\paragraph{Elliptic genus} Let us now compute the elliptic genus of the two theories including the addition of a fugacity for $\U(1)_d$. For convenience of the series expansion computation in $q$, we use the mixed R-symmetry $\U(1)_R^{\text{mix}}$ written in table \ref{table:GaugeWZDuality_table}, chosen so that the R-charges are strictly positive. For the WZ theory, we find:
\begin{dmath}
    \mathcal{I}^{\text{WZ}}_{\text{JK}}=1+q^{\frac{1}{12}}bd\mathbf{3}_{f}+q^{\frac{1}{6}}b^{2}d^{2}\mathbf{6}_{f}+... \\ +q\left(3+c^{6}+\frac{c^{8}d^{4}}{b^{4}}-\frac{cd^{2}\mathbf{2}_{h}}{b^{2}}-\frac{c^{7}d^{2}\mathbf{2}_{h}}{b^{2}}-b^{2}c^{5}d^{4}\mathbf{2}_{h}+\mathbf{3}_{h}+\mathbf{8}_{f}-b^{4}c^{4}d^{2}\mathbf{8}_{f}+\frac{b^{2}\mathbf{2}_{h}\mathbf{8}_{f}}{cd^{2}}+b^{4}c^{4}d^{8}\mathbf{10}_{f}-\frac{b^{2}d^{4}\mathbf{2}_{h}\mathbf{10}_{f}}{c}+b^{8}c^{2}d^{4}\mathbf{28}_{f}-\frac{b^{4}d^{2}\mathbf{35}_{f}}{c^{2}}+b^{6}c^{3}d^{6}\mathbf{2}_{h}\mathbf{35}_{f}-b^{10}cd^{8}\mathbf{2}_{h}\mathbf{55}_{f}-b^{8}c^{2}d^{10}\mathbf{80}_{f}+b^{12}d^{12}\mathbf{91}_{f}\right) +O(q^{\frac{13}{12}})
\end{dmath}
For the gauge theory, we get:
\begin{dmath}
    \mathcal{I}^{\text{gauge}}_{\text{JK}}=1+q^{\frac{1}{12}}b d \mathbf{3}_{f}+q^{\frac{1}{6}}b^{2}d^{2}\mathbf{6}_{f}+...\\+q\left(3+c^{6}-\frac{1}{d^{3}}+d^{3}+\frac{c^{8}d^{4}}{b^{4}}-b^{2}c^{5}d\mathbf{2}_{h}-\frac{cd^{2}\mathbf{2}_{h}}{b^{2}}-\frac{c^{7}d^{2}\mathbf{2}_{h}}{b^{2}}+\mathbf{3}_{h}+\mathbf{8}_{f}-b^{4}c^{4}d^{5}\mathbf{8}_{f}+\frac{b^{2}d^{4}\mathbf{2}_{h}\mathbf{8}_{f}}{c}+b^{4}c^{4}d^{2}\mathbf{10}_{f}-\frac{b^{2}d\mathbf{2}_{h}\mathbf{10}_{f}}{c}+b^{8}c^{2}d^{10}\mathbf{28}_{f}-\frac{b^{4}d^{5}\mathbf{35}_{f}}{c^{2}}+b^{6}c^{3}d^{6}\mathbf{2}_{h}\mathbf{35}_{f}-b^{10}cd^{11}\mathbf{2}_{h}\mathbf{55}_{f}-b^{8}c^{2}d^{7}\mathbf{80}_{f}+b^{12}d^{12}\mathbf{91}_{f}\right) +O(q^{\frac{13}{12}})
\end{dmath}
The ... denotes terms with powers of q between $\frac{1}{6}$ and $1$. For brevity, these terms are omitted here and are collected in appendix \ref{appendix:q-expansion-resultsNc2Nf3}.

We can make the following observations regarding the two elliptic genera. First, one can notice that they agree if the fugacity $d$ is set to one. Moreover, the equality also holds if we set $d^3=1$, but not otherwise. This is consistent with the anomaly matching and suggests that the duality cannot hold unless $\U(1)_d$ is broken to $\mathbb{Z}_3$. Next, it is interesting to examine the order $q$ terms in the elliptic genus. As previously mentioned, this is where we expect to find the contribution of conserved currents and marginal operators. Indeed, note that both genera contain the contribution $+q(3+3_{h}+8_{f})$ which we can identify as the conserved currents for the $\U(1)_b \times \U(1)_c \times \U(1)_d \times \SU(2)_h \times \SU(3)_f$ flavor symmetry. There are various additional positive terms at order $q$ in the elliptic genus, but these do not appear to be conserved currents as they are not charge conjugation invariant and so cannot be in the adjoint of a flavor symmetry. 

We next consider the negatively charged terms at order $q$, which could be marginal operators. We begin with the WZW side. Here there are several negatively charged operators, but they all carry charges also under the $\U(1)_b \times \U(1)_c \times \SU(2)_h \times \SU(3)_f$ part of the flavor symmetry. As such, introducing them in the superpotential will break additional symmetries. This is consistent with the observation that there are no marginal operators one can write that break $\U(1)_d$ to $\mathbb{Z}_3$. Similarly, we can consider the gauge theory side, however, here we find a contribution of the form $-d^{-3}q$. This is consistent with a marginal operator and has just the right charges to break $\U(1)_d$ to $\mathbb{Z}_3$. How is this reconciled with the lack of observed marginal operator in the gauge theory? For this it is convenient to compare the exact calculation of the elliptic genus, evaluated using the JK prescription, against the naive contribution of the perturbative matter evaluated using the series expansion of the integrand. We find:
\begin{equation}
    \mathcal{I}^{\text{gauge}}_{\text{JK}} = \mathcal{I}^{\text{gauge}}_{\text{pert.}} - d^{-3}q + o(q) .
\end{equation}

We see then that the potential marginal operator we observed in the elliptic genus of the gauge theory indeed does not arise perturbatively, rather appearing through the JK prescription. The charges of this additional operator are consistent with that of a marginal operator and inserting it in the superpotential would then break $\U(1)_d$ to $\mathbb{Z}_3$. However, note that there is no evidence for a similar operator on the dual WZW side. Indeed, as this is not a gauge theory the perturbative and exact analysis of the elliptic genus agree. As such, there is no evidence that the $\U(1)_d$ symmetry of the WZW model can be broken to $\mathbb{Z}_3$ without breaking additional symmetries. This suggests that the duality is between the WZW model and the $\SU(2)$ gauge theory deformed by this additional superpotential. The non-perturbative superpotential then breaks the $\U(1)_d$ on the $\SU(2)$ side to $\mathbb{Z}_3$ which should enhance back to $\U(1)$ at low-energies forming the $\U(1)_d$ of the WZW side. Both the 't Hooft anomalies and the elliptic genus are consistent with this picture as they agree between the two sides once $\U(1)_d$ is reduced to $\mathbb{Z}_3$. This picture is also implied from the 4d origin of the duality. Recall that the dual side in 4d is a collection of $15$ free chiral fields. We can define a $\U(1)$ symmetry acting on all of them with charge $+1$ which would then descend to $\U(1)_d$ of the 2d WZW theory. Technically, this 4d symmetry is broken by a superpotential in 4d, but as it is marginally irrelevant, it should emerge in the IR.  

\paragraph{c-extremization} The above picture requires that the superpotential be relevant in whatever fixed point the $\SU(2)$ gauge theory flows to if it is not inserted. To further test this, we want to analyze the RG flow without it. Specifically, we want to determine the superconformal R-symmetry from which we can understand the dimension of the operator at the fixed point. This can be determined through the use of c-extremization. It is straightforward to perform the analysis and we find that $\U(1)_R^{\text{SC}}=\U(1)_R^{\text{mix}}+\frac{1}{2} \U(1)_b$ is the one extremizing the $c$ central charge. As it involves mixing of $\U(1)_R^{\text{mix}}$, which is the R-symmetry used in the above calculation of the elliptic genus, with $\U(1)_b$ only and as the additional operator is uncharged under $\U(1)_b$, it naively appears that its charges at the fixed point are still consistent with a marginal operator.

However, there is a subtlety involved in c-extremization common in supersymmetric gauge theories due to the presence of unbounded moduli. Specifically, the presence of a non-compact moduli space violates some of the assumptions involved in c-extremization, and the procedure needs to be adjusted \cite{Benini:2012cz,Benini:2013cda}. The main correction is that one must enforce that the R-charges of all fields parametrizing the non-compact directions be set to zero. For the case at hand, the moduli space is parametrized by the scalar fields $\hat{\Phi}$ and $\hat{P}$. These are subjected to the F-terms and D-terms conditions. As usual, the D-term conditions can be solved by forming gauge invariants and in this case we have three options: $\hat{\Phi}^2$, $\hat{P}^2$ and $\hat{\Phi}\hat{P}$. The F-term conditions enforce that the last operator, $\hat{\Phi}\hat{P}$ cannot acquire a vev. Thus, we appear to have a non-compact moduli space made of two branches, one spanned by $\hat{\Phi}^2$ and the other by $\hat{P}^2$. 

The above mentioned constraint now forces that $R[\hat{\Phi}]=R[\hat{P}]=0$. We can now repeat the c-extremization and we find that the R-symmetry that extremize $c$ under these conditions is the one under which the fields carry the charges: $R[\hat{\Phi}]=R[\hat{P}]=R[\hat{\Psi}]=0$, $R[\hat{\Gamma}]=1$. With these charges, we have that $Tr(U(1)^2_R) = 1$. The R-symmetry with these charges is given by: $\U(1)_R = \U(1)_R^{\text{mix}} + \frac{1}{6}\U(1)_b - \frac{1}{3}\U(1)_d$, and we note that now there is a mixing with $\U(1)_d$. As the additional operator carry negative $\U(1)_d$ charge, we conclude that if its $\U(1)_R^{\text{mix}}$ R-charge was $1$, which is the required value for a marginal operator, then its $\U(1)_R$ value is $2$. This R-charge is consistent with that of an irrelevant operator and we appear to conclude that turning it on would just lead us back to the same fixed point. As such, it is still not clear how the 2d duality holds, unless the analysis performed here is too naive and fails for some subtle reason, like emergence of accidental symmetries.   

\subsubsection{$N_c=2$, $N_f=4$}

\begin{figure}
    \centering
    \begin{tikzpicture}[
        flavor/.style={draw, rectangle, minimum size=10mm, inner sep=2pt, font=\Large},
        gauge/.style={draw, circle, minimum size=10mm, inner sep=2pt, font=\Large},
        blackdot/.style={fill=black, circle, minimum size=2mm, inner sep=0pt},
        >={Stealth[scale=1.2]}
    ]

        \node[font=\large] at (-0.2, 2.7) {$\mathcal{T}_A$:};
        
        \node[flavor] (4fA) at (0, 0) {$4_f$};
        \node[gauge] (2A) at (3, 0) {$2$};
        \node[flavor] (2gA) at (1.5, 2.2) {$2_g$};
        \node[flavor] (2hA) at (5.5, 0) {$2_h$};

        \draw[->, thick] (4fA) -- node[below, yshift=-2pt] {$\Phi$} (2A);
        \draw[->, thick] (2A) -- node[right, xshift=2pt] {$P$} (2gA);
        \draw[thick, dashed] (2gA) -- node[left, xshift=-2pt] {$\Gamma$} (4fA);
        \draw[thick, dashed] (2A) -- node[below, yshift=-2pt] {$\Psi$} (2hA);

        \node[font=\large] at (6.75, 1.1) {$\cong$};

        \node[font=\large] at (7.8, 2.7) {$\mathcal{T}_B$:};
        
        \node[flavor] (4fB) at (8, 0) {$4_f$};
        \node[gauge] (2B) at (11, 0) {$2$};
        \node[flavor] (2hB) at (9.5, 2.2) {$2_h$};
        \node[flavor] (2gB) at (13.5, 0) {$2_g$};

        \draw[->, thick] (4fB) -- node[below, yshift=-2pt] {$\hat{\Phi}$} (2B);
        \draw[->, thick] (2B) -- node[right, xshift=2pt] {$\hat{P}$} (2hB);
        \draw[thick, dashed] (2hB) -- node[left, xshift=-2pt] {$\hat{\Gamma}$} (4fB);
        \draw[thick, dashed] (2B) -- node[below, yshift=-2pt] {$\hat{\Psi}$} (2gB);

    \end{tikzpicture}
    \caption{Quiver diagram of an $\SU(2)$ self-duality emerging from a reduction of Seiberg duality on a sphere.}
    \label{fig:Quiver_self_duality}
\end{figure}
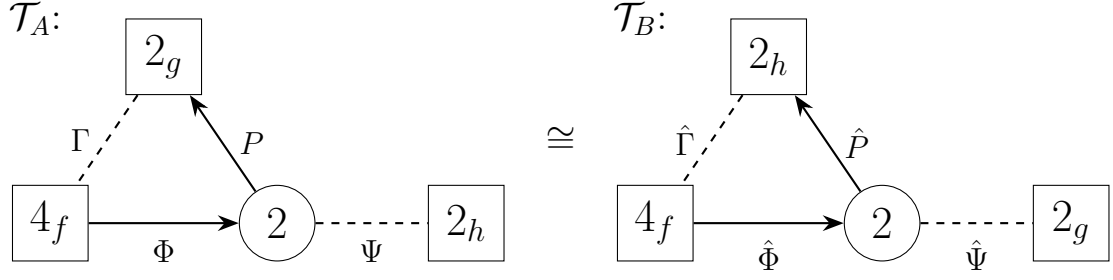

\begin{table}
\begin{centering}
\begin{tabular}{c|c|c c c c c c c c}

 & $\SU(2)_{\text{gauge}}$ & $\SU(2)_{h}$ & $\SU(2)_{g}$ & $\SU(4)_{f}$ & $\U(1)_{b}$ & $\U(1)_{c}$ & $\U(1)_{d}$ & $\U(1)_{R}^{4d}$ & $\U(1)_{R}^{\text{mix}}$ \\
\hline  
$\Phi$ & $\Box$ & $1$ & $1$ & $\Box$ & $\frac{1}{2}$ & $0$ & $\frac{1}{2}$ & $0$ & $\frac{1}{4}$\\

$P$ & $\Box$ & $1$ & $\Box$ & $1$ & $-\frac{1}{2}$ & $2$ & $\frac{1}{2}$ & $0$ & $\frac{1}{4}$\\

$\Psi$ & $\Box$ & $\Box$ & $1$ & $1$ & $-\frac{1}{2}$ & $-2$ & $\frac{1}{2}$ & $1$ & $\frac{1}{4}$\\
 
$\Gamma$ & $1$ & $1$ & $\Box$ & $\overline{\Box}$ & $0$ & $-2$ & $-1$ & $1$ & $\frac{1}{2}$\\

\end{tabular}
\par \vspace{1.5ex}
$\cong$ \ %
\begin{tabular}{c|c|c c c c c c c c}

 & $\SU(2)_{\text{gauge}}$ & $\SU(2)_{h}$ & $\SU(2)_{g}$ & $\SU(4)_{f}$ & $\U(1)_{b}$ & $\U(1)_{c}$ & $\U(1)_{d}$ & $\U(1)_{R}^{4d}$ & $\U(1)_{R}^{\text{mix}}$\tabularnewline
\hline 
$\hat{\Phi}$ & $\Box$ & $1$ & $1$ & $\overline{\Box}$ & $\frac{1}{2}$ & $0$ & $\frac{1}{2}$ & $0$ & $\frac{1}{4}$\\

$\hat{P}$ & $\Box$ & $\Box$ & $1$ & $1$ & $-\frac{1}{2}$ & $2$ & $\frac{1}{2}$ & $0$ & $\frac{1}{4}$\\

$\hat{\Psi}$ & $\Box$ & $1$ & $\Box$ & $1$ & $-\frac{1}{2}$ & $-2$ & $\frac{1}{2}$ & $1$ & $\frac{1}{4}$\\

$\hat{\Gamma}$ & $1$ & $\Box$ & $1$ & $\Box$ & $0$ & $-2$ & $-1$ & $1$ & $\frac{1}{2}$\\

\end{tabular}
\par\end{centering}
\caption{Table of gauge and symmetry group representations of the $\SU(2)$ gauge self-duality. $\U(1)_R^{4d}$ is the R-symmetry inherited from 4d. $\U(1)_R^{\text{mix}}=\U(1)_R^{4d} + \frac{1}{2} \U(1)_b + \frac{1}{4} \U(1)_c$ is a mixed R-symmetry taken so that the R-charges of the fields are strictly positive. As before, $\U(1)_R^{\text{mix}}$ does not involve mixing with $\U(1)_d$, compared to $\U(1)_R^{4d}$, so as to not interfere with the self-duality.}
\label{table:Self_duality_table}
\end{table}

Next we consider the case of $\SU(2)$ with four flavors. This case is self-dual, and so leads to a self-duality of an $\SU(2)$ gauge theory also in 2d. Quiver diagrams of the two dual theories are provided in figure \ref{fig:Quiver_self_duality} and a numeration of the charges of the operators under the global symmetries is given in table \ref{table:Self_duality_table}. Note that even though it is a self-duality, it still has non-trivial physical content as it involves the exchange of $\SU(2)_g \leftrightarrow \SU(2)_h$, $\SU(4)\rightarrow \overline{\SU(4)}$. Thus, it implies that the theory is invariant under the exchange of the two symmetries together with charge conjugation of the $\SU(4)$ flavor symmetry. This can be checked using the elliptic genus. As before, the non-abelian flavor symmetries, $\U(1)_b$ and $\U(1)_c$ come from 4d while $\U(1)_d$ emerges in 2d from the 4d axial symmetry broken by the anomaly. Both theories also have a superpotential of the form $W = \hat{\Gamma}\hat{P}\hat{\Phi}$ for theory $A$ and $W=\Gamma P \Phi$ for theory $B$.

Note that the 't Hooft anomalies of $\text{Tr}(\SU(2)^2_h)$ and $\text{Tr}(\SU(2)^2_g)$ agree, which is indeed consistent with the duality. Since this is a self-duality, this is the only check provided by 't Hooft anomalies and should essentially follow from the 4d self-duality.

\paragraph{Elliptic genus} To further check the self-duality, we compute the elliptic genus of the two theories refined by $\U(1)_d$. As before, for convenience of the series expansion computation in $q$, we use the mixed R-symmetry $\U(1)_R^{\text{mix}}$ written in table \ref{table:Self_duality_table} and chosen so that the R-charges are strictly positive. We find:
\begin{align}
    \mathcal{I}_{\text{JK}}^{A} &= 1+q^{\frac{1}{4}}\left(\frac{c^{4}d}{b}+bd\mathbf{6}_{f}\right)+q^{\frac{1}{2}}\left(-bc^{2}d^{2}\mathbf{2}_{g}\bar{\mathbf{4}}_{f}+\frac{c^{8}d^{2}}{b^{2}}+b^{2}d^{2}\mathbf{20}'_{f}-bc^{2}\mathbf{4}_{f}\mathbf{2}_{h}-c^{4}\mathbf{2}_{g}\mathbf{2}_{h}\right) \notag\\
    &\quad +q^{\frac{3}{4}}\biggl(-b^{2}c^{2}d\mathbf{2}_{h}\overline{\mathbf{20}}_{f}-\frac{\mathbf{2}_{g}\bar{\mathbf{4}}_{f}}{c^{2}d}+\frac{c^{12}d^{3}}{b^{3}}+b^{3}d^{3}\mathbf{50}_{f}-b^{2}c^{2}d^{3}\mathbf{20}_{f}\mathbf{2}_{g}-\frac{c^{8}d\mathbf{2}_{g}\mathbf{2}_{h}}{b} \notag\\
    &\qquad +bc^{4}d^{3}\mathbf{3}_{g}+\frac{bc^{4}\mathbf{3}_{h}}{d}-\frac{d\mathbf{2}_{g}\mathbf{2}_{h}}{b}-\frac{d\mathbf{4}_{f}\mathbf{2}_{h}}{c^{2}}\biggr) \notag\\
    &\quad +q\biggl(-b^{3}c^{2}d^{4}\mathbf{2}_{g}\overline{\mathbf{60}}_{f}+b^{2}c^{4}d^{4}\overline{\mathbf{10}}_{f}+bc^{6}d^{2}\mathbf{2}_{h}\bar{\mathbf{4}}_{f}-\frac{bd^{2}\mathbf{2}_{h}\overline{\mathbf{20}}_{f}}{c^{2}}+\frac{c^{16}d^{4}}{b^{4}}+b^{4}d^{4}\mathbf{105}_{f} \notag\\
    &\qquad -b^{3}c^{2}d^{2}\mathbf{60}_{f}\mathbf{2}_{h}-\frac{c^{12}d^{2}\mathbf{2}_{g}\mathbf{2}_{h}}{b^{2}}+\mathbf{6}_{f}\left(b^{2}c^{4}d^{4}\mathbf{3}_{g}+b^{2}c^{4}\mathbf{3}_{h}\right)+\mathbf{15}_{f}\left(b^{2}c^{4}d^{2}\mathbf{2}_{g}\mathbf{2}_{h}+1\right) \notag\\
    &\qquad -\frac{c^{4}d^{2}\mathbf{2}_{g}\mathbf{2}_{h}}{b^{2}}+b^{2}c^{4}\mathbf{10}_{f}+bc^{6}\mathbf{4}_{f}\mathbf{2}_{g}-\frac{b\mathbf{20}_{f}\mathbf{2}_{g}}{c^{2}}+c^{8}\mathbf{3}_{g}+c^{8}\mathbf{3}_{h}-\frac{1}{d^{4}}+\mathbf{3}_{g}+\mathbf{3}_{h}+3\biggr) \notag\\
    &\quad +O(q^{\frac{5}{4}})
\end{align}
for theory $A$, and:
\begin{align}
    \mathcal{I}_{\text{JK}}^{B} &= 1+q^{\frac{1}{4}}\left(\frac{c^{4}d}{b}+bd\mathbf{6}_{f}\right)+q^{\frac{1}{2}}\left(-bc^{2}\mathbf{2}_{g}\bar{\mathbf{4}}_{f}+\frac{c^{8}d^{2}}{b^{2}}+b^{2}d^{2}\mathbf{20}'_{f}-bc^{2}d^{2}\mathbf{4}_{f}\mathbf{2}_{h}-c^{4}\mathbf{2}_{g}\mathbf{2}_{h}\right) \notag\\
    &\quad +q^{\frac{3}{4}}\biggl(-b^{2}c^{2}d^{3}\mathbf{2}_{h}\overline{\mathbf{20}}_{f}-\frac{d\mathbf{2}_{g}\bar{\mathbf{4}}_{f}}{c^{2}}+\frac{c^{12}d^{3}}{b^{3}}+b^{3}d^{3}\mathbf{50}_{f}-b^{2}c^{2}d\mathbf{20}_{f}\mathbf{2}_{g}-\frac{c^{8}d\mathbf{2}_{g}\mathbf{2}_{h}}{b} \notag\\
    &\qquad +bc^{4}d^{3}\mathbf{3}_{h}+\frac{bc^{4}\mathbf{3}_{g}}{d}-\frac{d\mathbf{2}_{g}\mathbf{2}_{h}}{b}-\frac{\mathbf{4}_{f}\mathbf{2}_{h}}{c^{2}d}\biggr) \notag\\
    &\quad +q\biggl(-b^{3}c^{2}d^{2}\mathbf{2}_{g}\overline{\mathbf{60}}_{f}+b^{2}c^{4}\overline{\mathbf{10}}_{f}+bc^{6}\mathbf{2}_{h}\bar{\mathbf{4}}_{f}-\frac{b\mathbf{2}_{h}\overline{\mathbf{20}}_{f}}{c^{2}}+\frac{c^{16}d^{4}}{b^{4}}+b^{4}d^{4}\mathbf{105}_{f} \notag\\
    &\qquad -b^{3}c^{2}d^{4}\mathbf{60}_{f}\mathbf{2}_{h}-\frac{c^{12}d^{2}\mathbf{2}_{g}\mathbf{2}_{h}}{b^{2}}+\mathbf{6}_{f}\left(b^{2}c^{4}d^{4}\mathbf{3}_{h}+b^{2}c^{4}\mathbf{3}_{g}\right)+b^{2}c^{4}d^{4}\mathbf{10}_{f} \notag\\
    &\qquad +\mathbf{15}_{f}\left(b^{2}c^{4}d^{2}\mathbf{2}_{g}\mathbf{2}_{h}+1\right)-\frac{c^{4}d^{2}\mathbf{2}_{g}\mathbf{2}_{h}}{b^{2}}+bc^{6}d^{2}\mathbf{4}_{f}\mathbf{2}_{g}-\frac{bd^{2}\mathbf{20}_{f}\mathbf{2}_{g}}{c^{2}} \notag\\
    &\qquad +c^{8}\mathbf{3}_{g}+c^{8}\mathbf{3}_{h}-\frac{1}{d^{4}}+\mathbf{3}_{g}+\mathbf{3}_{h}+3\biggr)+O(q^{\frac{5}{4}})
\end{align}
for theory $B$.

Note that the elliptic genus refined with $\U(1)_d$ is not consistent with the self-duality. The self-duality does hold if we set $d=1$, as expected from the 4d duality, but is disrupted by the presence of $\U(1)_d$. As before, to better understand what is going on, we turn to study the order $q$ terms in the elliptic genus. First, we note that both genera indeed contain the contribution $+q(3+3_{h}+3_{g}+15_{f})$ which we can identify as the conserved currents for the $\U(1)_b \times \U(1)_c \times \U(1)_d \times \SU(2)_h \times \SU(2)_g \times \SU(4)_f$ flavor symmetry. As before, there are many other positive terms at order $q$ in the elliptic genus, but these do not appear to be conserved currents. 

We next consider the negatively charged terms at order $q$, which could be marginal operators. Here we are particularly concerned with marginal operators that are charged only under $\U(1)_d$, and indeed there is exactly one such operator. As before, this operator is not apparent from the perturbative matter, as is evident by comparing the elliptic genus against the naive perturbative counting. Specifically, we find:
\begin{align} \begin{split}
    \mathcal{I}_{\text{JK}}^{A} = \mathcal{I}_{\text{pert.}}^{A} - d^{-4}q +o(q) \\
    \mathcal{I}_{\text{JK}}^{B} = \mathcal{I}_{\text{pert.}}^{B} - d^{-4}q +o(q)
\end{split} \end{align}
so said operator is not built from the basic perturbative operators. 

As before, the existence of this operator provides a potential remedy for the self-duality. Specifically, if we deform both theories by the superpotential, we break $\U(1)_d$ and the elliptic genus should be consistent with the self-duality. Note, however, that turning on this operator would break $\U(1)_d$ to $\mathbb{Z}_4$ which naively is insufficient to make the two genera agree. Nevertheless, it is possible to match the two if the $\mathbb{Z}_4$ remnant of $\U(1)_d$ do not simply map to one another. Specifically, we have an $\SU(4)_f$ flavor symmetry which contains a $\mathbb{Z}_4$ element given by its center that commutes with all other symmetries. Denoting the fugacity for the $\mathbb{Z}_4$ remnant of $\U(1)_d$ on the two sides by $d$ and $\hat{d}$, we note that the elliptic genera match if we map: $d\rightarrow \hat{d}$, $4_f \rightarrow \hat{d}^2\overline{4}_f$, given that $d^4=\hat{d}^4=1$. That is, the symmetry mapping is that the $\mathbb{Z}_4$ on one side is mapped to a diagonal combination of the $\mathbb{Z}_4$ on the other side and the $-I$ central element of $\SU(4)$.     

\paragraph{c-extremization} Next, we want to study the RG flow of the theory without the added superpotential to better understand its behavior once it is turned on. For that we consider the superconformal R-symmetry of the theory if we allow mixing with $\U(1)_d$, which can be determined using c-extremization. As before, if we allow unconstrained mixing, we find that $\U(1)_R^{\text{SC}}=\U(1)_R^{\text{mix}}+\frac{1}{2} \U(1)_b$ is the one extremizing the $c$ central charge. Again, as it involves mixing of $\U(1)_R^{\text{mix}}$, which is the R-symmetry used in the above calculation of the elliptic genus, with $\U(1)_b$ only and as the additional operator is uncharged under $\U(1)_b$, it naively appears that its charges at the fixed point are still consistent with a marginal operator.

However, there is the issue of the unbound moduli. The structure of the moduli space is similar to the previous case, where we have the gauge invariants $\hat{\Phi}^2$, $\hat{P}^2$ and $\hat{\Phi}\hat{P}$, with the last one forced to zero by the F-terms. Overall, we again find two non-compact branches parametrized by $\hat{\Phi}^2$ and $\hat{P}^2$. Demanding that $R[\hat{\Phi}]=R[\hat{P}]=0$, the c-extremization determine the R-symmetry to be the one where $R[\hat{\Phi}]=R[\hat{P}]=R[\hat{\Psi}]=0$, $R[\hat{\Gamma}]=1$. With these charges, we again have that $Tr(\U(1)^2_R) = 1$. The R-symmetry giving these charges is simply: $\U(1)_R = \U(1)_R^{\text{mix}} - \frac{1}{2}\U(1)_d$, and we note that now there is a mixing with $\U(1)_d$. As the additional operator carry negative $\U(1)_d$ charge, we conclude that if its $\U(1)_R^{\text{mix}}$ R-charge was $1$, which is the required value for a marginal operator, then its $\U(1)_R$ value is $3$. This R-charge is consistent with that of an irrelevant operator and we appear to conclude that turning it on would just lead us back to the same fixed point. As such, it is still not clear how the 2d self-duality holds, unless the analysis performed here is too naive and fails for some subtle reason, like emergence of accidental symmetries.   

\subsubsection{$N_c=2$, $N_f=5$}

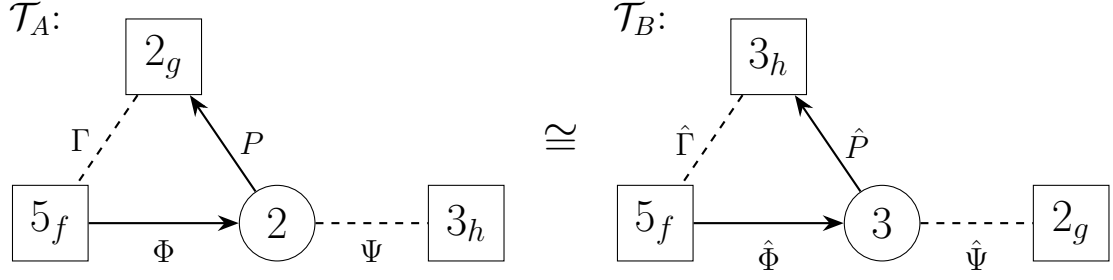
\begin{figure}
    \centering
    \begin{tikzpicture}[
        flavor/.style={draw, rectangle, minimum size=10mm, inner sep=2pt, font=\Large},
        gauge/.style={draw, circle, minimum size=10mm, inner sep=2pt, font=\Large},
        blackdot/.style={fill=black, circle, minimum size=2mm, inner sep=0pt},
        >={Stealth[scale=1.2]}
    ]

        \node[font=\large] at (-0.2, 2.7) {$\mathcal{T}_A$:};
        
        \node[flavor] (5fA) at (0, 0) {$5_f$};
        \node[gauge] (2A) at (3, 0) {$2$};
        \node[flavor] (2gA) at (1.5, 2.2) {$2_g$};
        \node[flavor] (3hA) at (5.5, 0) {$3_h$};

        \draw[->, thick] (5fA) -- node[below, yshift=-2pt] {$\Phi$} (2A);
        \draw[->, thick] (2A) -- node[right, xshift=2pt] {$P$} (2gA);
        \draw[thick, dashed] (2gA) -- node[left, xshift=-2pt] {$\Gamma$} (5fA);
        \draw[thick, dashed] (2A) -- node[below, yshift=-2pt] {$\Psi$} (3hA);

        \node[font=\Large] at (6.75, 1.1) {$\cong$};

        \node[font=\large] at (7.8, 2.7) {$\mathcal{T}_B$:};
        
        \node[flavor] (5fB) at (8, 0) {$5_f$};
        \node[gauge] (3B) at (11, 0) {$3$};
        \node[flavor] (3hB) at (9.5, 2.2) {$3_h$};
        \node[flavor] (2gB) at (13.5, 0) {$2_g$};

        \draw[->, thick] (5fB) -- node[below, yshift=-2pt] {$\hat{\Phi}$} (3B);
        \draw[->, thick] (3B) -- node[right, xshift=2pt] {$\hat{P}$} (3hB);
        \draw[thick, dashed] (3hB) -- node[left, xshift=-2pt] {$\hat{\Gamma}$} (5fB);
        \draw[thick, dashed] (3B) -- node[below, yshift=-2pt] {$\hat{\Psi}$} (2gB);

    \end{tikzpicture}
    \caption{Quiver diagram of an $\SU(2)$ - $\SU(3)$ gauge duality emerging from a reduction of Seiberg duality on a sphere.}
    \label{fig:SU3_duality_quiver}
\end{figure}

\begin{table}
\begin{centering}
\begin{tabular}{c|c|c c c c c c c c}

 & $\SU(2)_{\text{gauge}}$ & $\SU(3)_{h}$ & $\SU(5)_{f}$ & $\SU(2)_{g}$ & $\U(1)_{b}$ & $\U(1)_{c}$ & $\U(1)_{d}$ & $\U(1)_{R}^{\text{SC}}$ & $\U(1)_{R}^{\text{mix}}$\tabularnewline

\hline 
$\Phi$ & $\Box$ & $1$ & $\Box$ & $1$ & $\frac{1}{2}$ & $0$ & $1$ & $0$ & $\frac{2}{5}$\tabularnewline

$P$ & $\Box$ & $1$ & $1$ & $\Box$ & $-\frac{1}{2}$ & $3$ & $1$ & $0$ & $\frac{1}{5}$\tabularnewline

$\Psi$ & $\Box$ & $\Box$ & $1$ & $1$ & $-\frac{1}{2}$ & $-2$ & $1$ & $0$ & $\frac{1}{5}$\tabularnewline

$\Gamma$ & $1$ & $1$ & $\overline{\Box}$ & $\Box$ & $0$ & $-3$ & $-2$ & $1$ & $\frac{2}{5}$\tabularnewline

\end{tabular}
\par \vspace{1.5ex}
$\cong$ \ %
\begin{tabular}{c|c|c c c c c c c c}

 & $\SU(3)_{\text{gauge}}$ & $\SU(3)_{h}$ & $\SU(5)_{f}$ & $\SU(2)_{g}$ & $\U(1)_{b}$ & $\U(1)_{c}$ & $\U(1)_{d}$ & $\U(1)_{R}^{\text{SC}}$ & $\U(1)_{R}^{\text{mix}}$\tabularnewline
\hline 
$\Phi$ & $\overline{\Box}$ & $1$ & $\overline{\Box}$ & $1$ & $\frac{1}{3}$ & $0$ & $-1$ & $0$ & $\frac{4}{15}$\tabularnewline

$P$ & $\Box$ & $\overline{\Box}$ & $1$ & $1$ & $-\frac{1}{3}$ & $2$ & $-1$ & $0$ & $\frac{2}{15}$\tabularnewline

$\Psi$ & $\Box$ & $1$ & $1$ & $\Box$ & $-\frac{1}{3}$ & $-3$ & $-1$ & $0$ & $\frac{2}{15}$\tabularnewline

$\Gamma$ & $1$ & $\Box$ & $\Box$ & $1$ & $0$ & $-2$ & $2$ & $1$ & $\frac{3}{5}$\tabularnewline

\end{tabular}
\par\end{centering}
\caption{Table of gauge and symmetry group representations for the $\SU(2)$ - $\SU(3)$ gauge duality}
\label{table:SU3_duality_table}
\end{table}

Next we consider the case of $\SU(2)$ with five flavors. Now the dual theory is $\SU(3)$ with five flavors, so both sides of the duality are gauge theories. The 2d reduction of this 4d Seiberg duality leads to a similar 2d duality between an $\SU(2)$ and an $\SU(3)$ gauge theory. Quiver diagrams of the two dual theories are depicted in figure \ref{fig:SU3_duality_quiver}. A table listing the matter content and charges under the symmetries is given in \ref{table:SU3_duality_table}. Like the previous cases, both theories contain the superpotential $W=\Gamma P \Phi$. 

The $\SU(5)_f$ and $\SU(2)_g \times \SU(3)_h\times \U(1)_c$ come from the 4d $\SU(5)_L\times \SU(5)_R$ rotating the fundamental and anti-fundamental chiral fields, while $\U(1)_b$ descends from the 4d baryon number symmetry. However, $\U(1)_d$ does not appear in 4d, as it is broken by ABJ type anomalies, and so only emerges in 2d. 

One can check that the 't Hooft anomalies of the symmetries originating from 4d match between the proposed dual sides. However, there is a mismatch in most of the 't Hooft anomalies involving $\U(1)_d$, save for that for $\U(1)^{SC}_R$ which vanishes on both sides. The mismatching anomalies are:
\bea
&& Tr(\U(1)^2_d)_{\SU(2)} = -32 \neq Tr(\U(1)^2_d)_{\SU(3)} = -42 \; , \\ \nonumber 
&& Tr(\U(1)_d \U(1)_b)_{\SU(2)} = 6 \neq Tr(\U(1)_d \U(1)_b)_{\SU(3)} = -4 \; , \\ \nonumber
&& Tr(\U(1)_d \U(1)_c)_{\SU(2)} = -36 \neq Tr(\U(1)_d \U(1)_c)_{\SU(3)} = 24 \; . \\ \nonumber
\eea

Note though that in all cases the mismatch is a multiple of 10. This again can be attributed to the fact that the 4d axial symmetry is broken to $\mathbb{Z}_{10}$ by the ABJ anomaly so the anomalies should match if $\U(1)_d$ is broken to $\mathbb{Z}_{10}$. As before, this naturally suggests that the duality should really be thought of as between two versions of these theories where $\U(1)_d$ is broken to at least $\mathbb{Z}_{10}$. However, there does not appear to be any perturbative superpotential that can implement this breaking leading to the now familiar puzzle involving this duality. To better understand this issue we again turn to the elliptic genus.

\paragraph{Elliptic genus} Let us now compute the elliptic genus of the two theories including the addition of a fugacity for $\U(1)_d$. For convenience of the series expansion computation in $q$, we use the mixed R-symmetry $\U(1)_R^{\text{mix}}$ written in table \ref{table:SU3_duality_table} and chosen so that the R-charges are strictly positive. We find: 

\begin{align}
    \mathcal{I}_{\text{JK}}^{\SU(2)} &= 1+q^{\frac{1}{5}}\frac{c^{6}d^{2}}{b}+q^{\frac{2}{5}}\left(\frac{c^{12}d^{4}}{b^{2}}+bd^{2}\mathbf{10}_{f}\right)-q^{\frac{1}{2}}c^{5}\bar{\mathbf{3}}_{h}\mathbf{2}_{g}+q^{\frac{3}{5}}\left(\frac{c^{18}d^{6}}{b^{3}}-bc^{2}\mathbf{5}_{f}\bar{\mathbf{3}}_{h}\right) \notag\\
    &\quad -q^{\frac{7}{10}}\left(\frac{c^{11}d^{2}}{b}\bar{\mathbf{3}}_{h}\mathbf{2}_{g}+\frac{cd^{2}}{b}\mathbf{3}_{h}\mathbf{2}_{g}+bc^{3}d^{4}\overline{\mathbf{10}}_{f}\mathbf{2}_{g}+\frac{1}{c^{3}d^{2}}\bar{\mathbf{5}}_{f}\mathbf{2}_{g}\right) \notag\\
   &\quad +q^{\frac{4}{5}}\left(b^{2}d^{4}\overline{\mathbf{50}}_{f}+\frac{bc^{4}\bar{\mathbf{6}}_{h}}{d^{2}}+\frac{c^{24}d^{8}}{b^{4}}-\frac{d^{2}\mathbf{5}_{f}\mathbf{3}_{h}}{c^{2}}\right)-q^{\frac{9}{10}}\left(\frac{c^{17}d^{4}}{b^{2}}\bar{\mathbf{3}}_{h}\mathbf{2}_{g}+\frac{c^{7}d^{4}}{b^{2}}\mathbf{3}_{h}\mathbf{2}_{g}\right) \notag\\
   &\quad +q\left(-b^{2}c^{2}d^{2}\overline{\mathbf{40}}_{f}\bar{\mathbf{3}}_{h}+\mathbf{3}_{g}\left(bc^{6}d^{6}\bar{\mathbf{5}}_{f}+c^{10}\mathbf{3}_{h}+1\right)+c^{10}\bar{\mathbf{6}}_{h}+\frac{c^{30}d^{10}}{b^{5}}-\frac{1}{d^{10}}+\mathbf{24}_{f}+\mathbf{8}_{h}+3\right) \notag\\
    &\quad +O(q^{\frac{11}{10}})
\end{align}
\begin{align}
   \mathcal{I}_{\text{JK}}^{\SU(3)} &= 1+q^{\frac{1}{5}}\frac{c^{6}}{bd^{3}}+q^{\frac{2}{5}}\left(\frac{c^{12}}{b^{2}d^{6}}+\frac{b}{d^{3}}\mathbf{10}_{f}\right)-q^{\frac{1}{2}}c^{5}\bar{\mathbf{3}}_{h}\mathbf{2}_{g}+q^{\frac{3}{5}}\left(\frac{c^{18}}{b^{3}d^{9}}-\frac{bc^{2}}{d^{5}}\mathbf{5}_{f}\bar{\mathbf{3}}_{h}\right) \notag\\
   &\quad -q^{\frac{7}{10}}\left(\frac{c^{11}}{bd^{3}}\bar{\mathbf{3}}_{h}\mathbf{2}_{g}+\frac{c}{bd^{3}}\mathbf{3}_{h}\mathbf{2}_{g}+\frac{bc^{3}}{d}\overline{\mathbf{10}}_{f}\mathbf{2}_{g}+\frac{1}{c^{3}d^{2}}\bar{\mathbf{5}}_{f}\mathbf{2}_{g}\right) \notag\\
   &\quad +q^{\frac{4}{5}}\left(\frac{b^{2}}{d^{6}}\overline{\mathbf{50}}_{f}+\frac{bc^{4}}{d^{7}}\bar{\mathbf{6}}_{h}+\frac{c^{24}}{b^{4}d^{12}}-\frac{d^{2}\mathbf{5}_{f}\mathbf{3}_{h}}{c^{2}}\right)-q^{\frac{9}{10}}\left(\frac{c^{17}}{b^{2}d^{6}}\bar{\mathbf{3}}_{h}\mathbf{2}_{g}+\frac{c^{7}}{b^{2}d^{6}}\mathbf{3}_{h}\mathbf{2}_{g}\right) \notag\\
   &\quad +q\left(-\frac{b^{2}c^{2}\overline{\mathbf{40}}_{f}\bar{\mathbf{3}}_{h}}{d^{8}}+\mathbf{3}_{g}\left(bdc^{6}\bar{\mathbf{5}}_{f}+c^{10}\mathbf{3}_{h}+1\right)+c^{10}\bar{\mathbf{6}}_{h}+\frac{c^{30}}{b^{5}d^{15}}-d^{10}+\mathbf{24}_{f}+\mathbf{8}_{h}+3\right) \notag\\
   &\quad +O(q^{\frac{11}{10}})
\end{align}

We can make the following observations regarding the two elliptic genera. First, one can notice that they agree if the fugacity $d$ is set to one. Moreover, the equality also holds if we set $d^{10}=1$, assuming we map the fugacities as $d_{\SU(2)} \rightarrow d_{\SU(3)}$, $b_{\SU(2)}\rightarrow b_{\SU(3)} d^5_{\SU(3)}$. This is consistent with the anomaly matching\footnote{Here there is a subtlety as the symmetry mapping involves a mixing with a $\mathbb{Z}_2$ element of $U(1)_b$ on the $\SU(3)$ side. However, this does not impede the anomaly matching as the 't Hooft anomalies of $U(1)_b$ are all even.} and suggests that the duality cannot hold unless $\U(1)_d$ is broken to a $\mathbb{Z}_{10}$.

Next, it is interesting to examine the order $q$ terms in the elliptic genus, specifically for the presence of conserved currents and marginal operators. Indeed, note that both genera contain the contribution $+q(3+\boldsymbol{24}_{f}+\boldsymbol{8}_{h}+\boldsymbol{3}_{g})$ which we can identify as the conserved currents for the $\U(1)_b \times \U(1)_c \times \U(1)_d \times \SU(5)_f \times \SU(3)_h\times \SU(2)_g$ global symmetry. As before, we observe additional positive terms at order $q$ that do not appear to be conserved currents. 

Next we consider the negatively charged terms at order $q$, which could be marginal operators. There are two such terms, of which only one is charged only under $\U(1)_d$. Introducing it to the superpotential will then break $\U(1)_d$ with no further casualties among the symmetries. Interestingly, this apparent marginal operator is not constructed from the basic fields of the gauge theory. This can be noted by comparing the elliptic genus with the perturbative counting. Specifically, we find:
\begin{align} \begin{split}
    \mathcal{I}_{\text{JK}}^{\SU(2)} = \mathcal{I}_{\text{pert.}}^{\SU(2)} - d^{-10}q +o(q) \\
    \mathcal{I}_{\text{JK}}^{\SU(3)} = \mathcal{I}_{\text{pert.}}^{\SU(3)} - d^{10}q +o(q)
\end{split} \end{align}
which supports the claim that the additional operator does not appear from the perturbative matter. The charges of the operator then suggest that deforming by it breaks the $\U(1)_d$ symmetry to $\mathbb{Z}_{10}$ on both sides, which as we have seen is what is required for the matching of the elliptic genera and 't Hooft anomalies. As before, this suggests that the correct way to understand the duality is as one between the theories deformed by this superpotential.

\paragraph{c-extremization} As before, it is interesting to examine the flow using the procedure of c-extremization. Specifically, we consider the flow with $\U(1)_d$ unbroken, try to determine the dimension of the operators at the fixed point and use this to estimate to behavior of the RG flow once the superpotential term breaking $\U(1)_d$ is inserted. Like in the previous examples, the theory has a moduli space with non-compact directions and we may need to set the R-charges of the chiral fields to zero to accommodate for that. Below we present the calculation in both cases and comment on the results. 

We begin with the unconstrained extremization. In this case we find that the superconformal R-symmetry is $\U(1)_R^{\text{SC}}=\U(1)_R^{\text{mix}}+\frac{2}{5} \U(1)_b$. Note that there is no mixing with $\U(1)_d$ and so the potential marginal operator observed above is just as marginal with this R-symmetry. 

However, if we enforce that the R-charges of $\Phi$ and $P$ are zero, then we instead find that the superconformal R-symmetry is the $\U(1)_R^{\text{SC}}$ appearing in \ref{table:SU3_duality_table}, giving R-charge $0$ to all fields save $\Gamma$. In terms of $\U(1)_R^{\text{mix}}$ used to compute the elliptic genus, it is given by: $\U(1)_R^{\text{SC}}=\U(1)_R^{\text{mix}}-\frac{3}{10} \U(1)_d-\frac{1}{5} \U(1)_b$ for the $\SU(2)$ case, and by $\U(1)_R^{\text{SC}}=\U(1)_R^{\text{mix}}+\frac{1}{5} \U(1)_d-\frac{1}{5} \U(1)_b$ for the $\SU(3)$ case. In both cases one see that the mixing with $\U(1)_d$ is such that the additional operator becomes irrelevant.

\subsection{2d reduction of S-confining $\USp$ gauge theory}

\begin{figure}
    \centering
    \begin{tikzpicture}[
        flavor/.style={draw, rectangle, minimum size=8mm, inner sep=4pt, font=\large},
        gauge/.style={draw, circle, minimum size=12mm, inner sep=2pt, font=\large},
        blackdot/.style={fill=black, circle, minimum size=2mm, inner sep=0pt},
        >={Stealth[scale=1.2]}
    ]

        \node[font=\large] at (-1.5, 2.5) {$\mathcal{T}_A$:};
        
        \node[gauge] (2nA) at (0, 0) {$2n$};
        \node[flavor] (2n3A) at (3, 0) {$2n+3$};
        \node[blackdot, label={right:\,$\psi$}] (dotA) at (0, 2) {};

        \draw[<-, thick] (2nA) -- node[above, yshift=2pt] {$Q$} (2n3A);
        \draw[thick, dashed] (2nA) -- (dotA);

        \node[font=\Large] at (5.5, 0.5) {$\cong$};

        \node[font=\large] at (6.5, 2.5) {$\mathcal{T}_B$:};
        
        \node[blackdot] (dotB) at (8, 0) {};
        \node[flavor] (2n3B) at (11, 0) {$2n+3$};
        
        \node[draw, minimum width=4mm, minimum height=8mm, inner sep=0pt] (phiB) at (11, 2) {};
        \draw (phiB.west) -- (phiB.east);
        \node[right=2pt] at (phiB.east) {$\Phi_M$};

        \draw[->, thick] (2n3B) -- (phiB);
        \draw[thick, dashed] (dotB) -- node[above] {$\Psi$} (2n3B);

    \end{tikzpicture}
    \caption{Quiver diagram of a gauge-WZ duality with antisymmetric matter. The circle denote this time $\text{USp}$ gauge group.}
    \label{fig:duality_quiver_3}
\end{figure}
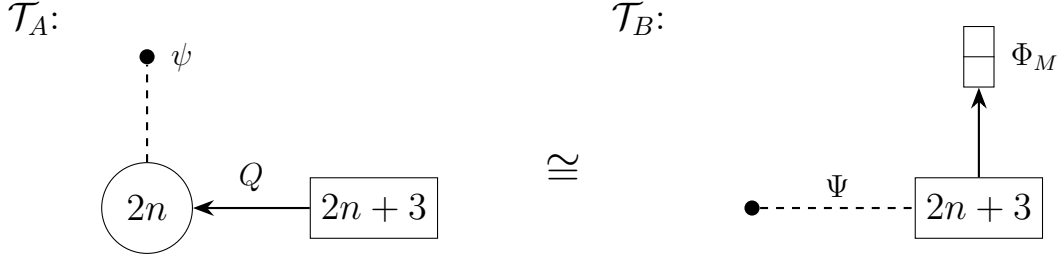

Additional examples of 2d dualities can be generated from 4d Seiberg dualities for more general groups or from more intricate 4d dualities, see \cite{Sacchi:2020pet,Jiang:2024ifv,Amariti:2024usp,Amariti:2025jvi} for some examples. Here we shall consider a specific example that was recently reviewed in \cite{Amariti:2025jvi}. This example comes from the reduction of an S-confining 4d $\mathcal{N}=1$ $\USp(2n)$ gauge theory. As such, one side of the 2d duality involves a $\USp(2n)$ gauge theory, while the other is a WZW model. Figure \ref{fig:duality_quiver_3} provides quiver diagrams of the dual theories and table \ref{table:GaugeWZDuality_table1} illustrates their charges under the various symmetries. 

\begin{table}

\begin{centering}

\begin{tabular}{c|c|c c c c}

 & $\USp(2n)_{\text{gauge}}$ & $\SU(2n+3)$ & $\U(1)_{b}$ & $\U(1)_{a}$ & $\U(1)^{4d}_{R}$ \\
\hline 

$Q$ & $\ydiagram{1}$ & $\ydiagram{1}$ & $1$ & $1$ & $0$\\

$\psi$ & $\ydiagram{1}$ & $1$ & $-2n-3$ & $-1$ & $1$\\

\end{tabular} 
\par \vspace{1.5ex}
$\cong$ \ %
\begin{tabular}{c|c c c c}

 & $\SU(2n+3)$ & $\U(1)_{b}$ & $\U(1)_{a}$ & $\U(1)^{4d}_{R}$ \\
\hline 
$\Phi_M$ & $\ydiagram{1,1}$ & $2$ & $2$ & $0$ \\

$\Psi$ & $\ydiagram{1}$ & $-2n-2$ & $0$ & $1$ \\

\end{tabular}
\par\end{centering}
\caption{Table of gauge and symmetry group representations of the gauge - WZ duality. $\U(1)_R^{4d}$ is the R-symmetry inherited from 4d.
}
\label{table:GaugeWZDuality_table1}
\end{table}

The mapping of operators between the dual sides is that the basic gauge invariant operators on the gauge theory side, $Q^2$ and $\psi Q$, map to the basic fields of the WZW model: $Q^2 \leftrightarrow \Phi_M$, $\psi Q \leftrightarrow \Psi$. Note that the WZW side also has a superpotential of the form $W=\Phi^{n+1}_M \Psi$. This is needed to match the non-abelian flavor symmetry, as without it the WZW model would have $\SU((n+1)(2n+3))\times \SU(2n+3)$ symmetry from independent rotations of $\Phi_M$ and $\Psi$. On the topic of symmetries, the $\SU(2n+3)$ and $\U(1)_b$ descend from 4d while $\U(1)_a$ is accidental, emerging from the 4d axial symmetry that is broken by the ABJ anomaly. Note that the superpotential on the WZW side also breaks $\U(1)_a$ to $\mathbb{Z}_{2n+2}$, which is also the discrete remnant of the 4d axial symmetry. However, there is no superpotential on the gauge theory side, and furthermore, there does not appear to be any superpotential one can write from the basic fields without breaking additional symmetries. As such, the presence of $\U(1)_a$ seems to impede the duality.  

As before, all 't Hooft anomalies not involving $\U(1)_a$ match across the duality as expected from the 4d origin, but those involving $\U(1)_a$ do not match. Specifically, we find: 

\bea
&& \text{Tr}(\U(1)^2_a)_{WZW} = 4(n+1)(2n+3) \neq \text{Tr}(\U(1)^2_a)_{gauge} = 4n(n+1) \; , \\ \nonumber 
&& \text{Tr}(\U(1)_a \U(1)_b)_{WZW} = 4(n+1)(2n+3) \neq \text{Tr}(\U(1)_a \U(1)_b)_{gauge} = 0 \; , \\ \nonumber
&& \text{Tr}(\U(1)_a \U(1)_R)_{WZW} = -2(n+1)(2n+3) \neq \text{Tr}(\U(1)_a \U(1)_R)_{gauge} = -4n(n+1) \; ,
\eea
where here we have also considered $\U(1)_a$ for the WZW model, even though it will be broken by the superpotential. One can note that the anomalies match if $\U(1)_a$ is broken to $\mathbb{Z}_{2n+2}$, as the difference is always a multiple of $2(n+1)$. One option then is that the $\mathbb{Z}_{2n+2}$ symmetry of the WZW model is enhanced back to $\U(1)_a$ at the IR, which then matches the $\U(1)_a$ symmetry on the gauge theory side. Another option is that there exist a superpotential on the gauge theory side breaking the $\U(1)_a$ symmetry and the duality is between the WZW model and the gauge theory deformed by this superpotential. We shall explore both options by evaluating the elliptic genus for low values of $n$.  

\subsubsection{$n=1$}

We begin with the $n=1$ case. We shall evaluate the elliptic genus of both the gauge theory and the WZW model refined by a fugacity for $\U(1)_a$. Note that this implies that we do not turn on the superpotential on the WZW side. Using the mixed R-symmetry $\U(1)_{R_{\text{mixed}}} = \U(1)^{4d}_{R} + \frac{1}{6} \U(1)_b$, we find:

\begin{equation}\label{EG:su2w4F}
\begin{split}
\mathcal{I}_{\text{JK}}^{\text{gauge}}
&= 1 + a^2b^2\,\mathbf{10}_u\,q^{1/6} + a^4b^4\,\overline{\mathbf{50}}_u\,q^{1/3}
   + a^2b^6\bigl(a^4\,\mathbf{175}''_u - \mathbf{5}_u\bigr)\,q^{1/2} \\
&\quad + \Bigl(a^8b^8\,\overline{\mathbf{490}}_u - a^4b^8\,\overline{\mathbf{40}}_u
   - \frac{\mathbf{5}_u}{b^4}\Bigr)\,q^{2/3} \\
&\quad + a^2\Bigl(b^{10} - \frac{\overline{\mathbf{40}}_u}{b^2} - a^4b^{10}\,\mathbf{175}'_u
   + a^8b^{10}\,\mathbf{1176}'_u\Bigr)\,q^{5/6} \\
&\quad + \Bigl(2 - \frac{1}{a^4} + \mathbf{24}_u - a^4\,\mathbf{175}'_u
   + b^{12}\bigl[a^4\,\mathbf{10}_u - a^8\,\mathbf{560}'_u + a^{12}\,\mathbf{2520}''''_u\bigr]\Bigr)\,q \\
&\quad + O\bigl(q^{7/6}\bigr)
\end{split}
\end{equation}

\begin{equation}\label{EG:su2wzw}
\begin{split}
\mathcal{I}_{\text{JK}}^{\text{WZ}}
&= 1 + a^2b^2\,\mathbf{10}_u\,q^{1/6}
   + b^4\bigl((a^4-1)\,\overline{\mathbf{5}}_u + a^4\,\overline{\mathbf{50}}_u\bigr)\,q^{1/3} \\
&\quad + a^2b^6\bigl(-\mathbf{5}_u + (a^4-1)\,\mathbf{45}_u + a^4\,\mathbf{175}''_u\bigr)\,q^{1/2} \\
&\quad + \Bigl(-\frac{\mathbf{5}_u}{b^4}
   + b^8\bigl[-(a^4-1)\,\overline{\mathbf{10}}_u + a^4(a^4-1)\,\overline{\mathbf{15}}_u
   - a^4\,\overline{\mathbf{40}}_u \\
&\qquad\qquad + a^4(a^4-1)\,\overline{\mathbf{210}}_u + a^8\,\overline{\mathbf{490}}_u\bigr]\Bigr)\,q^{2/3} \\
&\quad + \Bigl(a^2b^{10}\bigl[1 - (a^4-1)\bigl(\mathbf{24}_u + \mathbf{75}_u\bigr) - a^4\,\mathbf{175}'_u \\
&\qquad\qquad + a^4(a^4-1)\bigl(\overline{\mathbf{126}}_u + \overline{\mathbf{700}}'_u\bigr)
   + a^8\,\mathbf{1176}'_u\bigr] \\
&\qquad\quad - \frac{1}{b^2}(a^2\,\overline{\mathbf{40}}_u + (a^2-\frac{1}{a^2})\,\overline{\mathbf{10}}_u)\Bigr)\,q^{5/6} \\
&\quad + \Bigl(2 - a^4 - (a^4-2)\,\mathbf{24}_u - (a^4-1)\,\mathbf{75}_u - a^4\,\mathbf{175}'_u \\
&\qquad\quad + b^{12}\bigl[(2a^4-1)\,\mathbf{10}_u
   - a^4(a^4-1)\bigl(\mathbf{40}_u + \mathbf{175}_u + \mathbf{315}_u\bigr) \\
&\qquad\qquad + a^8(a^4-1)\bigl(\mathbf{35}_u + \mathbf{560}_u + \mathbf{1890}'_u\bigr)
   - a^8\,\mathbf{560}'_u + a^{12}\,\mathbf{2520}''''_u\bigr]\Bigr)\,q \\
&\quad + O\bigl(q^{7/6}\bigr)
\end{split}
\end{equation}

As before, we would be particularly concerned with the operators contributing at order $q$ which contain conserved currents and marginal operators. As we do not want to break $\U(1)_b$ and conserved currents should not be charged under it, we will be especially interested in $\U(1)_b$ neutral operators contributing at order $q$. In the gauge theory side these are: $q(\boldsymbol{24}+2-a^{-4}-a^{4}\boldsymbol{175}')$. The positive terms are consistent with being the conserved currents for the $\SU(5)\times \U(1)_b \times \U(1)_a$ flavor symmetry. The last negative term can be identified with the operator $Q^5\psi$. This is indeed a marginal operator that can be inserted into the superpotential, though it would break the $\SU(5)$ global symmetry. Finally, we have the operator $-q a^{-4}$. This has the right charges to be a marginal operator charged only under $\U(1)_a$. Note that it cannot be built from the perturbative fields. Indeed, we can compare the index $\mathcal{I}_{\text{JK}}^{\text{gauge}}$ against the perturbative counting of gauge invariants, finding:

\begin{equation}
\mathcal{I}_{\text{JK}}^{\text{gauge}}=\mathcal{I}_{\text{pert}.}^{\text{gauge}}-a^{-4}q+O\left(q^{2},b^{4}\right) .
\end{equation} 

This suggests that this operator has a non-perturbative origin exactly as in the previous examples. Similarly, turning it in the superpotential should break $\U(1)_a$ to $\mathbb{Z}_4$, and the duality would then appear to work. Specifically, note that if we set $a^4=1$ in \eqref{EG:su2w4F} and \eqref{EG:su2wzw} then the two genera match, and we have also noted that the anomalies match in this case.  

We can similarly consider the $\U(1)_b$ neutral operators contributing at order $q$ on the WZW side. These give: $q(2+\boldsymbol{75}+2\cdot\boldsymbol{24}-a^{4}\left(1+\boldsymbol{75}+\boldsymbol{24}+\boldsymbol{175'}\right))$. The positive terms are consistent with the conserved currents for an $\SU(10)_{\Phi}\times \SU(5)_{\Psi}\times \U(1)_b \times \U(1)_a$ flavor symmetry, where the $\SU(5)$ refined in the index is the diagonal of $\SU(5)_{\Phi}\times \SU(5)_{\Psi}$ with $\SU(5)_{\Phi} \subset \SU(10)_{\Phi}$ such that $\boldsymbol{10}_{\SU(10)_{\Phi}} \rightarrow \boldsymbol{10}_{\SU(5)_{\Phi}}$. This is since without the superpotential on the WZW side, the non-abelian global symmetry is actually $\SU(10)\times \SU(5)$. The negative terms are consistent with the operator $\Phi^2_M \Psi$. Note that this is a marginal operator in the $a^4\boldsymbol{55}_{\SU(10)_{\Phi}} \boldsymbol{5}_{\SU(5)_{\Psi}}$. Turning it into the superpotential would break $\SU(10)_{\Phi}\times \SU(5)_{\Psi}$ to the diagonal $\SU(5)$ refined in the elliptic genus and break $\U(1)_a$ to $\mathbb{Z}_4$. This can be seen by the presence of the singlet in the elliptic genus corresponding to the invariant part under the diagonal $\SU(5)$, and by the fact that once we set $a^4=1$ in \eqref{EG:su2wzw} some of the conserved currents get canceled by the marginal operators, leaving only the conserved currents for $\SU(5)\times \U(1)_b$. Note that we do not find any indication for an additional current corresponding to the $\U(1)_a$ of the gauge theory. Of course, this does not rule out its presence, as it could be canceled by another operator as indeed happens in \eqref{EG:su2w4F}. Nevertheless, the 4d picture combined with the presence of an operator on the gauge theory side with the right charges to break $\U(1)_a$ suggests the proper interpretation is that the duality should be understood with the gauge theory perturbed by the additional superpotential.   

\paragraph{c-extremization} We can again try to get a better understanding of the RG flow by using c-extremization to study the RG flow without the superpotential. With no superpotential turned on, the theory has a non-trivial moduli space spanned by the chiral fields on both sides, so we insist that their R-charge is set to zero. It is straightforward to show that in that case the superconformal R-symmetry would just be the one where $R[Q]=R[\psi]=0$ on the gauge theory side and $R[\Phi]=R[\Psi]=0$ on the WZW side. In terms of the symmetries in \ref{table:GaugeWZDuality_table1}, it is given by: $\U(1)_R = \U(1)_R^{4d} + \frac{1}{4}\U(1)_b - \frac{1}{4}\U(1)_a$. Due to the mixing with $\U(1)_a$ the marginal operators noted above will have different R-charges with respect to that R-symmetry. On the WZW side the superpotential carry $\U(1)_a$ charge of $4$ and so would carry R-charge $0$ under the new R-symmetry. As such, it should be relevant compared to that fixed point and inserting it into the superpotential should initiate an RG flow to a new fixed point.

On the other hand, on the gauge theory side, the marginal operator carry $\U(1)_a$ charge of $-4$ and so would carry R-charge $2$ under the new R-symmetry. As such, it appears to be irrelevant compared to that fixed point and should just flow to zero. We encounter then the same issue appearing in the previous examples.

One can also perform the c-extremization procedure without the constraint on the R-charges of the scalar fields. In that case, one finds the superconformal R-symmetry to be one where $R[Q]=1, R[\psi]=0$ on the gauge theory side. In terms of the symmetries in \ref{table:GaugeWZDuality_table1}, it is given by $\U(1)_R = \U(1)_R^{4d} + \U(1)_a$. Under this R-symmetry, the additional operator is now relevant. However, note that the resulting central charge is negative so this should not be the true superconformal R-symmetry.  

\subsubsection{$n=2$}

We can treat the $n=2$ case in a similar manner. As before, we evaluate the elliptic genus of the two sides refined by a fugacity for $\U(1)_a$. Using mixed R-charge $\U(1)_{R_{\text{mixed}}} = \U(1)^{4d}_{R} + \frac{1}{8} \U(1)_b$, we find:

\begin{equation}
\begin{split}
\mathcal{I}_{\text{JK}}^{\text{WZ}}
&= 1 + a^2b^2\,\mathbf{21}_u\,q^{1/8}
   + a^4b^4\bigl(\overline{\mathbf{35}}_u + \overline{\mathbf{196}}_u\bigr)\,q^{1/4} \\
&\quad + b^6\bigl((a^6-1)\,\overline{\mathbf{7}}_u
   + a^6(\overline{\mathbf{588}}_u + \overline{\mathbf{1176}}_u)\bigr)\,q^{3/8}  + ... \\
&\quad + \Bigl(2 - a^6 - (a^6-2)\,\mathbf{48}_u - (a^6-1)\,\mathbf{392}_u
   - a^6(\mathbf{784}_u + \mathbf{2940}_u + \mathbf{3528}_u + \mathbf{4704}''_u) \\
&\qquad\quad + a^4b^{16}\bigl[2\,\mathbf{21}_u
   - (a^6-1)(\mathbf{224}_u + \mathbf{490}_u + \mathbf{735}_u + \mathbf{3360}_u) \\
&\qquad\qquad - a^6(\mathbf{2940}'_u + \mathbf{9408}_u + \mathbf{17640}_u + \mathbf{19404}_u + \mathbf{20160}_u) \\
&\qquad\qquad + a^6(a^6-1)(\mathbf{756}_u + \mathbf{4725}_u + \mathbf{37044}_u
   + \mathbf{116424}'''_u + \mathbf{138600}_u) \\
&\qquad\qquad + a^{12}(\mathbf{24696}_u + \mathbf{346500}_u + \mathbf{429429}_u
   + \mathbf{952875}_u + \mathbf{1057056}_u)\bigr]\Bigr)\,q \\
&\quad + O\bigl(q^{9/8}\bigr)
\end{split}
\end{equation}

\begin{equation}
\begin{split}
\mathcal{I}_{\text{JK}}^{\text{gauge}}
&= 1 + a^2b^2\,\mathbf{21}_u\,q^{1/8}
   + a^4b^4\bigl(\overline{\mathbf{35}}_u + \overline{\mathbf{196}}_u\bigr)\,q^{1/4}
   + a^6b^6\bigl(\overline{\mathbf{588}}_u + \overline{\mathbf{1176}}_u\bigr)\,q^{3/8} + ... \\
&\quad + \Bigl(2 - \frac{1}{a^6} + \mathbf{48}_u + 2a^4b^{16}\,\mathbf{21}_u
   - a^6(\mathbf{784}_u + \mathbf{2940}_u + \mathbf{3528}_u + \mathbf{4704}''_u) \\
&\qquad\quad - a^{10}b^{16}(\mathbf{2940}'_u + \mathbf{9408}_u + \mathbf{17640}_u
   + \mathbf{19404}_u + \mathbf{20160}_u) \\
&\qquad\quad + a^{16}b^{16}(\mathbf{24696}_u + \mathbf{346500}_u + \mathbf{429429}_u
   + \mathbf{952875}_u + \mathbf{1057056}_u)\Bigr)\,q \\
&\quad + O\bigl(q^{9/8}\bigr)
\end{split}
\end{equation}

Here, the ... denotes  terms of order between $O(q^{\frac{3}{8}})$ and $O(q)$  which are omitted for brevity, but can be fully admired in \ref{appendix:q-expansion-resultsAntiSymN2}.

The two indices match up to the computed order provided that $a^6 \to 1$, but differ otherwise. As such, the elliptic genera and anomalies match if $\U(1)_a$ is broken to $\mathbb{Z}_6$. As done so far, to better understand this we turn to the order $q$ terms in the elliptic genus, where we shall concentrate on $\U(1)_b$ neutral operators. On the gauge theory side these are: $q(\boldsymbol{48}_u+2-a^{-6}-a^{6}(\mathbf{784}_u + \mathbf{2940}_u + \mathbf{3528}_u + \mathbf{4704}''_u))$. The positive terms are consistent with being the conserved currents for the $\SU(7)\times \U(1)_b \times \U(1)_a$ flavor symmetry. The negative terms going as $a^{6}$ can be identified with the operator $Q^7\psi$. This is indeed a marginal operator that can be inserted into the superpotential, though it would break the $\SU(7)$ global symmetry. Finally, we have the operator $-q a^{-6}$, which has the right charges to be a marginal operator charged only under $\U(1)_a$. As before, we note that it cannot be built from the perturbative fields, as can be confirmed by comparing the index $\mathcal{I}_{\text{JK}}^{\text{gauge}}$ against the perturbative counting of gauge invariants: 
\begin{equation}
    \mathcal{I}_{\text{JK}}^{\text{gauge}} = \mathcal{I}_{\text{pert.}}^{\text{gauge}}-a^{-6}q+O\left(q^{9/8}\right)
\end{equation}
Inserting this operator in the superpotential should then break $\U(1)_a$ to $\mathbb{Z}_6$ which as we have seen is required for the anomalies and elliptic genera to agree. 

We can perform a similar analysis on the WZW side. Here, there are more $\U(1)_b$ neutral operators that contribute at order $q$: $q(\mathbf{392}_u+2\boldsymbol{48}_u+2-a^{6}(1+\boldsymbol{48}_u+\mathbf{392}_u+\mathbf{784}_u + \mathbf{2940}_u + \mathbf{3528}_u + \mathbf{4704}''_u))$. The positive terms are consistent with the conserved currents for an $\SU(21)_{\Phi}\times \SU(7)_{\Psi}\times \U(1)_b \times \U(1)_a$ flavor symmetry, where the $\SU(7)$ refined in the index is the diagonal of $\SU(7)_{\Phi}\times \SU(7)_{\Psi}$ with $\SU(7)_{\Phi} \subset \SU(21)_{\Phi}$ such that $\boldsymbol{21}_{\SU(21)_{\Phi}} \rightarrow \boldsymbol{21}_{\SU(7)_{\Phi}}$. This is as without the superpotential on the WZW side, the non-abelian global symmetry is actually $\SU(21)\times \SU(7)$. The negative terms are consistent with the operator $\Phi^3_M \Psi$. Note that this is a marginal operator in the $a^6\boldsymbol{1771}_{\SU(21)_{\Phi}} \boldsymbol{7}_{\SU(7)_{\Psi}}$. Turning it into the superpotential would break $\SU(21)_{\Phi}\times \SU(7)_{\Psi}$ to the diagonal $\SU(7)$ refined in the elliptic genus and break $\U(1)_a$ to $\mathbb{Z}_6$. This can be seen by the presence of the singlet in the elliptic genus corresponding to the invariant part under the diagonal $\SU(7)$, and by the fact that once we set $a^6=1$ some of the conserved currents get canceled by the marginal operators, leaving only the conserved currents for $\SU(7)\times \U(1)_b$. As before, there are no indications for an additional current corresponding to the $\U(1)_a$ of the gauge theory. While this does not prove anything by itself, when combined with the 4d picture and the presence of the marginal operator on the gauge theory side, it suggests that the duality should be understood as between the WZW model and the gauge theory with both perturbed by the appropriate superpotential so as to break $\U(1)_a$ to $\mathbb{Z}_6$.

We can further study the proposed flow using the c-extremization prescription. The analysis follows quite similarly to the $n=1$ case. We again find that the superconformal R-symmetry is the one giving the charges $R[Q]=R[\psi]=0$ on the gauge theory side and $R[\Phi]=R[\Psi]=0$ on the WZW side. In terms of the symmetries in \ref{table:GaugeWZDuality_table1}, it is given by: $\U(1)_R = \U(1)_R^{4d} + \frac{1}{6}\U(1)_b - \frac{1}{6}\U(1)_a$. As both the charge of the additional operator and the mixing with $\U(1)_a$ is negative, we see that the operator is irrelevant with respect to the apparent fixed point.

The above result assumes that the R-charge of $Q$ is constrained to be zero. It is possible to do the c-extremization procedure without this constraint, in which case we again find the superconformal R-symmetry to be  $\U(1)_R = \U(1)_R^{4d} + \U(1)_a$, which is the one where $R[Q]=1, R[\psi]=0$ on the gauge theory side. Under this R-symmetry, the additional operator is now relevant, though the central charge is negative. 

\subsection{Isometry of the $S^2$}
\label{section:isometry}
So far we considered the $S^2$ reduction of a 4d $\mathcal{N}=1$ duality where we have chosen to twist using an R-charge where the chiral fields only get the R-charges $0$, $1$ or $2$. As previously mentioned, this case has some particularly nice properties that make it especially appealing. However, with these R-charges each chiral field reduces to either a single 2d chiral, if its R-charge is $0$, a 2d Fermi, if its R-charge is $2$, or does not survive in 2d. In contrast, a 4d chiral field of R-charge $R<0$ reduces to $1-R$ 2d chirals while a 4d chiral field of R-charge $R>2$ reduces to $R-1$ 2d Fermi multiplets \cite{Gadde:2015wta}. 

Cases involving 4d chiral fields of R-charge bigger than $2$ or negative then manifest a new feature, where a single 4d chiral begets multiple 2d fields. This then leads to a 2d symmetry rotating the multiple copies of said 4d chiral. We can ask whether this symmetry has an origin in the 4d construction or it is entirely accidental in 2d. It seems that some part of it is manifest in the construction but most of it is accidental. Specifically, the $S^2$ has an $\SU(2)$ isometry which in certain cases may appear as a symmetry of the low-energy theory. What is interesting about this symmetry is that the 2d anomalies involving it can be computed from the 4d anomaly polynomial \cite{Bah:2019rgq,Hosseini:2020vgl}, just like the anomalies of symmetries descending from 4d ones (see \cite{Hwang:2021xyw} for examples of this in the context of the $S^2$ reduction of 6d SCFTs). In particular, if the anomaly does not vanish then this $\SU(2)$ symmetry must appear in the low-energy theory.

What is observed in certain models then is that the symmetry rotating multiple copies of the same field is related to the $\SU(2)$ isometry symmetry of the $S^2$. Specifically, any $\SU(N)$ symmetry group has an $\SU(2)$ subgroup whose embedding is such that the fundamental representation of $\SU(N)$ gets mapped to the $N$ dimensional representation of the $\SU(2)$ \footnote{This embedding also plays a role in nilpotent orbits and punctures of 6d A-type $(2,0)$ theory where it is mapped to the maximal nilpotent orbit and maximal puncture, respectively.}. The isometry symmetry is then diagonally embedded with this embedding in each of the $\SU(N)$ rotating multiple copies of the same field. This is supported for instance by matching anomalies of this symmetry against the 4d expectations \cite{Nardoni:2024sos}.    

As such, 2d symmetries rotating multiple copies of 2d fields originating from the same 4d field are mostly accidental, save for some diagonal combination which is mapped to the $\SU(2)$ isometry of $S^2$. These additional symmetries then may interfere with 2d dualities that are motivated from the $S^2$ reduction of 4d dualities, in the same manner as in the cases examined above. For instance, the mapping of 't Hooft anomalies between 4d and 2d guarantees that the anomaly associated with the $\SU(2)$ isometry of the $S^2$ would agree between the two sides, but the 2d theories allow for individual rotations of the various multiples copies and the anomalies involved for each symmetry may not agree across the duality, see \cite{Nardoni:2024sos} for some examples. It is natural then to expect that like the previous cases, these additional symmetries are broken by a superpotential so that only the 4d symmetries and $\SU(2)$ isometry remain. Here we shall explore this conjecture using the same approach.

This case entails a few computational issues. First, as previously mentioned, it is possible for the 2d reduction of the 4d theory to yield a disjoint union of theories, coming from the sum over magnetic fluxes on the $S^2$. This can be avoided if the sum truncates so that only the zero flux sector remains (again we refer the reader to \cite{Gadde:2015wta} for the details). This turns out to happen when there are no negative R-charges. As it is not immediately clear how to think of the duality in a direct sum of theories, we shall focus only on cases where the sum truncates. Unfortunately, if one wants the sum to truncate on both sides of the duality, one is forced to only use R-charge of 0,1 and 2, as was indeed the case in the previous examples. However, to examine cases manifesting the $S^2$ isometry in 2d, we are forced to consider cases with R-charges that are bigger than $2$. This implies then that the dual side has fields with negative R-charges and the sum on that side may not truncate. This makes dealing with the dual side both technically and conceptually demanding, and we reserve this for future study.

Next we turn to considering specific cases. All cases are based on a 4d $\mathcal{N}=1$ $\SU(2)$ gauge theory with $N_f$ flavors. 

\subsubsection{$N_f=4$}

We begin with the case of 4d $\mathcal{N}=1$ $\SU(2)$ gauge theory with 8 fundamental chirals. This theory has an $\SU(8)$ global symmetry rotating the 8 chiral fields. We can also introduce a $\U(1)_A$ axial symmetry, under which the eight chirals have charge $1$, but it is broken to $\mathbb{Z}_{8}$ by the ABJ anomaly. This theory supports an anomaly free R-symmetry under which one chiral has R-charge $4$ and the rest have R-charge $0$. This makes this case the simplest (on account of smallest $N_f$) for our purposes. We illustrate the 4d theory with the chosen R-charges below.

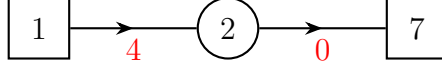
\begin{center}
\begin{tikzpicture}[node distance=2.5cm, auto, thick, scale=1, every node/.style={transform shape}]
    \node[rectangle, draw, minimum size=0.8cm] (SU1F) {$1$};
    \node[circle, draw, minimum size=0.8cm, right of=SU1F] (SU2G) {$2$};
    \node[rectangle, draw, minimum size=0.8cm, right of=SU2G] (SU7F) {$7$};

    \draw[mid arrow] (SU1F) -- node[below] {$\textcolor{red}{4}$} (SU2G);
    \draw[mid arrow] (SU2G) -- node[below] {$\textcolor{red}{0}$} (SU7F);
\end{tikzpicture}
\end{center}

We can consider the compactification of this theory on $S^2$ to get a 2d $(0,2)$ theory. To preserve supersymmetry we perform a topological twist with the R-symmetry introduced above. The compactification to 2d can be done following the general prescription given in \cite{Gadde:2015wta}. The 7 chiral multiplets with R-charge $0$ descend to 7 $(0,2)$ chiral multiplets, while the remaining chiral descends to 3 $(0,2)$ Fermi multiplets. We then get the following $(0,2)$ theory:
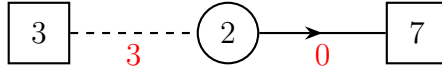
\begin{center}
\begin{tikzpicture}[node distance=2.5cm, auto, thick, scale=1, every node/.style={transform shape}]
    \node[rectangle, draw, minimum size=0.8cm] (SU3F) {$3$};
    \node[circle, draw, minimum size=0.8cm, right of=SU3F] (SU2G) {$2$};
    \node[rectangle, draw, minimum size=0.8cm, right of=SU2G] (SU7F) {$7$};

    \draw[dashed] (SU3F) -- node[below] {$\textcolor{red}{3}$} (SU2G);
    \draw[mid arrow] (SU2G) -- node[below] {$\textcolor{red}{0}$} (SU7F);
\end{tikzpicture}
\end{center}
The 4d theory has $\SU(8)$ global symmetry which is broken by the twisting to $\SU(7)\times \U(1)_B$. The 2d theory has global symmetry $\SU(3) \times \SU(7) \times \U(1)_B \times \U(1)_A$. We see that the symmetry of the 2d theory is enhanced in two ways. First, the axial symmetry which was broken by an ABJ anomaly in 4d is no longer anomalous in 2d and gets upgraded to a genuine symmetry. Additionally, we get an $\SU(3)$ global symmetry rotating the multiple copies of the 4d chiral which was given R-charge $4$. This symmetry is expected to contain the $\SU(2)_{\text{ISO}}$ symmetry of the $S^2$, where the embedding should be: $\SU(2)_{\text{ISO}}\cong \SO(3)\subset \SU(3)$.

The table of symmetry and charges is:
\begin{table}[H]
    \centering
    \begin{tabular}{c|c|c c c c c c}
        & $\SU(2)_{\text{gauge}}$ & $\SU(3)$ & $\SU(7)$ & $\U(1)_R$ & $\U(1)_{R_{mix}}$& $\U(1)_{B}$ & $\U(1)_{A}$ \\ \hline
        $\psi$ & $\ydiagram{1}$ & $\ydiagram{1}$ & $\mathbf{1}$ & $3$ & $\frac{2}{3}$ & $+7$ & $+1$ \\
        $\phi$ & $\ydiagram{1}$ & $\mathbf{1}$ & $\overline{\ydiagram{1}}$ & $0$  & $\frac{1}{3}$ & $-1$ & $+1$ \\
    \end{tabular}
    \caption{Table of gauge and symmetry charges of the field content of the $(0,2)$ 2d theory. Here $\U(1)_{R_{mix}}:=\U(1)_R - \frac{1}{3} \U(1)_B$ is a mixed R-charge chosen so that the charges are between zero and one.}
\end{table}

The presence of the additional symmetries then raises similar questions as before, and we can wonder whether a superpotential breaking the accidental symmetries can be found. To confirm this or rule it out, we turn to the elliptic genus.   

\paragraph{The elliptic genus}
We can compute the elliptic genus of this model using the techniques outlined thus far. We find:
\begin{align} \begin{split}
\mathcal{I}_{\text{JK}} & =1+q^{\frac{1}{3}}\left[a^{2}\boldsymbol{\overline{21}_{f}}b^{-2}+\boldsymbol{\bar{6}_{h}}a^{-2}b^{-14}-\boldsymbol{\bar{7}_{f}}\boldsymbol{\bar{3}_{h}}b^{-8}-a^{-10}b^{-14}\right]\\
 & +q^{\frac{2}{3}}\Big[-a^{2}\boldsymbol{\overline{112}_{f}}\boldsymbol{\bar{3}_{h}}b^{-10}-\boldsymbol{8_{h}}\boldsymbol{\bar{7}_{f}}a^{-2}b^{-22}+\boldsymbol{\overline{21}_{f}}\boldsymbol{\bar{6}_{h}}b^{-16}\\
 & +\boldsymbol{3_{h}}\boldsymbol{\overline{28}_{f}}b^{-16}+a^{4}\boldsymbol{196_{f}}b^{-4}+\boldsymbol{6_{h}}a^{-4}b^{-28}\Big]\\
 & +q\Big[-\boldsymbol{882_{f}}\boldsymbol{\bar{3}_{h}}a^{4}b^{-12}-a^{2}b^{6}\boldsymbol{3_{h}}\boldsymbol{\bar{7}_{f}}+a^{2}\boldsymbol{196_{f}}\boldsymbol{\bar{6}_{h}}b^{-18}-\boldsymbol{3_{h}}\boldsymbol{\bar{7}_{f}}a^{-4}b^{-36}\\
 & +\boldsymbol{6_{h}}\boldsymbol{\overline{21}_{f}}a^{-2}b^{-30}+\boldsymbol{\overline{28}_{f}}\boldsymbol{\bar{3}_{h}}a^{-2}b^{-30}-\boldsymbol{7_{f}}\boldsymbol{\bar{3}_{h}}a^{-2}b^{-6}-\boldsymbol{8_{h}}\boldsymbol{\overline{112}_{f}}b^{-24}\\
 & -\boldsymbol{\overline{84}_{f}}b^{-24}+\boldsymbol{1176_{f}}a^{6}b^{-6}+a^{2}\boldsymbol{378_{f}}\boldsymbol{3_{h}}b^{-18}+\boldsymbol{7_{f}}\boldsymbol{3_{h}}a^{-10}b^{-6}\\
 & -\boldsymbol{21_{f}}a^{-12}b^{-12}+a^{-6}b^{-42}-\boldsymbol{6_{h}}a^{-8}+\boldsymbol{8_{h}}+\boldsymbol{48_{f}}+2\Big] + O(q^{\frac{4}{3}})
\end{split} \end{align}
We note here the presence of the conserved current $\textbf{48}_f+\textbf{8}_h+2$ in the order $q$ terms of the index, corresponding to the $\SU(7) \times \SU(3) \times \U(1)_A \times \U(1)_B$ global symmetry of the $2d$ theory. We also observe the term $-q\textbf{6}_ha^{-8}$, whose charges are consistent with a marginal operator and is charged only under the 2d accidental symmetries. Assuming it is indeed a superpotential operator, inserting it into the superpotential then would preserve the symmetries of the 4d theory, including the R-symmetry, but would partially break the 2d accidental symmetries. The latter are broken to the subgroup that preserves this operator. For $\U(1)_a$ this is its $\mathbb{Z}_8$ subgroup. For $\SU(3)$ there are several options depending on the chosen component, but here we shall concentrate on the component that preserves the $\SO(3)$ subgroup. That such a component exists follows as the $\textbf{6}$ is the symmetric tensor of $\SU(3)$ and the subgroup of $\SU(N)$ preserving a symmetric matrix is $\SO(N)$.  

We see then that the elliptic genus implies the presence of an operator that could precisely break the accidental 2d symmetries leaving behind only the symmetries expected from the compactification picture. Nevertheless, it is not apparent how this operator can be built in the gauge theory. To better understand this we look at the difference between the exact elliptic genus and the perturbative calculation:

\begin{align} \begin{split}
\mathcal{I}_{\text{JK}} & =\mathcal{I}_{\text{pert.}} -q^{\frac{1}{3}}a^{-10}b^{-14}+q\left[\boldsymbol{7_{f}}\boldsymbol{3_{h}}a^{-10}b^{-6}-\boldsymbol{21_{f}}a^{-12}b^{-12}-\boldsymbol{6_{h}}a^{-8}\right] + o(q)
\end{split} \end{align}

We indeed see that this operator is not present in the perturbative index. It somehow arises non-perturbatively from the JK prescription.

\paragraph{c-extremization} As in the previous cases, it is interesting to check what is the expected superconformal R-symmetry and use this to estimate the relevancy properties of the additional operator. Like in the previous cases, there is a moduli space spanned by $\phi^2$ so we again encounter the problem of apparent non-compact moduli. Evaluating as before, while demanding that $\phi$ has R-charge $0$, we find the R-symmetry: $\U(1)_R^{\text{SC}}=\U(1)_{R_{mix}}-\frac{3}{8} \U(1)_A -\frac{1}{24} \U(1)_B$. Under this R-symmetry both $\phi$ and $\psi$ have free R-charges ($R=0$). Since the $\U(1)_A$ mixing and the charge of the additional operator are both negative, it is expected to be irrelevant with respect to this fixed point. We note that if one performs the c-extremization without the moduli constraint, the superconformal R-symmetry comes out to be: $\U(1)_R^{\text{SC}}=\U(1)_{R_{mix}}+\frac{1}{2} \U(1)_A -\frac{1}{6} \U(1)_B$. Since the mixing is positive, the additional operator is relevant with respect to this fixed point. Nevertheless, note that this R-symmetry gives $\psi$ R-charge $0$ and $\phi$ R-charge $1$, so the resulting central charge is negative.

\subsubsection{$N_f=5$}

Consider the 4d $\mathcal{N}=1$ gauge theory consisting of an $\SU(2)$ vector multiplet, 5 fundamental and 5 antifundamental chiral fields. As the fundamentals and antifundamentals of $\SU(2)$ are the same, the $\SU(5) \times \SU(5) \times \U(1)_B$ symmetry enhance to $\SU(10)$. We can also introduce a $\U(1)_A$ axial symmetry, under which all ten chirals have charge $1$, but it is broken to $\mathbb{Z}_{10}$ by the ABJ anomaly. We'll denote this theory by $\mathcal{T}_A^{4d}$ and depict it using the quiver below.

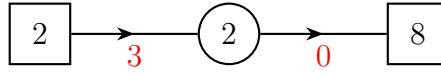
\begin{center}
\begin{tikzpicture}[node distance=2.5cm, auto, thick, scale=1, every node/.style={transform shape}]
    \node[rectangle, draw, minimum size=0.8cm] (SU2F) {$2$};
    \node[circle, draw, minimum size=0.8cm, right of=SU2F] (SU2G) {$2$};
    \node[rectangle, draw, minimum size=0.8cm, right of=SU2G] (SU8F) {$8$};

    \draw[mid arrow] (SU2F) -- node[below] {$\textcolor{red}{3}$} (SU2G);
    \draw[mid arrow] (SU2G) -- node[below] {$\textcolor{red}{0}$} (SU8F);
\end{tikzpicture}
\end{center}

Let us now define an R-symmetry assigning to 8 chirals R-charge 0 and to 2 chirals R-charge of 3. One can check that this is an anomaly free R-symmetry. Note, that this R-symmetry does not commute with the full $\SU(10)$ symmetry, rather only with its $\SU(2) \times \SU(8) \times \U(1)_B$ subgroup.

Next we consider reducing it to 2d on the $S^2$. To preserve supersymmetry we perform a topological twist using the R-symmetry defined above. The compactification to 2d can be done following the general prescription given in \cite{Gadde:2015wta}. The 8 chiral multiplets will descend into 8 $(0,2)$ chiral $(0,2)$ multiplets, while the 2 remaining chirals will descend into 4 $(0,2)$ Fermi multiplets. The quiver diagram of such a theory is:

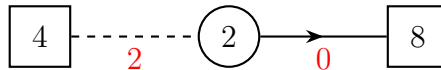
\begin{center}
\begin{tikzpicture}[node distance=2.5cm, auto, thick, scale=1, every node/.style={transform shape}]
    \node[rectangle, draw, minimum size=0.8cm] (SU4F) {$4$};
    \node[circle, draw, minimum size=0.8cm, right of=SU4F] (SU2G) {$2$};
    \node[rectangle, draw, minimum size=0.8cm, right of=SU2G] (SU8F) {$8$};

    \draw[dashed] (SU4F) -- node[below] {$\textcolor{red}{2}$} (SU2G);
    \draw[mid arrow] (SU2G) -- node[below] {$\textcolor{red}{0}$} (SU8F);
\end{tikzpicture}
\end{center}

The global symmetry of the resulting 2d theory is $\SU(4) \times \SU(8) \times \U(1)_B \times \U(1)_A$. We see that the 2d theory gets a symmetry enhancement in multiple ways. First the 4d $\SU(2)$ gets enhanced to $\SU(4)$. As previously mentioned, this comes about as each of the two 4d chirals with R-charge 3 got doubled in the 2d reduction. This $\SU(4)$ should contain the $\SU(2)$ global symmetry of the 4d theory and the $\SU(2)_{\text{ISO}}$ symmetry of the $S^2$. The preceding discussion suggests that these are embedded as: $\SU(2)\times \SU(2)_{\text{ISO}} = \text{Spin}(4) \subset \SU(4)$. Second, the axial symmetry which had an ABJ anomaly in 4d is no longer anomalous in 2d and gets updated to a genuinely true symmetry.

Below, we write down the table of gauge and flavor charges for the multiplets of theory $\mathcal{T}_A^{2d}$:

\begin{table}[H]
    \centering
    \begin{tabular}{c|c|c c c c c c}
        & $\SU(2)_{\text{gauge}}$ & $\SU(4)$ & $\SU(8)$ & $\U(1)_R$ & $\U(1)_{R_{mix}}$& $\U(1)_{B}$ & $\U(1)_{A}$ \\ \hline
        $\psi$ & $\ydiagram{1}$ & $\ydiagram{1}$ & $\mathbf{1}$ & $2$ & $\frac{2}{3}$ & $+4$ & $+1$ \\
        $\phi$ & $\ydiagram{1}$ & $\mathbf{1}$ & $\overline{\ydiagram{1}}$ & $0$  & $\frac{1}{3}$ & $-1$ & $+1$ \\
    \end{tabular}
    \caption{Table of gauge and symmetry charges of the field content of the $(0,2)$ 2d theory $\mathcal{T}_A^{2d}$. Here $\U(1)_{R_{mix}}=\U(1)_R - \frac{1}{3} \U(1)_B$ is a mixed R-charge chosen so that the charges are between zero and one.}
    \label{TableNF=5P}
\end{table}

As before, we want to check whether it is possible to break the additional 2d symmetries while preserving the ones expected from the 4d compactification. To do this we next compute the elliptic genus.

\paragraph{Elliptic genus computation}
Computing the elliptic genus of $\mathcal{T}_A^{2d}$ we find:

\begin{align} \begin{split}
\mathcal{I}_{\text{JK}} & =1+q^{\frac{1}{3}}\left[b^{-8}a^{-2}\boldsymbol{\overline{10}_{h}}-a^{-12}b^{-8}+\boldsymbol{\overline{28}_{f}}a^{2}b^{-2}-\boldsymbol{\overline{8}_{f}}\boldsymbol{\overline{4}_{h}}b^{-5}\right]\\
 & +q^{\frac{2}{3}}\Big[b^{-10}\boldsymbol{\overline{28}_{f}}\boldsymbol{\overline{10}_{h}}-b^{-7}a^{2}\boldsymbol{\overline{168}_{f}}\boldsymbol{\bar{4}_{h}}+a^{-4}b^{-16}\boldsymbol{20'_{h}}-a^{-2}b^{-13}\boldsymbol{\overline{8}_{f}}\boldsymbol{20_{h}}\\
 & +b^{-10}\boldsymbol{\overline{36}_{f}}\boldsymbol{6_{h}}+a^{4}b^{-4}\boldsymbol{\overline{336}_{f}}\Big]\\
 & +q\Big[-a^{-4}b^{-21}\boldsymbol{\overline{8}_{f}}\boldsymbol{\overline{20}_{h}}+a^{2}b^{-12}\boldsymbol{\overline{336}_{f}}\boldsymbol{\overline{10}_{h}}+\left(-a^{4}b^{-9}\boldsymbol{1680_{f}}-a^{-2}b^{-3}\boldsymbol{8_{f}}\right)\boldsymbol{\bar{4}_{h}}\\
 & +\left(a^{-6}b^{-24}-a^{-10}\right)\boldsymbol{10_{h}}+a^{-2}b^{-18}\boldsymbol{\overline{28}_{f}}\boldsymbol{20'_{h}}+\left(a^{-2}b^{-18}\boldsymbol{\overline{36}_{f}}+1\right)\boldsymbol{15_{h}}\\
 & -b^{-15}\boldsymbol{\overline{168}_{f}}\boldsymbol{20_{h}}+a^{2}b^{-12}\boldsymbol{\overline{630}_{f}}\boldsymbol{6_{h}}+b^{-6}\left(a^{6}\boldsymbol{2520'_{f}}-a^{-14}\boldsymbol{28_{f}}\right)\\
 & +\left(-b^{-15}\boldsymbol{\overline{120}_{f}}-a^{2}b^{3}\boldsymbol{\overline{8}_{f}}+a^{-12}b^{-3}\boldsymbol{8_{f}}\right)\boldsymbol{4_{h}}+\boldsymbol{63_{f}}+2\Big] + O(q^{\frac{4}{3}})
\end{split} \end{align}

We can clearly see the contribution of the conserved current $2 + \textbf{63}_f + \textbf{15}_h$ of the symmetry $\U(1)_A \times \U(1)_B \times \SU(8)_f \times \SU(4)_h$ in the order $q$ terms in the expression.

We now compare the superconformal index with the perturbative computation: 

\begin{align} \begin{split}
    \mathcal{I}_{\text{JK}}	=\mathcal{I}_{\text{pert.}}-a^{-12}b^{-8}q^{\frac{1}{3}}+q\left(-a^{-14}b^{-6}\boldsymbol{28_{f}}+a^{-12}b^{-3}\boldsymbol{8}_{f}\boldsymbol{4_{h}}-a^{-10}\boldsymbol{10_{h}}\right) + o(q)
\end{split} \end{align}

We note the presence of the term $-q a^{-10} \textbf{10}_h$. This term has the right charges to be a marginal operator, that is it would preserve the R-symmetry if introduced into the superpotential. However it is charged under $\U(1)_a$ and $\SU(4)$, which would then be broken upon its introduction to the superpotential, to the subgroup preserving this operator. For $\U(1)_a$ this is its $\mathbb{Z}_{10}$ subgroup. For the $\SU(4)$, the $\textbf{10}$ is its symmetric tensor representation, and it should contain a component that is fixed by the $SO$ subgroup. This should then lead to the breaking of $\SU(4) \to \SO(4) \cong \SU(2)_{4d} \times \SU(2)_{\text{ISO}}$. Overall, we indeed find the presence of an operator with the right charges to break the 2d accidental symmetries while preserving the ones expected from the compactification.

\paragraph{c-extremization} We can again try to determine the superconformal R-symmetry using c-extremization. The results are structurally similar to the previous case. Performing the c-extremization while demanding that $\phi$ has R-charge $0$ leads to the R-symmetry under which both $\phi$ and $\psi$ have R-charge zero. The explicit R-symmetry achieving this is: $\U(1)_R^{\text{SC}}=\U(1)_{R_{mix}}-\frac{2}{5} \U(1)_A -\frac{1}{15} \U(1)_B$, and we again conclude that the additional operator is irrelevant due to the negative $\U(1)_A$ mixing. Alternatively, performing the unconstrained mixing we find the R-symmetry under which $\psi$ has R-charge $0$ and $\phi$ has R-charge $1$. Explicitly, it is given by: $\U(1)_R^{\text{SC}}=\U(1)_{R_{mix}}+\frac{2}{5} \U(1)_A -\frac{4}{15} \U(1)_B$. The positive $\U(1)_A$ mixing implies that the additional operator is relevant but the resulting central charge is negative. 

\paragraph{Other choices of R-charge} In the above discussion we made a specific choice of R-symmetry to twist by in the reduction to 2d. Consider instead doing the same reduction but with a slightly different R-symmetry. Specifically, we shall still assign R-charge $0$ to 8 chiral fields, but for the last two we shall assign R-charge $4$ to one and R-charge $2$ to the other. This is also an anomaly free R-charge. This choice of R-charge preserves an $\SU(8) \times \U(1)_{\tilde{B}}\times \U(1)_{\tilde{C}}$ subgroup of the $\SU(10)$ symmetry. A depiction of a quiver diagram of this theory is:
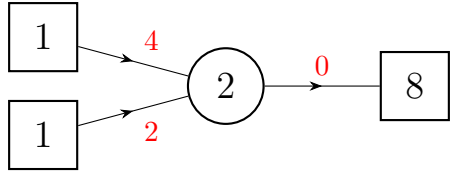
\begin{center}
\begin{tikzpicture}[
    gauge/.style={circle, draw, thick, minimum size=10mm, inner sep=0pt, font=\large},
    flavor/.style={rectangle, draw, thick, minimum size=9mm, inner sep=0pt, font=\large},
    edge label/.style={font=\small, auto}
]

\node[gauge] (G2) at (0,0) {$2$};
\node[flavor] (F8)  at (0:2.5)   {$8$};
\node[flavor] (F14) at (165:2.5) {$1$};
\node[flavor] (F12) at (195:2.5) {$1$};

\draw[mid arrow] (G2) -- node[above, edge label] {$\textcolor{red}{0}$} (F8);

\draw[mid arrow] (F14) -- node[edge label] {$\textcolor{red}{4}$} (G2);
\draw[mid arrow] (F12) -- node[swap, edge label] {$\textcolor{red}{2}$} (G2);

\end{tikzpicture}
\end{center}

Carrying out the reduction on the $S^2$ to 2d, the 8 chiral multiplets again descend to 8 $(0,2)$ chiral multiplets, the chiral with R-charge $2$ descend to 1 Fermi multiplet, and the chiral with R-charge $4$ descend to 3 Fermi multiplets. A quiver diagram of this 2d theory is:

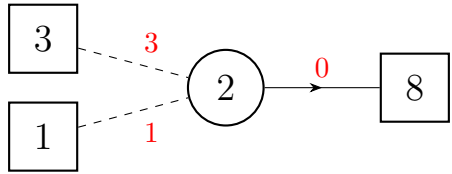
\begin{center}
\begin{tikzpicture}[
    gauge/.style={circle, draw, thick, minimum size=10mm, inner sep=0pt, font=\large},
    flavor/.style={rectangle, draw, thick, minimum size=9mm, inner sep=0pt, font=\large},
    edge label/.style={font=\small, auto}
]

\node[gauge] (G2) at (0,0) {$2$};

\node[flavor] (F8) at (0:2.5)   {$8$};
\node[flavor] (F3) at (165:2.5) {$3$};
\node[flavor] (F1) at (195:2.5) {$1$};

\draw[mid arrow] (G2) -- node[above, edge label] {$\textcolor{red}{0}$} (F8);

\draw[dashed] (F3) -- node[edge label] {$\textcolor{red}{3}$} (G2);
\draw[dashed] (F1) -- node[swap, edge label] {$\textcolor{red}{1}$} (G2);

\end{tikzpicture}
\end{center}

The charges of the 2d fields under the various symmetries is summarized in the table below.

\begin{table}[H]
    \centering
    \begin{tabular}{c|c|c c c c c c c}
        & $\SU(2)_{\text{gauge}}$ & $\SU(3)$ & $\SU(8)$ & $\U(1)_R$ & $\U(1)_{R_{mix}}$& $\U(1)_{\tilde{B}}$ & $\U(1)_{\tilde{A}}$ & $\U(1)_{\tilde{C}}$ \\ \hline
        $\psi_3$ & $\ydiagram{1}$ & $\ydiagram{1}$ & $\mathbf{1}$ & $3$ & $\frac{2}{3}$ & $+4$ & $+1$ & $$+1$$ \\
        $\psi_1$ &  $\ydiagram{1}$ & $\mathbf{1}$ & $\mathbf{1}$ & $1$ & $\frac{2}{3}$ & $+4$ & $+1$ & $-1$ \\
        $\phi$ & $\ydiagram{1}$ & $\mathbf{1}$ & $\overline{\ydiagram{1}}$ & $0$  & $\frac{1}{3}$ & $-1$ & $+1$ & $0$ \\
    \end{tabular}
    \caption{Table of gauge and symmetry charges of the field content of the $(0,2)$ 2d theory resulting from the $S^2$ reduction of the 4d theory with a slightly different choice of R-charges. Here $\U(1)_{R_{mix}}=\U(1)_R - \frac{1}{3} \U(1)_{\tilde{B}} - \U(1)_{\tilde{C}}$ is a mixed R-charge.}
\end{table}

Note that the resulting 2d theory is completely identical to the one we got previously. However, in the previous discussion, the symmetries that were inherited from 4d were the $\SU(8)\times \U(1)_B$ as well as $\SU(2)_{4d} \times \SU(2)_{\text{ISO}}$ that appear as a subgroup of the $\SU(4)$ symmetry of the quiver. In the current description, while we still retain the $\SU(8)\times U(1)_{\tilde{B}}$, we also expect a $\U(1)_{\tilde{C}} \times \SU(2)_{\text{ISO}}$. The latter should appear as a subgroup of the $\SU(4)$, where for $\SU(2)_{\text{ISO}}$, the expected embedding is $\SO(3)\subset \SU(3) \subset \SU(4)$. One can naturally wonder if this can be observed in the elliptic genus result.  

To do this we first convert the fugacities used there to those associated with symmetries of the current model. For this we note that the 4 Fermi multiplets and $8$ chiral fields should carry the charges: $\tilde{a} \tilde{b}^4 (\tilde{c} \boldsymbol{3_{\tilde{h}}} + \tilde{c}^{-1}) = \tilde{a} \tilde{b}^4 \sqrt{\tilde{c}} (\sqrt{\tilde{c}} \boldsymbol{3_{\tilde{h}}} + \tilde{c}^{-\frac{3}{2}})\rightarrow a b^4\boldsymbol{4_{h}}$, ,$\frac{\tilde{a}}{\tilde{b}} \boldsymbol{\overline{8}_{\tilde{f}}} \rightarrow \frac{a}{b} \boldsymbol{\overline{8}_{f}}$ where we used tilde to distinguish the fugacities associated with the current model. We see then that the proper mapping is: $\boldsymbol{8_{\tilde{f}}} \rightarrow \boldsymbol{8_{f}}$, $\sqrt{\tilde{c}} \boldsymbol{3_{\tilde{h}}} + \tilde{c}^{-\frac{3}{2}} \rightarrow \boldsymbol{4_{h}}$, $\tilde{b} \rightarrow b \tilde{c}^{-\frac{1}{10}}$, $\tilde{a} \rightarrow a \tilde{c}^{-\frac{1}{10}}$. 

With the elliptic genus we noted the presence of the operator contributing as $-a^{-10} \boldsymbol{10_{h}}$. This operator has the right contribution to be a marginal operator whose insertion to the superpotential could break the accidental 2d symmetries while preserving the ones coming from 4d. We shall next show that this is true also for the case considered here, but using a different component. For this we first map to the symmetries of the current model: $-a^{-10} \boldsymbol{10_{h}} \rightarrow -\tilde{a}^{-10} \tilde{c}^{-1} (\tilde{c}\boldsymbol{6_{\tilde{h}}} + \tilde{c}^{-1}\boldsymbol{3_{\tilde{h}}} + \tilde{c}^{-3})$. Note the presence of the component $-\tilde{a}^{-10} \boldsymbol{6_{\tilde{h}}}$. Introducing it into the superpotential, it is possible to break $\U(1)_{\tilde{A}} \rightarrow\mathbb{Z}_{10}$ and $\SU(3)\rightarrow \SO(3)$. This precisely breaks the 2d accidental symmetries while preserving the ones originating from 4d.

We see then that the additional operator arising in the elliptic genus allows us to preserve different symmetries, which in the above two examples correspond to two different 4d reductions leading to a similar 2d theory. The above also suggests that the resulting two 2d theories, while quite similar, are actually different, where they differ by the presence of this superpotential term. 

\paragraph{Illustrating the different components} Finally, we provide a short illustration of the different symmetry breaking patterns depending on the component of the operator one uses. As the operator in question has no perturbative origin, we shall adopt a toy example to illustrate the possibilities. 

Consider a collection of 4 chiral fields $Q_i$. We take these as transforming in the $\boldsymbol{4}$ of an $\SU(4)$ global symmetry, and can also include a $\U(1)$ symmetry rotating all the $Q_i$'s. The operator $Q_i Q_j$ then is the symmetric matrix of the $Q$'s and so transforms in the $\boldsymbol{10}$ of $\SU(4)$.

Consider the component $Q^2_1$. Introducing this component to the superpotential should break the $\SU(4)$ down to the $\SU(3)$ rotating $Q_{1,2,3}$. Additionally, it breaks the $\U(1)$ rotating all the $Q_i$'s and the $U(1)$ commutant of $\SU(3)$ in $\SU(4)$ to the combination that keeps $Q_1$ fixed. This can be seen in the above decomposition of $\SU(4) \rightarrow \SU(3)\times \U(1)_c$: $\boldsymbol{10} \rightarrow c\boldsymbol{6} + c^{-1}\boldsymbol{3} + c^{-3}$. Here the last term, $ c^{-3}$, corresponds to the component $Q^2_1$. If we use the fugacity $d$ for the $\U(1)$ acting diagonally on all the $Q_i$'s, then the charges of this components are: $d^2c^{-3}$, and we see that it leads to the breaking of $d=c^{\frac{3}{2}}$.

We can now illustrate the components required for the other breakings. Specifically, consider the combination $Q^2_1+Q^2_2+Q^2_3+Q^2_4$. This is invariant under the $\SO(4)$ subgroup of $\SU(4)$, and so would break $\SU(4)$ to that subgroup. Indeed if we decompose $\SU(4)\rightarrow \SO(4)$, we see that: $\boldsymbol{10} \rightarrow (\boldsymbol{1},\boldsymbol{1})+(\boldsymbol{3},\boldsymbol{3})$, and the singlet corresponds to this combination. Including the $\U(1)_d$ we see that it is also broken, here to $\mathbb{Z}_2$.

Finally, we can consider the combination $Q^2_2+Q^2_3+Q^2_4$. This is preserved under the $\SO(3)$ subgroup of $\SU(3)$. Additionally, it would break $\U(1)_d$ and $\U(1)_c$ to the combination that keeps these fields fixed. In the above decomposition, this corresponds to a component of $d^2 c\boldsymbol{6}$, and so preserves the combination $c=d^{-2}$. Similarly in the above model, this component would break $\U(1)_{\tilde{A}}\times \SU(4)\rightarrow \mathbb{Z}_{10}\times \U(1)_{\tilde{C}}\times \SO(3)$.

\subsubsection{$N_f=8$}

As our final example, consider the 4d $\mathcal{N}=1$ gauge theory consisting of an $\SU(2)$ vector multiplet and 16  chiral fields in the fundamental representation. Like the previous examples, this 4d theory has a non-abelian symmetry rotating  the chiral fields, here $\SU(16)$, and a $\U(1)_A$ axial symmetry, which is broken to a discrete group by the ABJ anomaly, here to $\mathbb{Z}_{16}$.

As before, we would like to consider the compactification of this theory on $\mathbb{S}^2$. To preserve supersymmetry we shall include a topological twist in a chosen R-symmetry, which we take to be one under which $12$ chiral fields have R-charge $0$ and $4$ have R-charge $3$. One can check that this R-symmetry is anomaly free. The theory with the chosen R-symmetry is depicted in the following quiver diagram:

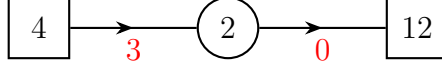
\begin{center}
\begin{tikzpicture}[node distance=2.5cm, auto, thick, scale=1, every node/.style={transform shape}]
    \node[rectangle, draw, minimum size=0.8cm] (SU4F) {$4$};
    \node[circle, draw, minimum size=0.8cm, right of=SU2F] (SU2G) {$2$};
    \node[rectangle, draw, minimum size=0.8cm, right of=SU2G] (SU12F) {$12$};

    \draw[mid arrow] (SU4F) -- node[below] {$\textcolor{red}{3}$} (SU2G);
    \draw[mid arrow] (SU2G) -- node[below] {$\textcolor{red}{0}$} (SU12F);
\end{tikzpicture}
\end{center}

Note that with this choice of R-symmetry the dimensional reduction would break the $\SU(16)$ to $\SU(12)\times \SU(4)\times \U(1)_B$. Performing the dimensional reduction we find the following 2d $(0,2)$ theory:

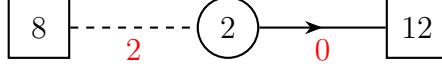
\begin{center}
\begin{tikzpicture}[node distance=2.5cm, auto, thick, scale=1, every node/.style={transform shape}]
    \node[rectangle, draw, minimum size=0.8cm] (SU8F) {$8$};
    \node[circle, draw, minimum size=0.8cm, right of=SU2F] (SU2G) {$2$};
    \node[rectangle, draw, minimum size=0.8cm, right of=SU2G] (SU12F) {$12$};

    \draw[dashed] (SU8F) -- node[below] {$\textcolor{red}{2}$} (SU2G);
    \draw[mid arrow] (SU2G) -- node[below] {$\textcolor{red}{0}$} (SU12F);
\end{tikzpicture}
\end{center}

The charges and symmetry table is:
\begin{table}[H]
    \centering
    \begin{tabular}{c|c|c c c c c c}
        & $\SU(2)_{\text{gauge}}$ & $\SU(8)$ & $\SU(12)$ & $\U(1)_R$ & $\U(1)_{R_{mix}}$& $\U(1)_{B}$ & $\U(1)_{A}$ \\ \hline
        $\psi$ & $\ydiagram{1}$ & $\ydiagram{1}$ & $\mathbf{1}$ & $2$ & $\frac{1}{2}$ & $+3$ & $+1$ \\
        $\phi$ & $\ydiagram{1}$ & $\mathbf{1}$ & $\overline{\ydiagram{1}}$ & $0$  & $\frac{1}{2}$ & $-1$ & $+1$ \\
    \end{tabular}
    \caption{Table of gauge and symmetry charges of the field content of the $(0,2)$ 2d theory. Here $\U(1)_{R_{mix}}:=\U(1)_R - \frac{1}{2} \U(1)_B$ is a mixed R-charge chosen so that the charges are between zero and one.}
\end{table}

As before, the model fully manifests the $\U(1)_A$ symmetry that in 4d is broken by anomalies. Additionally, the 2d model has an $\SU(8)$ global symmetry, where only the $\SU(2)_{\text{ISO}}\times \SU(4)$ subgroup (embedded such that $\boldsymbol{8}_{\SU(8)}\rightarrow \boldsymbol{2}_{\SU(2)}\boldsymbol{4}_{\SU(4)}$) is visible in 4d. This then raises the question of whether these additional 2d symmetries can be broken so that only the ones originating in the compactification remain. To better understand this we examine the elliptic genus of this model.  

\paragraph{The elliptic genus}

Following the standard procedure for calculating the elliptic genus, we obtain:

\begin{align} \begin{split}
\mathcal{I}_{\text{JK}} & =1-a^{-20}b^{-12}\\
 & +q^{\frac{1}{2}}\left[\boldsymbol{\overline{66}_{f}}a^{2}b^{-2}+\boldsymbol{\overline{36}_{h}}a^{-2}b^{-6}-\boldsymbol{\overline{12}_{f}}\boldsymbol{\bar{8}_{h}}b^{-4}+\boldsymbol{12_{f}}\boldsymbol{8_{h}}a^{-20}b^{-8}-\boldsymbol{66_{f}}a^{-22}b^{-10}-\boldsymbol{36_{h}}a^{-18}b^{-6}\right]\\
 & +q\Big[a^{4}\boldsymbol{\overline{1716}_{f}}b^{-4}-a^{2}b^{2}\boldsymbol{8_{h}}\boldsymbol{\overline{12}_{f}}-a^{2}\boldsymbol{\overline{572}_{f}}\boldsymbol{\bar{8}_{h}}b^{-6}+\boldsymbol{12_{f}}\boldsymbol{\bar{8}_{h}}a^{-22}b^{-14}+\boldsymbol{8_{h}}\boldsymbol{\overline{12}_{f}}a^{-18}b^{-10}\\
 & -\boldsymbol{\overline{12}_{f}}\boldsymbol{\overline{168}_{h}}a^{-2}b^{-10}-\boldsymbol{12_{f}}\boldsymbol{\bar{8}_{h}}a^{-2}b^{-2}+\boldsymbol{\overline{336}_{h}}a^{-4}b^{-12}+\boldsymbol{\overline{66}_{f}}\boldsymbol{\overline{36}_{h}}b^{-8}+\boldsymbol{\overline{78}_{f}}\boldsymbol{\overline{28}_{h}}b^{-8}\\
 & +\boldsymbol{572_{f}}\boldsymbol{8_{h}}a^{-22}b^{-6}-\boldsymbol{78_{f}}\boldsymbol{28_{h}}a^{-20}b^{-4}-\boldsymbol{66_{f}}\boldsymbol{36_{h}}a^{-20}b^{-4}+\boldsymbol{12_{f}}\boldsymbol{168_{h}}a^{-18}b^{-2}\\
 & -\boldsymbol{1716_{f}}a^{-24}b^{-8}-\boldsymbol{63_{h}}a^{-20}b^{-12}-\left(2+\boldsymbol{143_{f}}\right)a^{-20}b^{-12}-\boldsymbol{336_{h}}a^{-16}+\boldsymbol{63_{h}}+\boldsymbol{143_{f}}+2\Big] \\
 &+ O(q^{\frac{5}{4}})
\end{split} \end{align}

For future reference we also note the comparison with the perturbative residue:
\begin{align} \begin{split}
\mathcal{I}_{\text{JK}}& =\mathcal{I}_{\text{pert.}} -a^{-20}b^{-12}+q^{\frac{1}{2}}\left(\boldsymbol{12_{f}}\boldsymbol{8_{h}}a^{-20}b^{-8}-\boldsymbol{66_{f}}a^{-22}b^{-10}-\boldsymbol{36_{h}}a^{-18}b^{-6}\right)\\
 & +q\Big[\boldsymbol{12_{f}}\boldsymbol{\bar{8}_{h}}a^{-22}b^{-14}+\boldsymbol{8_{h}}\boldsymbol{\overline{12}_{f}}a^{-18}b^{-10}+\boldsymbol{572_{f}}\boldsymbol{8_{h}}a^{-22}b^{-6}\\
 & -\boldsymbol{78_{f}}\boldsymbol{28_{h}}a^{-20}b^{-4}-\boldsymbol{66_{f}}\boldsymbol{36_{h}}a^{-20}b^{-4}+\boldsymbol{12_{f}}\boldsymbol{168_{h}}a^{-18}b^{-2}\\
 & -\boldsymbol{1716_{f}}a^{-24}b^{-8}-\boldsymbol{63_{h}}a^{-20}b^{-12}-\left(2+\boldsymbol{143_{f}}\right)a^{-20}b^{-12}-\boldsymbol{336_{h}}a^{-16}\Big] \\
 & + o(q)
\end{split} \end{align}

We note several things about the results. First one can note the contribution of an operator at $q^0$. This operator is fermionic, hence why there is only one such operator and the expansion in $q$ is well defined. The original choice of $\U(1)_{R_{mix}}$ was done to ensure that all perturbative operators have positive R-charge so as to avoid such terms. However, this term does not come from the perturbative matter and so can evade this constraints. We note that this R-symmetry is not necessarily the superconformal one so this does not signal an issue with the theory.

As usual, we turn our main focus to the terms at order $q$. First we note the presence of the terms $+q(\boldsymbol{63_{h}}+\boldsymbol{143_{f}}+2)$ that appear also in the perturbative index. These are the contributions of the conserved currents of the $\SU(12)\times \SU(8)\times \U(1)_B \times \U(1)_A$ global symmetry. Next we consider the negatively charged terms, with special consideration to those charged only under $\U(1)_A$ and the $\SU(8)$. Indeed there is exactly one term that fits these criteria: $-q\boldsymbol{336_{h}}a^{-16}$. We also see that this operator does not arise from the perturbative matter.

This operator indeed have the right charges to break the additional 2d symmetries while preserving the 4d ones. To see this, recall that the 4d $\SU(2)_{\text{ISO}}\times \SU(4)$ is embedded in $\SU(8)$ as $\boldsymbol{8}_{\SU(8)}\rightarrow \boldsymbol{2}_{\SU(2)}\boldsymbol{4}_{\SU(4)}$. The decomposition of the $\boldsymbol{336}$ of $\SU(8)$ under this embedding is:

\be
\boldsymbol{336}_{\SU(8)} \rightarrow \boldsymbol{35}_{\SU(4)} + \boldsymbol{45}_{\SU(4)} \boldsymbol{3}_{\SU(2)} + \boldsymbol{20'}_{\SU(4)} \boldsymbol{5}_{\SU(2)} + \boldsymbol{20'}_{\SU(4)} + \boldsymbol{15}_{\SU(4)} \boldsymbol{3}_{\SU(2)} + 1 .
\ee

We note the presence of the singlet, which suggests that there is a component of this operator that is fixed by the $\SU(2)\times \SU(4)$ subgroup. One should note the presence of the component with charges $\boldsymbol{15}_{\SU(4)} \boldsymbol{3}_{\SU(2)}$. These are important as the adjoint of $\SU(8)$ decomposes as $\boldsymbol{63}_{\SU(8)}\rightarrow \boldsymbol{15}_{\SU(4)} + \boldsymbol{3}_{\SU(2)} + \boldsymbol{15}_{\SU(4)} \boldsymbol{3}_{\SU(2)}$. Once the symmetry is broken the additional conserved currents in the $\boldsymbol{15}_{\SU(4)} \boldsymbol{3}_{\SU(2)}$ must somehow vanish from the index. However, the index cannot change under continuous deformations so either these shift due to a change in R-charge or they are canceled by a different term. The former cannot happen here as there are no $\U(1)$ commutants whose mixing could shift this term\footnote{This was also the case in most of our previous examples in this subsubsection, save for the model with the alternative R-charges for $N_f=5$. In that case, not all the extra currents can cancel against the terms from the marginal operator, but that case preserves a $\U(1)$ commutant, $\U(1)_{\tilde{c}}$. One can check that all terms not charged under this $\U(1)$ are canceled and the remaining ones could shift due to mixing of $\U(1)_{\tilde{c}}$ with the R-symmetry.}, and thus these terms must cancel against terms coming from the marginal operator. The fact the marginal operator has the right terms to cancel the extra currents is an important consistency check that this breaking is indeed possible.    

To summarize then, this operator contributes at order $-q$ and so has the right charges to be a superpotential operator that preserves the R-symmetry. Additionally, it should preserve the $\SU(12)$ and $\U(1)_B$ symmetries as it is uncharged under them. Note that the R-symmetry here is $\U(1)_{R_{mix}}$ but since the operator should preserve $\U(1)_B$ as well, we see that it should preserve the 4d R-symmetry. It is charged under $\U(1)_A$ and the $\SU(8)$, so these would be broken depending on the component of the operator inserted into the superpotential. Furthermore, we have argued that there should be a component of the operator breaking $\SU(8)\rightarrow \SU(2)_{ISO}\times \SU(4)$ such that the embedding is $\boldsymbol{8}_{\SU(8)}\rightarrow \boldsymbol{2}_{\SU(2)}\boldsymbol{4}_{\SU(4)}$. Furthermore, the charge of the operator under $\U(1)_A$ suggests that it is broken to $\mathbb{Z}_{16}$. This symmetry breaking pattern is precisely what is required to break the accidental 2d symmetries while preserving the ones with origin in the compactification picture.

\paragraph{c-extremization} The results of c-extremization for this model are again similar to the previous case. Performing the c-extremization while demanding that $\phi$ has R-charge $0$ leads to the R-symmetry under which both $\phi$ and $\psi$ have R-charge zero. The explicit R-symmetry achieving this is: $\U(1)_R^{\text{SC}}=\U(1)_{R_{mix}}-\frac{1}{2} \U(1)_A$, and we again conclude that the additional operator is irrelevant due to the negative $\U(1)_A$ mixing. Alternatively, performing the unconstrained mixing we find the R-symmetry under which $\psi$ has R-charge $0$ and $\phi$ has R-charge $1$. Explicitly, it is given by: $\U(1)_R^{\text{SC}}=\U(1)_{R_{mix}}+\frac{1}{4} \U(1)_A -\frac{1}{4} \U(1)_B$. The positive $\U(1)_A$ mixing implies that the additional operator is relevant but the resulting central charge is negative. 

\paragraph{Related choices of R-charges}

\begin{figure}[htbp]
    \centering
    
    \begin{minipage}{0.45\textwidth}
        \centering
        \begin{tikzpicture}[
            gauge/.style={circle, draw, thick, minimum size=10mm, inner sep=0pt, font=\large},
            flavor/.style={rectangle, draw, thick, minimum size=9mm, inner sep=0pt, font=\large},
            edge label/.style={font=\small, auto}
        ]
        
        \node[gauge] (G2) at (0,0) {$2$};
        
        \node[flavor] (F12)  at (0:2.5)   {$12$};
        \node[flavor] (F2_4) at (165:2.5) {$2$};
        \node[flavor] (F2_2) at (195:2.5) {$2$};
        
        \draw[mid arrow] (G2) -- node[above, edge label] {$\textcolor{red}{0}$} (F12);
        
        \draw[mid arrow] (F2_4) -- node[edge label] {$\textcolor{red}{4}$} (G2);
        \draw[mid arrow] (F2_2) -- node[swap, edge label] {$\textcolor{red}{2}$} (G2);
        
        \end{tikzpicture}
    \end{minipage}\hfill
    \begin{minipage}{0.45\textwidth}
        \centering
        \begin{tikzpicture}[
            gauge/.style={circle, draw, thick, minimum size=10mm, inner sep=0pt, font=\large},
            flavor/.style={rectangle, draw, thick, minimum size=9mm, inner sep=0pt, font=\large},
            edge label/.style={font=\small, auto}
        ]
        
        \node[gauge] (G2) at (0,0) {$2$};
        
        \node[flavor] (F12)  at (15:2.5)  {$12$};
        \node[flavor] (F1_4) at (165:2.5) {$1$};
        \node[flavor] (F1_2) at (195:2.5) {$1$};
        \node[flavor] (F2)   at (345:2.5) {$2$};
        
        \draw[mid arrow] (G2)   -- node[edge label]       {$\textcolor{red}{0}$} (F12);
        \draw[mid arrow] (F1_4) -- node[edge label]       {$\textcolor{red}{4}$} (G2);
        \draw[mid arrow] (F1_2) -- node[swap, edge label] {$\textcolor{red}{2}$} (G2);
        \draw[mid arrow] (F2)   -- node[edge label]       {$\textcolor{red}{3}$} (G2);
        
        \end{tikzpicture}
    \end{minipage}
    
    \caption{Quiver diagram of a 4d theory with $N_f = 8$ with a different non-anomalous R-charge choice.}
    \label{fig:combined_quivers}
\end{figure}
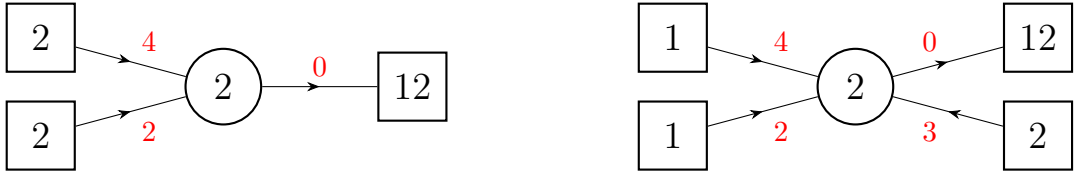

Consider the compactification of the same 4d theory, but where we perform the topological twisting with a different R-symmetry. For instance, we can choose an R-symmetry under which two chiral fields have R-charge $4$, 2 have R-charge $2$ and the rest have R-charge $0$. Alternatively, we can take only one chiral with R-charge $4$ and $2$ and keep two chirals with R-charge $3$ and the rest with R-charge $0$. We illustrate these two theories in figure \ref{fig:combined_quivers}.

Carrying out the reduction, one can see that we end up with a similar looking 2d $(0,2)$ theory, consisting of an $\SU(2)$ vector multiplet, $12$ fundamental chiral fields and $8$ fundamental Fermi fields. The resulting models are naively identical though the compactification preserves different subgroups of the 4d symmetries. We can then ask whether these options can be realized by using different components of the marginal operator. Here we shall explore this issue.   

We begin with the slightly simpler case where we have two sets of chiral fields with R-charges $4$ and $2$. A quiver for that theory is:

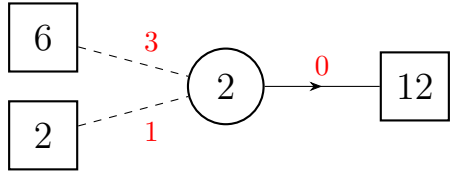
\begin{center}
\begin{tikzpicture}[
    gauge/.style={circle, draw, thick, minimum size=10mm, inner sep=0pt, font=\large},
    flavor/.style={rectangle, draw, thick, minimum size=9mm, inner sep=0pt, font=\large},
    edge label/.style={font=\small, auto}
]

\node[gauge] (G2) at (0,0) {$2$};

\node[flavor] (F12) at (0:2.5)   {$12$};
\node[flavor] (F6)  at (165:2.5) {$6$};
\node[flavor] (F2)  at (195:2.5) {$2$};

\draw[mid arrow] (G2) -- node[above, edge label] {$\textcolor{red}{0}$} (F12);

\draw[dashed] (F6) -- node[edge label] {$\textcolor{red}{3}$} (G2);
\draw[dashed] (F2) -- node[swap, edge label] {$\textcolor{red}{1}$} (G2);

\end{tikzpicture}
\end{center}

The charges of the operators under the relevant symmetries are shown in the table below:

\begin{table}[H]
    \centering
    \begin{tabular}{c|c|c c c c c c c c}
        & $\SU(2)_{\text{gauge}}$ & $\SU(6)$ & $\SU(2)$ & $\SU(12)$ & $\U(1)_R$ & $\U(1)_{R_{mix}}$& $\U(1)_{\tilde{B}}$ & $\U(1)_{\tilde{A}}$ & $\U(1)_{\tilde{C}}$ \\ \hline
        $\psi_4$ & $\ydiagram{1}$ & $\ydiagram{1}$ & $\mathbf{1}$ & $\mathbf{1}$ & $3$ & $\frac{1}{2}$ & $+3$ & $+1$ & $$+1$$ \\
        $\psi_2$ &  $\ydiagram{1}$ & $\mathbf{1}$ & $\ydiagram{1}$ & $\mathbf{1}$ & $1$ & $\frac{1}{2}$ & $+3$ & $+1$ & $-1$ \\
        $\phi$ & $\ydiagram{1}$ & $\mathbf{1}$ & $\mathbf{1}$ & $\overline{\ydiagram{1}}$ & $0$  & $\frac{1}{2}$ & $-1$ & $+1$ & $0$ \\
    \end{tabular}
    \caption{Table of gauge and symmetry charges of the field content of the $(0,2)$ 2d theory resulting from the $S^2$ reduction of the 4d theory with a slightly different choice of R-charges. Here $\U(1)_{R_{mix}}=\U(1)_R - \frac{1}{2} \U(1)_{\tilde{B}} - \U(1)_{\tilde{C}}$ is a mixed R-charge.}
\end{table}

Here, $\U(1)_R$ is the 4d R-symmetry, and the $\U(1)_{\tilde{B}}\times \U(1)_{\tilde{C}} \times \SU(2) \times \SU(12)$ global symmetries all originate from 4d. Additionally, the $\SU(2)\times \SO(3)_{\text{ISO}} \subset \SU(6)$, with the embedding such that $\boldsymbol{6}_{\SU(6)}\rightarrow \boldsymbol{2}_{\SU(2)}\boldsymbol{3}_{\SO(3)_{ISO}}$, are also symmetries arising from the compactification. The $\SU(8)$ rotating the $8$ Fermi multiplets and $\U(1)_{\tilde{A}}$ are accidental enhancements in 2d and we want to check whether these can be broken with the additional superpotential. To check this we need to translate between the charges of the current theory to those of the previous case for which we evaluated the elliptic genus.

First, we note that $\U(1)_{R_{mix}}$ is also an R-symmetry of this model that involves mixing of the $4d$ R-symmetry with only symmetries originating in 4d. Therefore, we expect the additional term to preserve it and so can rely on the previous result for the elliptic genus. Next we need to determine how to map the symmetries between the models. For this we note that the fields should be mapped as:

\bea
& & \tilde{a} \tilde{b}^3 \tilde{c} \boldsymbol{6}_{\tilde{h}} + \tilde{a} \tilde{b}^3 \tilde{c}^{-1}\boldsymbol{2}_{\tilde{g}} = \tilde{a} \tilde{b}^3 \sqrt{\tilde{c}} (\sqrt{\tilde{c}}\boldsymbol{6}_{\tilde{h}} + \tilde{c}^{-\frac{3}{2}}\boldsymbol{2}_{\tilde{g}}) \rightarrow a b^3 \boldsymbol{8}_{h} , \\ \nonumber
& & \frac{\tilde{a}}{\tilde{b}} \boldsymbol{\overline{12}}_{\tilde{f}} \rightarrow \frac{a}{b} \boldsymbol{\overline{12}}_{f} .
\eea

This determines that the mapping should be:

\be
\boldsymbol{12}_{f} \rightarrow \boldsymbol{12}_{\tilde{f}} \; , \; \boldsymbol{8}_{h} \rightarrow \sqrt{\tilde{c}}\boldsymbol{6}_{\tilde{h}} + \tilde{c}^{-\frac{3}{2}}\boldsymbol{2}_{\tilde{g}} \; , \; \tilde{a} \rightarrow a \tilde{c}^{-\frac{1}{8}} \; , \; \tilde{b} \rightarrow b \tilde{c}^{-\frac{1}{8}} .
\ee

Previously we observed a non-perturbative operator contributing as $-q\boldsymbol{336_{h}}a^{-16}$ in the elliptic genus. We would like to convert the charges to those under the symmetries of the current model, for which we use the transformation above to find:

\be
\boldsymbol{336_{h}}a^{-16} \rightarrow \tilde{a}^{-16} \tilde{c}^{-2} ( \boldsymbol{70}_{\tilde{h}} \boldsymbol{2}_{\tilde{g}} + \tilde{c}^{-2} \boldsymbol{21}_{\tilde{h}} \boldsymbol{3}_{\tilde{g}} + \tilde{c}^{2} \boldsymbol{\overline{105'}}_{\tilde{h}} + \tilde{c}^{-4} \boldsymbol{6}_{\tilde{h}} \boldsymbol{2}_{\tilde{g}} + \tilde{c}^{-2} \boldsymbol{15}_{\tilde{h}} + \tilde{c}^{-6} ).
\ee

We note that these contain the term $\tilde{a}^{-16} \boldsymbol{\overline{105'}}_{\tilde{h}}$, which seems to provide the necessary breaking. Specifically, decomposing the $\boldsymbol{\overline{105'}}$ of $\SU(6)$ under the embedding  $\SU(2)\times \SO(3) \subset \SU(2)\times \SU(3) \subset \SU(6)$ (such that $\boldsymbol{6}_{\SU(6)}\rightarrow \boldsymbol{2}_{\SU(2)}\boldsymbol{3}_{\SU(3)} \rightarrow \boldsymbol{2}_{\SU(2)}\boldsymbol{3}_{\SO(3)}$), we find:

\bea
& & \boldsymbol{\overline{105'}} \rightarrow \boldsymbol{15'}_{\SU(3)} + \boldsymbol{15}_{\SU(3)}\boldsymbol{3}_{\SU(2)} + \boldsymbol{\overline{6}}_{\SU(3)}\boldsymbol{5}_{\SU(2)} + \boldsymbol{\overline{6}}_{\SU(3)} + \boldsymbol{3}_{\SU(3)}\boldsymbol{3}_{\SU(2)} \\ \nonumber &\rightarrow& 2+ \boldsymbol{9}_{\SU(3)} + 2\boldsymbol{5}_{\SU(3)} + \boldsymbol{3}_{\SU(2)} (\boldsymbol{7}_{\SU(3)} + \boldsymbol{5}_{\SU(3)} + 2\boldsymbol{3}_{\SU(3)}) + \boldsymbol{5}_{\SU(2)} (\boldsymbol{5}_{\SU(3)} + 1) .
\eea

We can indeed see the presence of the singlet, indicating the invariant components. Furthermore, we can examine the decomposition of the adjoint representation of $\SU(6)$:

\be
\boldsymbol{35} \rightarrow \boldsymbol{8}_{\SU(3)}\boldsymbol{3}_{\SU(2)} + \boldsymbol{8}_{\SU(3)} + \boldsymbol{3}_{\SU(2)} \rightarrow \boldsymbol{5}_{\SU(3)}\boldsymbol{3}_{\SU(2)} + \boldsymbol{3}_{\SU(3)}\boldsymbol{3}_{\SU(2)} + \boldsymbol{5}_{\SU(3)} + \boldsymbol{3}_{\SU(2)} + \boldsymbol{3}_{\SU(3)} .
\ee

One can see that the additional terms in the adjoint, beside the adjoint of $\SU(2)\times \SO(3)$ can cancel against terms coming from the $\boldsymbol{\overline{105'}}$ so the resulting index is consistent. The other terms appearing in the decomposition of the $\boldsymbol{336_{h}}$ are all charged under $\U(1)_{\tilde{C}}$ and so could shift due to mixing with this symmetry once the $\SU(8)$ is broken. Finally, the charge under $\U(1)_{\tilde{A}}$ implies that introducing this component would break this symmetry to $\mathbb{Z}_{16}$. Overall then, we see that this superpotential term appears to allow the breaking of $\SU(8)\times \U(1)_{\tilde{A}}\rightarrow \U(1)_{\tilde{C}}\times \SU(2)^2\times \SO(3)_{\text{ISO}}\times \mathbb{Z}_{16}$, which are exactly the symmetries expected from 4d.

Next we turn to the second case, where we have chiral fields with R-charges $4$, $3$ and $2$. A quiver for the resulting $2d$ theory is:

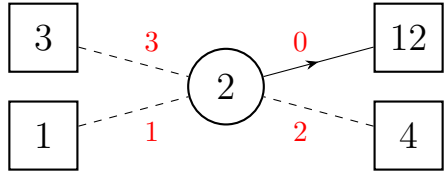
\begin{center}
\begin{tikzpicture}[
    gauge/.style={circle, draw, thick, minimum size=10mm, inner sep=0pt, font=\large},
    flavor/.style={rectangle, draw, thick, minimum size=9mm, inner sep=0pt, font=\large},
    edge label/.style={font=\small, auto}
]

\node[gauge] (G2) at (0,0) {$2$};

\node[flavor] (F12) at (15:2.5)  {$12$};
\node[flavor] (F3)  at (165:2.5) {$3$};
\node[flavor] (F1)  at (195:2.5) {$1$};
\node[flavor] (F4)  at (345:2.5) {$4$};

\draw[mid arrow] (G2) -- node[edge label]       {$\textcolor{red}{0}$} (F12);
\draw[dashed]    (F3) -- node[edge label]       {$\textcolor{red}{3}$} (G2);
\draw[dashed]    (F1) -- node[swap, edge label] {$\textcolor{red}{1}$} (G2);
\draw[dashed]    (F4) -- node[edge label]       {$\textcolor{red}{2}$} (G2);

\end{tikzpicture}
\end{center}

The charges of the operators under the relevant symmetries are shown in the table below:

\begin{table}[H]
    \centering
    \begin{tabular}{c|c|c c c c c c c c c}
        & $\SU(2)_{G}$ & $\SU(4)$ & $\SU(3)$ & $\SU(12)$ & $\U(1)_R$ & $\U(1)_{\hat{R}}$& $\U(1)_{\hat{B}}$ & $\U(1)_{\hat{A}}$ & $\U(1)_{\hat{C}}$ & $\U(1)_{\hat{D}}$ \\ \hline
        $\psi_4$ & $\ydiagram{1}$ & $\mathbf{1}$ & $\ydiagram{1}$ & $\mathbf{1}$ & $3$ & $\frac{1}{2}$ & $+3$ & $+1$ & $+1$ & $+1$ \\
        $\psi_3$ & $\ydiagram{1}$ & $\ydiagram{1}$ & $\mathbf{1}$ & $\mathbf{1}$ & $2$ & $\frac{1}{2}$ & $+3$ & $+1$ & $-1$ & $0$ \\
        $\psi_2$ &  $\ydiagram{1}$ & $\mathbf{1}$ & $\mathbf{1}$ & $\mathbf{1}$ & $1$ & $\frac{1}{2}$ & $+3$ & $+1$ & $+1$ & $-1$ \\
        $\phi$ & $\ydiagram{1}$ & $\mathbf{1}$ & $\mathbf{1}$ & $\overline{\ydiagram{1}}$ & $0$  & $\frac{1}{2}$ & $-1$ & $+1$ & $0$ & $0$ \\
    \end{tabular}
    \caption{Table of gauge and symmetry charges of the field content of the $(0,2)$ 2d theory resulting from the $S^2$ reduction of the 4d theory with a slightly different choice of R-charges. Here $\U(1)_{\hat{R}}=\U(1)_R - \frac{1}{2} \U(1)_{\hat{B}} - \U(1)_{\hat{D}}$ is a mixed R-charge, and $\SU(2)_{G}$ stands for the gauge symmetry.}
\end{table}

As before $\U(1)_R$ is the 4d R-symmetry, and the $\U(1)_{\hat{B}}\times \U(1)_{\hat{C}}\times \U(1)_{\hat{D}} \times \SU(12)$ global symmetries all originate from 4d. Additionally, there are the 4d $\SU(2)$ and the $\SO(3)_{\text{ISO}}$ that arise from the compactification picture. These are embedded in the $\SU(3)\times \SU(4)$ symmetries as $\SO(3)_{\text{ISO}} \subset \SU(3)$, $\SU(2)\times \SU(2)_{\text{ISO}} = \text{Spin}(4) \subset \SU(4)$. The $\SU(8)$ rotating the $8$ Fermi multiplets and $\U(1)_{\hat{A}}$ are accidental enhancements in 2d. Like in the previous cases, we suspect that these can be broken with an additional superpotential and want to check this using the previous results for the elliptic genus. we note that again $\U(1)_{\hat{R}}$ is also an R-symmetry of this model that involves mixing with the $4d$ R-symmetry of only symmetries originating in 4d so there should be no issue with utilizing the previous result.

Next we need to determine the mapping of symmetries between the models. To do this, we first consider the mapping of the fields:

\bea
& & \frac{\hat{a} \hat{b}^3 \hat{c}}{\hat{d}} + \hat{a} \hat{b}^3 \hat{c}\hat{d} \boldsymbol{3}_{\hat{g}} + \frac{\hat{a} \hat{b}^3}{\hat{c}}\boldsymbol{4}_{\hat{h}} = \hat{a} \hat{b}^3 \hat{d}^{\frac{1}{4}} (\frac{\hat{c}}{\hat{d}^{\frac{5}{4}}} + \hat{c}\hat{d}^{\frac{3}{4}} \boldsymbol{3}_{\hat{g}} + \frac{1}{\hat{c}\hat{d}^{\frac{1}{4}}}\boldsymbol{4}_{\hat{h}}) \rightarrow a b^3 \boldsymbol{8}_{h} , \\ \nonumber
& & \frac{\hat{a}}{\hat{b}} \boldsymbol{\overline{12}}_{\hat{f}} \rightarrow \frac{a}{b} \boldsymbol{\overline{12}}_{f} .
\eea

This determines that the mapping should be:

\be
\boldsymbol{12}_{f} \rightarrow \boldsymbol{12}_{\hat{f}} \; , \; \boldsymbol{8}_{h} \rightarrow \frac{\hat{c}}{\hat{d}^{\frac{5}{4}}} + \hat{c}\hat{d}^{\frac{3}{4}} \boldsymbol{3}_{\hat{g}} + \frac{1}{\hat{c}\hat{d}^{\frac{1}{4}}}\boldsymbol{4}_{\hat{h}} \; , \; a \rightarrow \hat{a} \hat{d}^{\frac{1}{16}} \; , \; b \rightarrow \hat{b} \hat{d}^{\frac{1}{16}} .
\ee

Previously we observed a non-perturbative operator contributing as $-q\boldsymbol{336_{h}}a^{-16}$ in the elliptic genus. We would like to convert the charges to those under symmetries of the current model. Using the transformation above, we find:

\bea
& & \boldsymbol{336_{h}}a^{-16} \rightarrow \hat{a}^{-16} \hat{d}^{-1} ( \hat{d} \boldsymbol{10}_{\hat{h}} \boldsymbol{6}_{\hat{g}} + \hat{d}^{-1} \boldsymbol{10}_{\hat{h}} \boldsymbol{3}_{\hat{g}} + \hat{d}^{-3} \boldsymbol{10}_{\hat{h}} + \hat{c}^2 \boldsymbol{4}_{\hat{h}} \boldsymbol{6}_{\hat{g}} + \hat{c}^2 \hat{d}^2 \boldsymbol{4}_{\hat{h}} \boldsymbol{8}_{\hat{g}} \nonumber  \\ \nonumber &+& \hat{c}^2 \hat{d}^{-2} \boldsymbol{4}_{\hat{h}} \boldsymbol{3}_{\hat{g}} + \hat{c}^2 \boldsymbol{4}_{\hat{h}} \boldsymbol{\overline{3}}_{\hat{g}} + \hat{c}^4 \hat{d} \boldsymbol{8}_{\hat{g}} + \hat{c}^4 \hat{d}^3 \boldsymbol{\overline{6}}_{\hat{g}} + \hat{c}^4 \hat{d}^{-1} \boldsymbol{6}_{\hat{g}} + \hat{c}^{-2} \boldsymbol{\overline{20}}_{\hat{h}} \boldsymbol{3}_{\hat{g}} + \hat{c}^{-2}\hat{d}^{-2} \boldsymbol{\overline{20}}_{\hat{h}} \\ &+& \hat{c}^{-4}\hat{d}^{-1} \boldsymbol{20'}_{\hat{h}} + \hat{d} \boldsymbol{6}_{\hat{h}} \boldsymbol{\overline{3}}_{\hat{g}} + \hat{d}^{-1} \boldsymbol{6}_{\hat{h}} \boldsymbol{3}_{\hat{g}}).
\eea

As before we want to preserve $\U(1)_{\hat{C}}$ and $\U(1)_{\hat{D}}$ and so look for terms uncharged under these. Indeed we find the following terms in the decomposition: $\hat{a}^{-16} (\boldsymbol{10}_{\hat{h}} \boldsymbol{6}_{\hat{g}} + \boldsymbol{6}_{\hat{h}} \boldsymbol{\overline{3}}_{\hat{g}})$. These two terms indeed appear to allow the desired breaking. Specifically, consider the first term. Decomposing $\SU(4)\rightarrow \SU(2)_1\times \SU(2)_2$ and $\SU(3)\rightarrow SO(3)_3$, we see that it decomposes to: $\boldsymbol{10}_{\hat{h}} \boldsymbol{6}_{\hat{g}}\rightarrow 1 + \boldsymbol{3}_{1}\boldsymbol{3}_{2} + \boldsymbol{5}_{3}+\boldsymbol{3}_{1}\boldsymbol{3}_{2}\boldsymbol{5}_{3}$, and we indeed observe a singlet. The second and third terms in the decomposition can then cancel against the broken currents, as $\boldsymbol{15}_{\hat{h}}\rightarrow \boldsymbol{3}_{1} + \boldsymbol{3}_{2} + \boldsymbol{3}_{1}\boldsymbol{3}_{2}$, $\boldsymbol{8}_{\hat{g}}\rightarrow \boldsymbol{3}_{3} + \boldsymbol{5}_{3}$.  

This gives part of the decomposition, but we also need to lock $\SU(2)_3$ to either $\SU(2)_1$ or $\SU(2)_2$. This cannot be done by this term, but can be done by the second term, $\boldsymbol{6}_{\hat{h}} \boldsymbol{\overline{3}}_{\hat{g}}$. To see this we note that it decomposes to $\boldsymbol{6}_{\hat{h}} \boldsymbol{\overline{3}}_{\hat{g}} \rightarrow \boldsymbol{3}_{1}\boldsymbol{3}_{3} + \boldsymbol{3}_{2}\boldsymbol{3}_{3}$, with the first term locking $\SU(2)_3$ to $\SU(2)_1$ and the second locking it to $\SU(2)_2$. Additionally as before, $\U(1)_{\hat{A}}$ should be broken to $\mathbb{Z}_{16}$ because of the charge of the operator. To summarize, it appears that this superpotential term allows the breaking of $\SU(8)\times \U(1)_{\hat{A}}\rightarrow \U(1)_{\hat{C}}\times \U(1)_{\hat{D}}\times \SU(2)\times \SO(3)_{\text{ISO}}\times \mathbb{Z}_{16}$, which are exactly the symmetries expected from 4d.

\subsubsection{Cases involving fields with R-charge $1$}

So far we have used R-symmetries under which no field gets R-charge $1$. This case is somewhat special as the 4d fields with unit R-charge do not lead to any 2d fields. As such, in these cases the axial 2d $\U(1)$ symmetry can be lifted to a genuine symmetry of the 4d theory by letting all 4d fields with non-unit R-charge carry charges as dictated by the 2d fields and give the ones with unit R-charge whatever charge is needed to cancel the anomalies. The problems that we previously encountered involving symmetries broken by ABJ anomalies do not arise in these cases, hence why they were not considered here.

However, we can still consider compactification with fields carrying both unit R-charge and $R>2$ ones. In that case, we may still have issues with accidental non-abelian symmetries rotating multiple copies of fields produced in the reduction, and one can wonder if these can be broken by a superpotential without losing any of the 4d symmetries as in the above examples. This however, does not appear to be the general behavior, as we shall now argue.

The main issue is that we can always add additional chirals with unit R-charge without interfering with the reduction. For instance, given any of the above models, we can add any even number (to not encounter a Witten anomaly) of chiral fields with unit R-charge, and still get the same 2d theory. The only difference would be that now the $\U(1)_A$ symmetry can be lifted to a genuine 4d symmetry. However, the non-perturbative operators encountered above all carry charges under the $\U(1)_A$ symmetry, which was necessary so that they could break it. Furthermore, we have not observed any operator with the right charges to be marginal and that is completely uncharged under the abelian symmetries. As such, it seems that one cannot break the additional non-abelian 2d symmetries without sacrificing some of the abelian ones.

In the cases outlined above, where we add chiral fields with R-charge $1$ to the models considered here, we see that either the duality should work with the additional symmetries or some of the 4d symmetries should be broken by some mechanism. As the dual frame is involved in these cases, we defer a more detailed study of the duality in these cases for future work. Furthermore, the results found so far and the above argument suggest that this should hold more generally whenever chiral fields with R-charge $1$ are involved in the reduction, unless the gauge anomaly cancellation somehow excludes the possibility of getting the same 2d theory via a 4d model with an anomalous flavor symmetry. 

Another option is that there are multiple operators with the right charges allowing different breaking patterns depending on the different 4d origin of the 2d symmetries. For instance, we have seen how similar 2d theories can result from different 4d reductions preserving different symmetries, and how these can be accommodated by using different components of the appropriate marginal operator. This does not seem applicable to the cases considered here if one adds to them chiral fields of R-charge $1$, and we have not observed an example of this in these cases, but cannot rule out the possibility.     

\section*{Acknowledgments}
We would like to thank Hee-Cheol Kim, Ki-Hong Lee and Shlomo Razamat for insightful comments. GZ is partially supported by the Israel Science Foundation under grant no. 759/23. The research of SM is supported in part by the Israel Science Foundation under grant no. 2159/22, and by BSF-NSF grant no. 2023769.

\begin{appendix}

\section{Comment on the methodology}
\label{appendix:methodology}
Let us give a couple of comments regarding the technical method employed in the computations.

\subsection{The perturbative residue}
In the main body of this paper, we compare the full elliptic genus with the expected perturbative contribution. This latter quantity, referred to as the "perturbative residue," is computed as follows. 

Let $Z_{\text{1-loop}}(z_i, a_j; q)$ denote the 1-loop integrand as a function of the gauge fugacities $z_i$ and the flavor fugacities $a_j$. 
Here $Z_{\text{1-loop}}$ is the product of the one-loop determinants of the
multiplets of the theory, given in section \ref{sec:EGa2dI}: a factor
$1/\theta(a_j q^{\frac{r}{2}};q)$ for every chiral multiplet,
$\theta(a_j q^{\frac{r+1}{2}};q)$ for every Fermi multiplet, and
$(q;q)^{2 r_{\mathfrak{g}}}_\infty \prod_{\alpha} \theta(z^{\alpha}_i;q)$ for
the vector multiplet combined with the Haar measure, with the gauge and flavor
fugacities substituted according to the representation carried by each field.

Expanding this 1-loop determinant around $q\to 0$, we obtain a series expansion: 
\begin{equation}
    Z_{\text{1-loop}}(z_i,a_j;q) = \sum_\alpha q^\alpha Z_\alpha(z_i,a_j) 
\end{equation}

The perturbative residue takes the gauge invariant combinations. It is then computed as:
\begin{equation}
    \mathcal{I}_{\text{pert.}}(a_j; q) = \sum_\alpha q^\alpha \  \underset{z_i=0}{\text{Res}} \ Z_\alpha(z_i,q_j)
\end{equation}

\subsection{Numerical determination of representations}
In many cases, the superconformal index is computationally hard to evaluate. In order to make computations faster, the method employed in this paper relies on a semi-numerical approach. We formulate an initial ansatz for the representation content and then verify it numerically.

We illustrate this procedure using representations of $\SU(N)$. Let $\chi_R(\boldsymbol{u})$ denote the character of a representation $R$, expressed as a function of the fugacities $\boldsymbol{u} = (u_1, \dots, u_N)$ parameterizing the maximal torus. By evaluating this character at a specific point in $\mathbb{R}^N$, we obtain a unique numerical signature for each representation. For example, by choosing integer values $u_k = k$ for $k \in \{1, \dots, N\}$ (with appropriate rescaling to satisfy the $\SU(N)$ constraint $\prod_{k=1}^N u_k = 1$), we compute a characteristic scalar value $v(R) = \chi_R(\boldsymbol{u})_{u_k = k}$.

Exploiting this property, we first compile a dictionary mapping a truncated set of low-dimensional representations $R$ to their corresponding signatures $v(R)$. By numerically evaluating the index at these same coordinates, we can match the resulting coefficients against our dictionary to uniquely identify the associated representation $R$ appearing at each order. This matching is computationally doable if we assume that the target representations contain a low number of irreducible representations. Finally, to ensure the uniqueness and correctness of this identification, we establish the validity of the proposed ansatz by evaluating and matching the characters at an independent set of points in $\mathbb{R}^N$.

\section{Detailed index expansions}
In this appendix, we collect the full expressions for indices that, for clarity, were not written out completely in the main text.
\subsection{Seiberg $\SU$ gauge duality with $N_c=2$, $N_f=3$} \label{appendix:q-expansion-resultsNc2Nf3}
The superconformal index for the Wess-Zumino model is up to order $O(q)$:
\begin{dmath}
    \mathcal{I}^{\text{WZ}}_{\text{JK}}=1+q^{\frac{1}{12}}bd\mathbf{3}_{f}+q^{\frac{1}{6}}b^{2}d^{2}\mathbf{6}_{f}+q^{\frac{1}{4}}\left(\frac{c^{2}d}{b}-\frac{bc\mathbf{2}_{h}}{d}+b^{3}d^{3}\mathbf{10}_{f}\right)\\+q^{\frac{1}{3}}\left(-b^{2}c\mathbf{2}_{h}\mathbf{3}_{f}+b^{4}d^{4}\mathbf{15}'_{f}\right)\\+q^{\frac{5}{12}}\left(-bc^{2}d^{3}\bar{\mathbf{3}}_{f}-b^{3}cd\mathbf{2}_{h}\mathbf{6}_{f}+b^{5}d^{5}\overline{\mathbf{21}}_{f}\right)\\+q^{\frac{1}{2}}\left(\frac{b^{2}c^{2}}{d^{2}}+\frac{c^{4}d^{2}}{b^{2}}-c^{3}\mathbf{2}_{h}-b^{2}c^{2}d^{4}\mathbf{8}_{f}-b^{4}cd^{2}\mathbf{2}_{h}\mathbf{10}_{f}+b^{6}d^{6}\mathbf{28}_{f}\right)\\+q^{\frac{7}{12}}\left(\frac{b^{3}c^{2}\mathbf{3}_{f}}{d}-b^{3}c^{2}d^{5}\mathbf{15}_{f}-b^{5}cd^{3}\mathbf{2}_{h}\mathbf{15}'_{f}+b^{7}d^{7}\mathbf{36}_{f}\right)\\+q^{\frac{2}{3}}\left(-\frac{\bar{\mathbf{3}}_{f}}{c^{2}d^{2}}+b^{2}c^{3}d^{2}\mathbf{2}_{h}\bar{\mathbf{3}}_{f}+b^{4}c^{2}\mathbf{6}_{f}-b^{6}cd^{4}\mathbf{2}_{h}\overline{\mathbf{21}}_{f}-b^{4}c^{2}d^{6}\overline{\mathbf{24}}_{f}+b^{8}d^{8}\overline{\mathbf{45}}_{f}\right)\\+q^{\frac{3}{4}}\left(\frac{bc^{4}}{d}+\frac{c^{6}d^{3}}{b^{3}}+bc^{4}d^{5}-\frac{d\mathbf{2}_{h}}{bc}-\frac{c^{5}d\mathbf{2}_{h}}{b}-\frac{b\mathbf{8}_{f}}{c^{2}d}+b^{3}c^{3}d^{3}\mathbf{2}_{h}\mathbf{8}_{f}+b^{5}c^{2}d\mathbf{10}_{f}-b^{7}cd^{5}\mathbf{2}_{h}\mathbf{28}_{f}-b^{5}c^{2}d^{7}\mathbf{35}_{f}+b^{9}d^{9}\mathbf{55}_{f}\right)\\+q^{\frac{5}{6}}\left(b^{2}c^{4}d^{6}\mathbf{3}_{f}-\frac{d^{2}\mathbf{2}_{h}\mathbf{3}_{f}}{c}-\frac{b^{2}\mathbf{15}_{f}}{c^{2}}+b^{4}c^{3}d^{4}\mathbf{2}_{h}\mathbf{15}_{f}+b^{6}c^{2}d^{2}\mathbf{15}'_{f}-b^{8}cd^{6}\mathbf{2}_{h}\mathbf{36}_{f}-b^{6}c^{2}d^{8}\mathbf{48}_{f}+b^{10}d^{10}\mathbf{66}_{f}\right)\\+q^{\frac{11}{12}}\left(-b^{3}c^{4}d\bar{\mathbf{3}}_{f}+\frac{b\mathbf{2}_{h}\bar{\mathbf{3}}_{f}}{cd^{3}}+b^{3}c^{4}d^{7}\mathbf{6}_{f}-\frac{bd^{3}\mathbf{2}_{h}\mathbf{6}_{f}}{c}+b^{7}c^{2}d^{3}\overline{\mathbf{21}}_{f}-\frac{b^{3}d\overline{\mathbf{24}}_{f}}{c^{2}}+b^{5}c^{3}d^{5}\mathbf{2}_{h}\overline{\mathbf{24}}_{f}-b^{9}cd^{7}\mathbf{2}_{h}\overline{\mathbf{45}}_{f}-b^{7}c^{2}d^{9}\overline{\mathbf{63}}_{f}+b^{11}d^{11}\overline{\mathbf{78}}_{f}\right)\\+q\left(3+c^{6}+\frac{c^{8}d^{4}}{b^{4}}-\frac{cd^{2}\mathbf{2}_{h}}{b^{2}}-\frac{c^{7}d^{2}\mathbf{2}_{h}}{b^{2}}-b^{2}c^{5}d^{4}\mathbf{2}_{h}+\mathbf{3}_{h}+\mathbf{8}_{f}-b^{4}c^{4}d^{2}\mathbf{8}_{f}+\frac{b^{2}\mathbf{2}_{h}\mathbf{8}_{f}}{cd^{2}}+b^{4}c^{4}d^{8}\mathbf{10}_{f}-\frac{b^{2}d^{4}\mathbf{2}_{h}\mathbf{10}_{f}}{c}+b^{8}c^{2}d^{4}\mathbf{28}_{f}-\frac{b^{4}d^{2}\mathbf{35}_{f}}{c^{2}}+b^{6}c^{3}d^{6}\mathbf{2}_{h}\mathbf{35}_{f}-b^{10}cd^{8}\mathbf{2}_{h}\mathbf{55}_{f}-b^{8}c^{2}d^{10}\mathbf{80}_{f}+b^{12}d^{12}\mathbf{91}_{f}\right)
\end{dmath}
The gauge theory dual has superconformal index (again, up to  order $O(q)$):
\begin{dmath}
    \mathcal{I}^{\text{gauge}}_{\text{JK}}=1+q^{\frac{1}{12}}bd\mathbf{3}_{f}+q^{\frac{1}{6}}b^{2}d^{2}\mathbf{6}_{f}+q^{\frac{1}{4}}\left(\frac{c^{2}d}{b}-bcd^{2}\mathbf{2}_{h}+b^{3}d^{3}\mathbf{10}_{f}\right)\\+q^{\frac{1}{3}}\left(-b^{2}cd^{3}\mathbf{2}_{h}\mathbf{3}_{f}+b^{4}d^{4}\mathbf{15}'_{f}\right)\\+q^{\frac{5}{12}}\left(-bc^{2}\bar{\mathbf{3}}_{f}-b^{3}cd^{4}\mathbf{2}_{h}\mathbf{6}_{f}+b^{5}d^{5}\overline{\mathbf{21}}_{f}\right)\\+q^{\frac{1}{2}}\left(\frac{c^{4}d^{2}}{b^{2}}+b^{2}c^{2}d^{4}-c^{3}\mathbf{2}_{h}-b^{2}c^{2}d\mathbf{8}_{f}-b^{4}cd^{5}\mathbf{2}_{h}\mathbf{10}_{f}+b^{6}d^{6}\mathbf{28}_{f}\right)\\+q^{\frac{7}{12}}\left(b^{3}c^{2}d^{5}\mathbf{3}_{f}-b^{3}c^{2}d^{2}\mathbf{15}_{f}-b^{5}cd^{6}\mathbf{2}_{h}\mathbf{15}'_{f}+b^{7}d^{7}\mathbf{36}_{f}\right)\\+q^{\frac{2}{3}}\left(-\frac{d\bar{\mathbf{3}}_{f}}{c^{2}}+b^{2}c^{3}d^{2}\mathbf{2}_{h}\bar{\mathbf{3}}_{f}+b^{4}c^{2}d^{6}\mathbf{6}_{f}-b^{6}cd^{7}\mathbf{2}_{h}\overline{\mathbf{21}}_{f}-b^{4}c^{2}d^{3}\overline{\mathbf{24}}_{f}+b^{8}d^{8}\overline{\mathbf{45}}_{f}\right)\\+q^{\frac{3}{4}}\left(\frac{bc^{4}}{d}+bc^{4}d^{2}+\frac{c^{6}d^{3}}{b^{3}}-\frac{d\mathbf{2}_{h}}{bc}-\frac{c^{5}d\mathbf{2}_{h}}{b}-\frac{bd^{2}\mathbf{8}_{f}}{c^{2}}+b^{3}c^{3}d^{3}\mathbf{2}_{h}\mathbf{8}_{f}+b^{5}c^{2}d^{7}\mathbf{10}_{f}-b^{7}cd^{8}\mathbf{2}_{h}\mathbf{28}_{f}-b^{5}c^{2}d^{4}\mathbf{35}_{f}+b^{9}d^{9}\mathbf{55}_{f}\right)\\+q^{\frac{5}{6}}\left(b^{2}c^{4}\mathbf{3}_{f}-\frac{\mathbf{2}_{h}\mathbf{3}_{f}}{cd}-\frac{b^{2}d^{3}\mathbf{15}_{f}}{c^{2}}+b^{4}c^{3}d^{4}\mathbf{2}_{h}\mathbf{15}_{f}+b^{6}c^{2}d^{8}\mathbf{15}'_{f}-b^{8}cd^{9}\mathbf{2}_{h}\mathbf{36}_{f}-b^{6}c^{2}d^{5}\mathbf{48}_{f}+b^{10}d^{10}\mathbf{66}_{f}\right)\\+q^{\frac{11}{12}}\left(-b^{3}c^{4}d^{4}\bar{\mathbf{3}}_{f}+\frac{bd^{3}\mathbf{2}_{h}\bar{\mathbf{3}}_{f}}{c}+b^{3}c^{4}d\mathbf{6}_{f}-\frac{b\mathbf{2}_{h}\mathbf{6}_{f}}{c}+b^{7}c^{2}d^{9}\overline{\mathbf{21}}_{f}-\frac{b^{3}d^{4}\overline{\mathbf{24}}_{f}}{c^{2}}+b^{5}c^{3}d^{5}\mathbf{2}_{h}\overline{\mathbf{24}}_{f}-b^{9}cd^{10}\mathbf{2}_{h}\overline{\mathbf{45}}_{f}-b^{7}c^{2}d^{6}\overline{\mathbf{63}}_{f}+b^{11}d^{11}\overline{\mathbf{78}}_{f}\right)\\+q\left(3+c^{6}-\frac{1}{d^{3}}+d^{3}+\frac{c^{8}d^{4}}{b^{4}}-b^{2}c^{5}d\mathbf{2}_{h}-\frac{cd^{2}\mathbf{2}_{h}}{b^{2}}-\frac{c^{7}d^{2}\mathbf{2}_{h}}{b^{2}}+\mathbf{3}_{h}+\mathbf{8}_{f}-b^{4}c^{4}d^{5}\mathbf{8}_{f}+\frac{b^{2}d^{4}\mathbf{2}_{h}\mathbf{8}_{f}}{c}+b^{4}c^{4}d^{2}\mathbf{10}_{f}-\frac{b^{2}d\mathbf{2}_{h}\mathbf{10}_{f}}{c}+b^{8}c^{2}d^{10}\mathbf{28}_{f}-\frac{b^{4}d^{5}\mathbf{35}_{f}}{c^{2}}+b^{6}c^{3}d^{6}\mathbf{2}_{h}\mathbf{35}_{f}-b^{10}cd^{11}\mathbf{2}_{h}\mathbf{55}_{f}-b^{8}c^{2}d^{7}\mathbf{80}_{f}+b^{12}d^{12}\mathbf{91}_{f}\right)
\end{dmath}

\subsection{$\USp$ gauge duality at $n=2$} \label{appendix:q-expansion-resultsAntiSymN2}
The index up to order $O(q)$ of the Wess-Zumino model is:
\begin{equation}
\begin{split}
\mathcal{I}_{\text{JK}}^{\text{WZ}}
&= 1 + a^2b^2\,\mathbf{21}_u\,q^{1/8}
   + a^4b^4\bigl(\overline{\mathbf{35}}_u + \overline{\mathbf{196}}_u\bigr)\,q^{1/4} \\
&\quad + b^6\bigl((a^6-1)\,\overline{\mathbf{7}}_u
   + a^6(\overline{\mathbf{588}}_u + \overline{\mathbf{1176}}_u)\bigr)\,q^{3/8} \\
&\quad + a^2b^8\bigl(-\mathbf{7}_u + (a^6-1)\,\mathbf{140}_u
   + a^6(\mathbf{490}'_u + \mathbf{4704}'_u + \mathbf{5292}''_u)\bigr)\,q^{1/2} \\
&\quad + \Bigl(-\frac{\mathbf{7}_u}{b^6}
   + a^4b^{10}\bigl[-(\mathbf{35}_u + \mathbf{112}_u) + (a^6-1)(\mathbf{210}_u + \mathbf{1260}_u) \\
&\qquad\qquad + a^6(\mathbf{7056}_u + \mathbf{19404}'_u + \mathbf{25200}_u)\bigr]\Bigr)\,q^{5/8} \\
&\quad + \Bigl(-\frac{a^2}{b^4}(\mathbf{35}_u + \mathbf{112}_u)
   + b^{12}\bigl[-(a^6-1)\,\overline{\mathbf{21}}_u
   - a^6(\overline{\mathbf{224}}_u + \overline{\mathbf{490}}_u + \overline{\mathbf{882}}_u) \\
&\qquad\qquad + a^6(a^6-1)(\overline{\mathbf{28}}_u + \overline{\mathbf{3402}}_u + \overline{\mathbf{7350}}'_u) \\
&\qquad\qquad + a^{12}(\overline{\mathbf{4116}}_u + \overline{\mathbf{50400}}_u
   + \overline{\mathbf{60984}}_u + \overline{\mathbf{103950}}'_u)\bigr]\Bigr)\,q^{3/4} \\
&\quad + \Bigl(a^2b^{14}\bigl[1 - (a^6-1)(\mathbf{48}_u + \mathbf{392}_u)
   - a^6(\mathbf{784}_u + \mathbf{2940}_u + \mathbf{3528}_u + \mathbf{4704}''_u) \\
&\qquad\qquad + a^6(a^6-1)(\overline{\mathbf{540}}_u + \overline{\mathbf{2646}}_u
   + \overline{\mathbf{26460}}_u + \overline{\mathbf{32340}}''_u) \\
&\qquad\qquad + a^{12}(\mathbf{52920}_u + \mathbf{169884}_u + \mathbf{247500}_u + \mathbf{355740}_u)\bigr] \\
&\qquad\quad - \frac{1}{b^2}(a^4(\overline{\mathbf{224}}_u + \overline{\mathbf{490}}_u + \overline{\mathbf{882}}_u)+(a^4-\frac{1}{a^2})\,\overline{\mathbf{21}}_u)\Bigr)\,q^{7/8} \\
&\quad + \Bigl(2 - a^6 - (a^6-2)\,\mathbf{48}_u - (a^6-1)\,\mathbf{392}_u
   - a^6(\mathbf{784}_u + \mathbf{2940}_u + \mathbf{3528}_u + \mathbf{4704}''_u) \\
&\qquad\quad + a^4b^{16}\bigl[2\,\mathbf{21}_u
   - (a^6-1)(\mathbf{224}_u + \mathbf{490}_u + \mathbf{735}_u + \mathbf{3360}_u) \\
&\qquad\qquad - a^6(\mathbf{2940}'_u + \mathbf{9408}_u + \mathbf{17640}_u + \mathbf{19404}_u + \mathbf{20160}_u) \\
&\qquad\qquad + a^6(a^6-1)(\mathbf{756}_u + \mathbf{4725}_u + \mathbf{37044}_u
   + \mathbf{116424}'''_u + \mathbf{138600}_u) \\
&\qquad\qquad + a^{12}(\mathbf{24696}_u + \mathbf{346500}_u + \mathbf{429429}_u
   + \mathbf{952875}_u + \mathbf{1057056}_u)\bigr]\Bigr)\,q 
\end{split}
\end{equation}

The dual theory has gauge group $\USp(2)$ and its superconformal index up to order $O(q)$ is:
\begin{equation}
\begin{split}
\mathcal{I}_{\text{JK}}^{\text{gauge}}
&= 1 + a^2b^2\,\mathbf{21}_u\,q^{1/8}
   + a^4b^4\bigl(\overline{\mathbf{35}}_u + \overline{\mathbf{196}}_u\bigr)\,q^{1/4}
   + a^6b^6\bigl(\overline{\mathbf{588}}_u + \overline{\mathbf{1176}}_u\bigr)\,q^{3/8} \\
&\quad + a^2b^8\bigl(-\mathbf{7}_u + a^6(\mathbf{490}'_u + \mathbf{4704}'_u + \mathbf{5292}''_u)\bigr)\,q^{1/2} \\
&\quad + \Bigl(-\frac{\mathbf{7}_u}{b^6} - a^4b^{10}(\mathbf{35}_u + \mathbf{112}_u)
   + a^{10}b^{10}(\mathbf{7056}_u + \mathbf{19404}'_u + \mathbf{25200}_u)\Bigr)\,q^{5/8} \\
&\quad + \Bigl(-\frac{a^2}{b^4}(\mathbf{35}_u + \mathbf{112}_u)
   - a^6b^{12}(\overline{\mathbf{224}}_u + \overline{\mathbf{490}}_u + \overline{\mathbf{882}}_u) \\
&\qquad\quad + a^{12}b^{12}(\overline{\mathbf{4116}}_u + \overline{\mathbf{50400}}_u
   + \overline{\mathbf{60984}}_u + \overline{\mathbf{103950}}'_u)\Bigr)\,q^{3/4} \\
&\quad + \Bigl(a^2b^{14}
   - \frac{a^4}{b^2}(\overline{\mathbf{224}}_u + \overline{\mathbf{490}}_u + \overline{\mathbf{882}}_u)
   - a^8b^{14}(\mathbf{784}_u + \mathbf{2940}_u + \mathbf{3528}_u + \mathbf{4704}''_u) \\
&\qquad\quad + a^{14}b^{14}(\mathbf{52920}_u + \mathbf{169884}_u
   + \mathbf{247500}_u + \mathbf{355740}_u)\Bigr)\,q^{7/8} \\
&\quad + \Bigl(2 - \frac{1}{a^6} + \mathbf{48}_u + 2a^4b^{16}\,\mathbf{21}_u
   - a^6(\mathbf{784}_u + \mathbf{2940}_u + \mathbf{3528}_u + \mathbf{4704}''_u) \\
&\qquad\quad - a^{10}b^{16}(\mathbf{2940}'_u + \mathbf{9408}_u + \mathbf{17640}_u
   + \mathbf{19404}_u + \mathbf{20160}_u) \\
&\qquad\quad + a^{16}b^{16}(\mathbf{24696}_u + \mathbf{346500}_u + \mathbf{429429}_u
   + \mathbf{952875}_u + \mathbf{1057056}_u)\Bigr)\,q 
\end{split}
\end{equation}

\end{appendix}

\end{document}